\documentclass[notitlepage, reprint, superscriptaddress, aps, prl, onecolumn, nofootinbib, nobibnotes, floatfix]{revtex4-2}

\usepackage{titlesec}
\titleformat{\section} 
{\large\bfseries\color{black}}
{\thesection}{1em}{}
\pdfoutput=1
\usepackage{color,graphicx}
\usepackage{comment}
\usepackage{datetime}
\usepackage{amsmath, mathtools, amssymb, bigints}
\usepackage{subcaption}
\usepackage{bm}
\usepackage{dsfont}
\usepackage{booktabs}
\usepackage{float}
\usepackage[font=small,labelfont=bf, justification=RaggedRight]{caption}
\usepackage[colorlinks=true,citecolor=blue, linkcolor=blue,urlcolor=blue]{hyperref}

\newcommand{\BM}{\text{BM}}

\newcommand{\MFL}{\text{MFL}}
\newcommand{\CL}{\text{CL}}

\usepackage[dvipsnames]{xcolor}

\begin{document}

\title{Universal Theory of Strange and Bad Metals from Spatially Random Interactions with Underdamped Quantum-Critical Bosons}

\author{Aaron Kleger}
\email{Aaron.R.Kleger.GR@dartmouth.edu}
\author{Rufus Boyack}
\email{Rufus.Boyack@dartmouth.edu}

\affiliation{Department of Physics and Astronomy, Dartmouth College, Hanover, New Hampshire 03755, USA}

\begin{abstract}
One of the  most important challenges in condensed matter physics is understanding the macroscopic properties of strange metals, particularly why their resistivity remains linear in temperature over several decades. We propose that a (2+1)d theory of fermions coupled to \emph{underdamped} quantum-critical bosons through a spatially disordered interaction provides this long-sought-after theoretical explanation. The key novelty of our work is to establish the appropriate separation of scales so that the bosons are underdamped. Using this framework, we obtain two distinct forms for the fermionic self-energy: a conformal-limit solution,  $\Sigma\sim\sqrt{\omega}$, and a marginal Fermi-liquid solution, $\Sigma\sim\omega\log(1/\omega)$, corresponding to regimes above and below the Mott-Ioffe-Regel limit, respectively. These solutions produce Planckian scattering, exhibit exact $T$-linear resistivity, $\omega/T$ scaling in the optical conductivity, as well as $\sim T\log(1/T)$ heat capacity at low temperatures. Moreover, our theory naturally unifies strange- and bad-metal phenomena within a single framework, suggesting that we are now approaching a new epoch in understanding the physics of strongly correlated matter. 
\end{abstract}

\maketitle

\section{Introduction}

Understanding unconventional superconductors remains one of the central and most important open problems in condensed matter physics~\cite{Varma2020}. Despite their diverse microscopic properties, cuprate~\cite{Ayres2021}, heavy-fermion~\cite{Bruin2013}, and iron-pnictide~\cite{DoironLeyraud2009} superconductors, as well as twisted bilayer graphene~\cite{Polshyn2019, Wu2019}, share the emergence of a `strange-metal' phase above the superconducting transition temperature. The strange-metal phase is famously characterized by distinctly non-Fermi-liquid physics~\cite{Lohneysen2007, Lee2018}, including $T$-linear resistivity~\cite{Varma2020}, $\omega/T$ scaling in the optical conductivity, and $T\log(1/T)$ heat capacity at low temperatures~\cite{Orenstein2000, McGreevy2010, Cha2020, Phillips2022, Patel2023, Savitsky2025}. Moreover, the connection between strange-metal behavior and unconventional superconductivity is supported by the observed scaling between the coefficient of the $T$-linear resistivity and the superconducting transition temperature~\cite{Hussey2011}, suggesting that the interactions responsible for the strange-metal phase may also be intimately connected to the pairing mechanism~\cite{Savitsky2025}. Indeed, a full understanding of the normal phase is believed to be one of the most decisive ways to understand the pairing mechanism in unconventional superconductors~\cite{Varma2020}. 

Several aspects of the strange-metal phase remain poorly understood. Most notably, the origin of the $T$-linear resistivity over such a broad range of temperatures has been elusive~\cite{Gurvitch1987, Mackenzie1996, Fournier1998, Nguyen2021}. Another significant unresolved question is why in certain materials the $T$-linear resistivity persists to high temperatures and eventually extends beyond the Mott-Ioffe-Regel (MIR) bound without saturation or, in certain cases, without any change in slope~\cite{Ioffe1960, Mott1972, Hussey2004}. Materials in this high-temperature regime, characterized by a resistivity $\rho\gtrsim \hbar/e^2$ in two spatial dimensions, are often referred to as `bad metals'~\cite{Emery1995, Gunnarsson2003, Deng2013, Vucicevic2015, Kokalj2017, Delacretaz2017, Lucas2017, Mousatov2019, Pustogow2021, Valentinis2026, Kleger2026}. The bad-metal regime lies beyond the MIR limit, where conventional Drude and Boltzmann transport theory breaks down. This raises another fundamental and longstanding question of how electronic excitations can scatter on length scales shorter than the interatomic lattice spacing and Fermi wavelength~\cite{Phillips2022}.

The significance of strange- and bad-metal behavior extends beyond unconventional superconductivity. Strange metals form a distinct non-Fermi-liquid phase of matter found across a wide class of materials that exhibit universal macroscopic properties. This phase often occurs~\cite{Mirarchi2022} between an ordered phase and a disordered Fermi-liquid phase and spans a broad parameter regime accessible by tuning a non-thermal parameter such as doping, pressure~\cite{Shen2022}, or external magnetic field~\cite{Bruin2013, Grigera2001, Lester2021}. Taken together, the characteristic properties of strange metals suggest~\cite{Valla1999, Marel2003} a connection to local quantum-criticality~\cite{Si2001, Gegenwart2008, Hu2024}, raising the possibility that strange metallicity reflects universal features of quantum-critical metals rather than a material-specific phenomenon. A central goal of this work is to develop a framework that captures the universal characteristics of strange and bad metals across a broad range of strongly correlated materials. Recent theoretical progress~\cite{Esterlis2019, Wang2020, WangChubukov2020, Patel2023, Valentinis2023, Valentinis2023b, Li2024, Pan2021, Gnezdilov2025, Patel2018b, Chowdhury2020, Valentinis2026} has begun to identify the common theoretical concepts underlying these systems, pointing toward a new paradigm~\cite{Esterlis2025} for understanding strongly interacting electronic matter. 

In what follows, we outline the universal empirical features that a theory of strange and bad metals must capture. We revisit the seminal framework of spatially random interactions between electrons and overdamped quantum-critical bosons~\cite{Patel2023}. In contrast to the current focus in the field~\cite{Valentinis2026, Patel2025, Kryhin2026, Guo2026}, we investigate the scenario of underdamped critical bosons and discover solutions that provide a unified description of strange- and bad-metal phenomena. In particular, we demonstrate that $T$-linear resistivity does not require Landau damping of the bosonic modes; instead, underdamped critical bosons are essential for obtaining the high-frequency and both the low and high-temperature signatures observed in cuprate superconductors~\cite{Michon2023, Marel2003, Kaminski2005, Valla1999, Zhu2008}. Our solutions continuously connect the low-temperature marginal Fermi liquid (MFL) regime to the high-temperature conformal solution of the Sachdev-Ye-Kitaev (SYK) model within a single electron-boson framework, thus providing a continuous cross-over~\cite{Chowdhury2022} from strange- to bad-metal behavior. These results provide a unified framework that explains the anomalous transport properties of strange and bad metals.

\section{Theory of a Quantum-Critical Metal}

\subsection{Model}

As discussed earlier, the strange- and bad-metal regimes are typically accessed by tuning a non-thermodynamic parameter, and their phase diagram is reminiscent of the quantum-critical fan that exists above a quantum-critical point. Furthermore, the experimentally observed scale-free $T$-linear resistivity~\cite{Phillips2022}, $\omega/T$ scaling of the optical conductivity, and logarithmic heat capacity $C\sim T\log(1/T)$ collectively suggest an important role of quantum criticality in these phases. A natural theoretical framework is therefore provided by an electronic Fermi-surface coupled to a critical boson through a Yukawa interaction. To describe the quantum-critical metal, we use the (2+1)d Yukawa-Sachdev-Ye-Kitaev (Y-SYK) model~\cite{Esterlis2019, Wang2020,Patel2023}. The imaginary-time Lagrangian is given by
\begin{equation} 
\mathcal{L}=\psi_{i \sigma}^{\dagger}\left(\partial_\tau+\varepsilon(-i \nabla)-\mu\right) \psi_{i \sigma}+\frac{v_{i j}(\boldsymbol{r})}{\sqrt{N}} \psi_{i \sigma}^{\dagger} \psi_{j \sigma}  +\frac{1}{2}\left(\left(\partial_\tau \phi_l\right)^2+c^2\left(\nabla \phi_l\right)^2+m_{B,0}^2 \phi_l^2\right)+\frac{g_{i j l}(\boldsymbol{r})}{\sqrt{NM}}  \phi_{l} \psi_{i \sigma}^{\dagger}  \psi_{j \sigma},
\label{eq:Lagrangian} 
\end{equation}
where we consider $N$ electrons $\psi_{i\sigma}$ coupled to $M$ bosons $\phi_l$ and the flavor $i,j =1, \cdots, N$; $l =1, \cdots, M$  and spin $\sigma=\uparrow \downarrow$ indices are summed over. The bare boson mass is denoted by $m_{B,0}$. Throughout, we will use Natural units: $\hbar=k_{B}=1$. 

The defining phenomenological signatures of strange and bad metals place strong constraints on the underlying electron-scattering mechanisms. In the Galilean-invariant continuum limit, finite DC conductivity requires breaking translation invariance~\cite{Maslov2017}, implying that the underlying electron-scattering processes must relax the electronic momentum. This requires the electron-boson interaction itself to break translation invariance, since fermionic potential disorder alone cannot account for the $T$-linear resistivity. Although the marginal Fermi-liquid (MFL) phase can be realized~\cite{Patel2023} with a translation-invariant interaction between fermions and bosons in the presence of fermion potential disorder, the resistivity is not linear in temperature. The simplest realization of a quantum-critical metal with $T$-linear resistivity is therefore a spatially disordered electron-boson interaction and potential~\cite{Sachdev2024}. We take the couplings $g_{ijl}$ and $v_{ij}$ to be sampled from Gaussian orthogonal ensembles with zero mean and nonzero local variances:
\begin{align}
\left\langle g_{i j l}(\bm{r}) g_{a b c}\left(\bm{r}^{\prime}\right)\right\rangle &= 2 g^2 \delta\left(\bm{r}-\bm{r}^{\prime}\right)\delta_{ia}\delta_{jb}\delta_{lc},  \label{eq:gVariance}\\ 
\langle v_{ij}(\bm{r}) v_{ab}(\bm{r}^{\prime})\rangle &= 2v^2\delta({\bm r}-{\bm r}')\delta_{ia}\delta_{jb}.
\label{eq:vVariance}
\end{align} 

In the large-$N,M$ limit the disorder-averaged model is exactly solvable. The normal-state saddle-point equations for the imaginary-time fermionic and bosonic self-energies are $\Sigma(\tau)=(g^2\mathcal{D}(\tau) +v^2)\mathcal{G}(\tau)$ and $\Pi(\tau)=-2(N/M)g^2\mathcal{G}(\tau)\mathcal{G}(-\tau)$, respectively. Due to the local nature of the variances in Eqs.~\eqref{eq:gVariance}-\eqref{eq:vVariance}, the momentum integrals for the fermionic and bosonic self-energies factorize, leading to manifestly momentum-independent self-energies. The effective fermionic and bosonic propagators $\mathcal{G}$ and $\mathcal{D}$ are independent of momentum and given by~\cite{Gnezdilov2025, Esterlis2021}:
\begin{align}
\label{eq:Geff-maintext}
\mathcal{G}\left(i \omega_n\right) &= \int \frac{1}{i \omega_n-\xi_{\bm{k}}-\Sigma\left(i \omega_n\right)} \frac{d^2 k}{(2 \pi)^2},  \\ \mathcal{D}\left(i \Omega_m\right) &= \int_0^{\omega_D / c} \frac{1}{\Omega_m^2+c^2q^2+m_{B,0}^2-\Pi\left(i \Omega_m\right)} \frac{q d q}{2 \pi}.  
\label{eq:Deff-maintext}
\end{align}
Equations~\eqref{eq:Geff-maintext} and~\eqref{eq:Deff-maintext} define the Dyson equations for the effective propagators. Here, $\omega_D$ and $c$ are analogous to the Debye frequency and the speed of sound, respectively. The ratio $\omega_D/c$ acts as a bosonic ultraviolet (UV) cutoff. In momentum space, the electronic dispersion relation is $\xi_{\bm{k}}=\varepsilon_{\bm{k}}-\mu;\;\varepsilon_{\bm{k}}\equiv k^2/(2m)\in[0,\Lambda]$, with bandwidth $\Lambda$ and chemical potential $\mu$.

\subsection{Tuning to the Quantum-Critical Point}

In order to access the quantum-critical regime, we introduce a constraint that fixes the boson density~\cite{Esterlis2021}. To determine the renormalized boson mass, $m_{B}(T)$, defined by $m_B^{2}(T)\equiv m_{B,0}^2-\Pi(0)$ (where $\left.\Pi(0)\equiv\Pi(i\Omega_{m})\right|_{\Omega_{m}=0}$), we equate the boson density at zero temperature with that at nonzero temperature $T$: 
\begin{equation} 
\frac{1}{2\pi}\int_{-\infty}^{\infty} \left. \mathcal{D}(i\Omega) d\Omega\right|_{m_B=0}= T\sum_{\Omega_m}\mathcal{D}(i\Omega_m)|_{m_B=m_B(T)}.
\label{eq:Boson-constraint}
\end{equation}
In the absence of dynamical renormalization from the bosonic self-energy, the boson mass at criticality is found to be $m_B=\text{arccosh}(3/2)T$ for $T\ll\omega_D$~\cite{Kleger2026}. This is in contrast to the solution $m_B\approx\sqrt{T\omega_D}$ (up to logarithms) observed in the conventional overdamped scenario~\cite{Esterlis2021}.

Performing the momentum integration in Eq.~\eqref{eq:Deff-maintext}, the effective bosonic propagator in the underdamped regime is given by $\mathcal{D}(i\Omega_m)=\log(1+\omega_D^2/(\Omega_m^2+m_B^2))/(4\pi c^2)$. At criticality, we observe a striking and unexpected correspondence between the underdamped effective bosonic propagator and the two-point correlation functions of finite-temperature conformal field theory~\cite{TsvelikBook}. In the imaginary-time domain, $|\tau|\in[0,1/T]$, we find that the effective propagator has the approximate low-energy ``conformal'' form~\cite{Ye1993, Parcollet1999, Sachdev2015} for temperatures $T\ll\omega_D$: 
\begin{equation}
\mathcal{D}(\tau)\approx\frac{1}{4\pi c^2}\left|\frac{\pi T}{\sin(\pi T\tau)}\right|. 
\end{equation}
The effective bosonic propagator therefore has the scaling dimension $\Delta_{\phi}\simeq1/2$ for $T\ll\omega_D$. At zero temperature, it takes the exact low-energy conformal form of local criticality. In this regime $\mathcal{D}(\tau)=(1-e^{-|\tau|\omega_D})/(4\pi c^2 |\tau|)\sim1/(4\pi c^2 |\tau|)$ for $|\tau|\omega_D\gg1$; thus $\Delta_{\phi}=1/2$ is exact in the infrared regime. As we will show, the emergence of conformal physics at criticality in the underdamped scenario plays a key role in obtaining scale-free $T$-linear resistivity both below and well above the MIR bound. This allows us to realize the distinct physics encoded in the emergent conformal solutions of the SYK model~\cite{Sachdev2015, Maldacena2016, Sarosi2017}.

\section{Strange and Bad-Metal Solutions Across the MIR Bound}

\subsection{Strange Metals}

Strange metals are characterized~\cite{Phillips2022} by $T$-linear resistivity below the Mott-Ioffe-Regel bound, restricting this regime to $\rho< \hbar/e^2$ in two spatial dimensions~\cite{Li2024}. Consequently, the electronic self-energy of strange metals satisfies the separation of scales $|\Sigma_R^{\prime\prime}(\omega)|\ll\Lambda, \varepsilon_F$ at low temperatures. When the separation of scales $\mu, \Lambda-\mu  \gg  |i\omega_n - \Sigma(i\omega_n)|$ is satisfied, the effective fermionic propagator in Eq.~\eqref{eq:Geff-maintext} can be evaluated by extending the momentum integration limits to positive and negative infinity. In this infinite-bandwidth limit, $\mathcal{G}\left(i \omega_n\right) = -i\pi\nu\operatorname{sgn}\left(\omega_n\right)$, where $\nu\equiv m/(2\pi)$ is the electron density of states per spin in 2d, and thus the effective propagator depends only on frequency and is independent of the fermionic self-energy~\cite{Patel2018, Gnezdilov2025}.

The dynamical part of the bosonic self-energy is defined by $\delta\Pi(i\Omega_m)\equiv\Pi\left(i \Omega_m\right)-\Pi(0)$. In analogy to the classical harmonic oscillator, we take the underdamped regime to be defined by $|\delta\Pi(i\Omega_m)|\ll\Omega_m^2 +m_B^2$, such that the bare critical terms in the effective bosonic propagator dominate over the dynamical part of the self-energy\footnote{In certain instances this definition coincides with the classical definition $c_d<2\sqrt{c^2q^2+m_B^2}$, where $c_d$ is the coefficient of the Landau damping term $\delta \Pi(i\Omega_m)=-c_d|\Omega_m|$.}. In the infinite-bandwidth limit, it is given by~\cite{Esterlis2021, Patel2023, Gnezdilov2025}: $\delta\Pi(i\Omega_m)=-2 \pi \nu^2 g^2(N/M)\left|\Omega_m\right|$, which physically corresponds to Landau damping of the bosonic modes~\cite{Patel2024} with dynamical critical exponent $z=2$~\cite{Esterlis2021, Patel2023}. As in Ref.~\cite{Gnezdilov2025}, we define the dimensionless electron-boson coupling constant $\lambda\equiv\nu g^2 /(4 \pi c^2)$. 

For an undamped boson tuned to quantum criticality, the zero-temperature fermionic self-energy is given by~\cite{Kleger2026}:
\begin{equation} 
\Sigma(i\omega)\simeq -2i\lambda\omega\log\left(\frac{e\omega_D}{|\omega|}\right), \quad |\omega| \ll \omega_{D}.
\label{eq:MFL-UD1-maintext}
\end{equation}
Importantly, the fermionic self-energy takes the form of a marginal Fermi liquid~\cite{Varma1989}, despite the absence of Landau damping in the bosonic propagator~\cite{Kleger2026}. The imaginary part of the retarded self-energy determines the single-particle scattering rate, $1/(2\tau(\omega))\equiv|\Sigma_R^{\prime\prime}(\omega)|$, which takes the form
\begin{equation}
\Sigma_R^{\prime\prime}(\omega) = -\pi \lambda
\begin{cases}
    |\omega|,\quad|\omega|<\omega_D,\\
    \omega_D,\quad|\omega|\geq\omega_D.
\end{cases}
\label{eq:MFL-form-0T}
\end{equation}
In the underdamped framework, the MFL UV cutoff is given by the bosonic UV cutoff $\omega_c=\omega_D$. The result in Eq.~\eqref{eq:MFL-form-0T} contains both the linear-in-energy and high-energy plateau found in the single-particle scattering rate determined from ARPES measurements of Bi-2212~\cite{Valla1999,Kaminski2005} and LSCO~\cite{Zhu2008}.

At nonzero temperatures, the fermionic self-energy exhibits exact $\omega/T$-scaling. For frequencies $|\omega|<\omega_D$, we find 
\begin{equation}
\Sigma_R^{\prime\prime}(\omega,T) = -\pi \lambda T \log\left(3+2\cosh(\omega/T)\right), \quad T\ll\omega_D.
\label{eq:Im-SE-SM-simplified-main-text}
\end{equation}
Thus, the electronic scattering rate exhibits exact Planckian characteristics, with both $T$-linear and $\omega/T$-scaling within the intermediate scaling regime. For $|\omega|\ll T\ll\omega_D$ we find that\footnote{The magnitude of $\Sigma_R^{\prime\prime}(0,T)$ can be tuned by varying the parametrization of the quantum critical boson mass; see Supplementary Materials.} $\Sigma^{\prime\prime}_R(\omega,T) = -\pi\log(5) \lambda T$ while for  $T\ll|\omega|\ll\omega_D$, $\Sigma_R^{\prime\prime}(\omega,T)\sim-\pi\lambda|\omega|$. To date, no quantum-critical electron-boson framework has been able to capture the exact $T$-linear and linear-in-frequency scaling regime without concomitant logarithmic deviations from MFL phenomenology.

\begin{figure}[h]
\centering\includegraphics[width=.4\columnwidth]{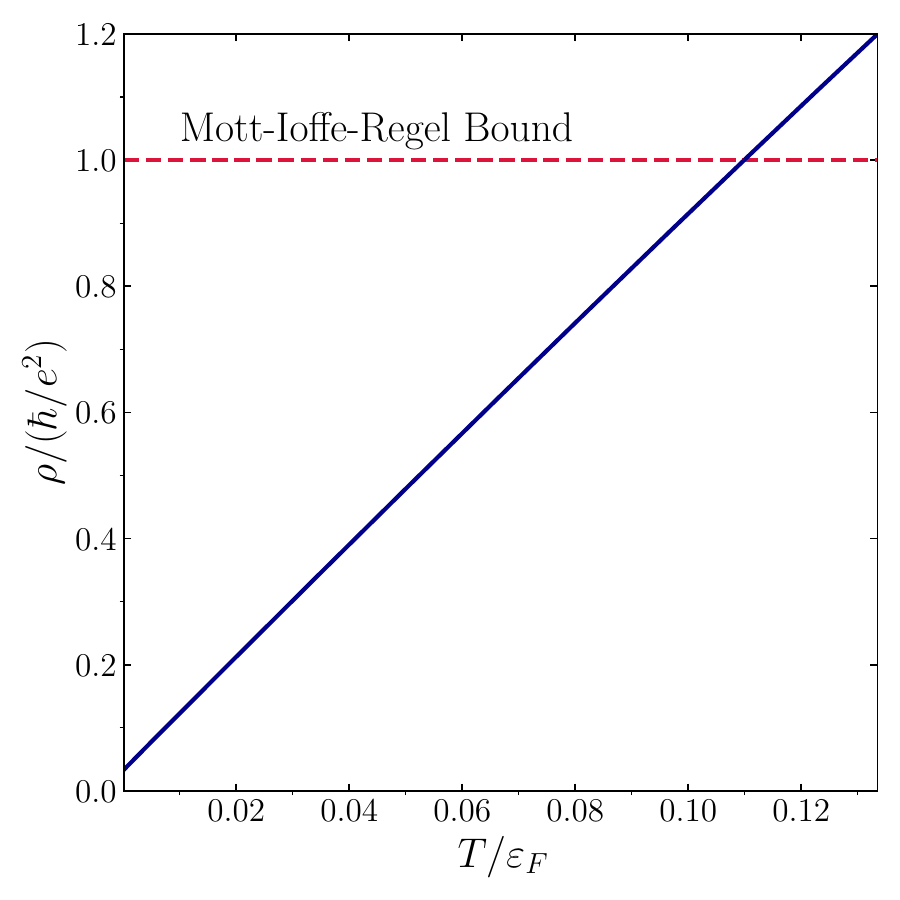}
\caption{Theoretical normal-state strange- and bad-metal resistivity as a function of temperature. The dashed line denotes the Mott-Ioffe-Regel (MIR) bound.}
\label{fig:Rho}
\end{figure}

\begin{figure}[h]
\centering\includegraphics[width=.8\columnwidth]{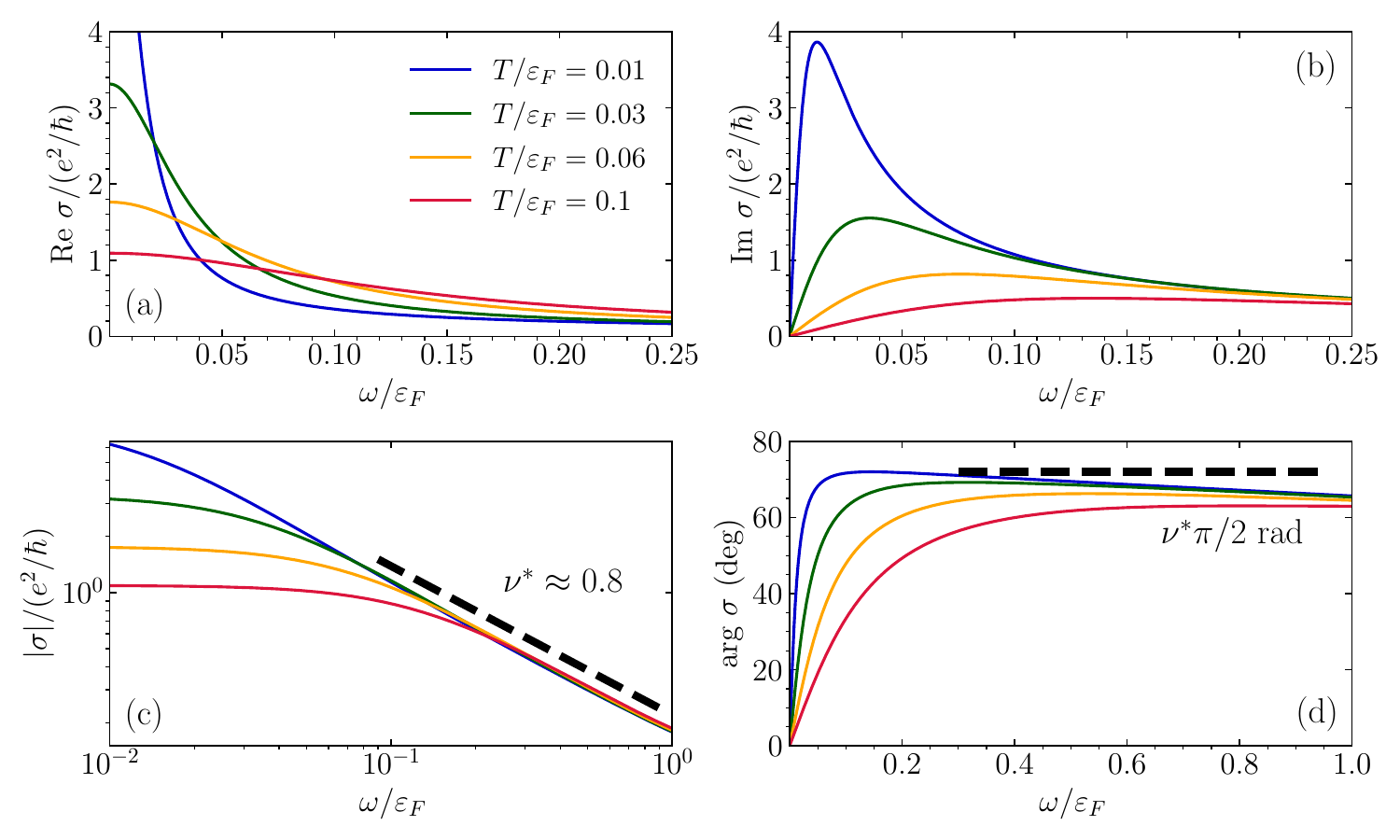}
\caption{Theoretical normal-state optical conductivity for various temperatures.} 
\label{fig:optical-conductivity}
\end{figure}

Because the Yukawa interaction explicitly breaks translation invariance, the fermionic self-energy in Eq.~\eqref{eq:Im-SE-SM-simplified-main-text} directly determines the transport scattering rate. Specifically, in the regime $T\ll\omega_D$, the scattering rate satisfies the scaling relation $1/\tau_{\text{tr}}(\omega)=Tf(\omega/T)$ for frequencies $|\omega|<\omega_D$. Indeed, for low temperatures, where $|\Sigma_R^{\prime\prime}(\omega)|\ll\Lambda,\varepsilon_F$, the electrical conductivity obeys a Drude-like relation $\sigma\propto\tau$, and is therefore scale-free as a result of the scaling-form of the microscopic single-particle scattering rate given in Eq.~\eqref{eq:Im-SE-SM-simplified-main-text}. The electrical resistivity and the heat capacity for this solution are $\rho_{\MFL}\simeq 2\pi\lambda (k_{B}T/\varepsilon_F )(h/e^2)$ and $C/T\sim (4\pi^2/3)\lambda\nu k_B^2 \log(\hbar\omega_D/(ek_BT))$ (we have omitted the higher-order in temperature Sommerfeld and critical-boson contributions) respectively. The optical scattering rate is given by $\hbar/\tau_{\text{tr}}\simeq4\pi\lambda k_B T$ in the DC limit. In Fig.~\ref{fig:Rho}, we present the resistivity of the strange metal, calculated using the solution derived in this section. Remarkably, the resistivity exceeds the MIR bound (discussed in the next subsection), entering the bad-metal regime without any discernible change in slope. For direct comparison with notable strange metals, we set $\lambda=0.23$ in Figs.~\ref{fig:Rho} and~\ref{fig:optical-conductivity}, which is deduced by comparison with the scattering rate determined from ARPES~\cite{Kaminski2005, Valla1999} and optical response~\cite{Michon2023} of the optimally doped Bi-2212 and overdoped LSCO cuprates respectively. The high-frequency cutoff in the single-particle scattering rate is $\omega_c\simeq0.3\!-\!0.5$ eV in the cuprate superconductors~\cite{Kaminski2005, Valla1999, Michon2023, Zhu2008, NormanChubukov2006}; accordingly, we set $\omega_D=\varepsilon_F$ to reflect the non-adiabaticity between the fermionic and bosonic energy scales. In Figs.~\ref{fig:Rho} and~\ref{fig:optical-conductivity} we introduce an elastic scattering rate $\Gamma_0=\pi \nu v^2= 5\cdot10^{-3}\omega_D$ arising from spatially random potential disorder to reflect the residual resistivity commonly found in strange metals. 

In Figs.~\ref{fig:optical-conductivity}(a) and \ref{fig:optical-conductivity}(b), we present the optical conductivity for the underdamped scenario. The modulus and phase of the optical conductivity are presented in Figs.~\ref{fig:optical-conductivity}(c) and \ref{fig:optical-conductivity}(d). We find that the modulus follows a sub-linear power law $|\sigma|\sim1/\omega^{\nu^*}$ with an exponent $\nu^*\approx0.8$, and the phase plateaus at a value slightly below $\arg \sigma \simeq \nu^* \pi/2$ over intermediate frequencies. Our result for the scaling exponent agrees with the intermediate to high-frequency scaling of the optical conductivity observed in Ref.~\cite{Michon2023}.

\subsection{Mott-Ioffe-Regel Limit} 

Here we develop a theoretical framework for how strange metals maintain $T$-linear resistivity through the MIR limit without saturation or, in certain cases, a change in slope. Although the notion of the Mott-Ioffe-Regel limit is well established, its precise definition as a bound on electronic transport remains ambiguous. Several common criteria can be formulated in terms of the mean free path $\ell$ (defined as the average distance traversed between collisions) of the electronic excitation~\cite{Hussey2004}: (i) $\ell\sim\lambda_F$, (ii) $\ell\sim a$, and (iii) $\ell\sim 1/k_F=\lambda_F/(2\pi)$, where $\lambda_F$, $a$, and $k_F$, are the Fermi wavelength, lattice spacing, and Fermi wavevector, respectively. In terms of a scattering time $\tau\equiv \ell/v_F$, the bounds are given by (i) $\Lambda\tau\sim 2\pi\hbar$, (ii) $\Lambda\tau\sim 2\hbar$, and (iii) $\Lambda\tau\sim \hbar$, respectively. Criterion (ii) is given for the case of particle-hole symmetry with the Fermi velocity satisfying $\hbar v_F/a=\varepsilon_F=\Lambda/2$. Each criterion for the scattering rate corresponds to a distinct resistivity bound, summarized in Table~\ref{tab:Comparison-table}.
\begin{table}[H]
\centering
\setlength{\tabcolsep}{8pt}
\begin{tabular}{ccccc}
\toprule
\textbf{$\ell$} & \textbf{$\Lambda\tau/\hbar$}\quad & \textbf{$\rho_{\text{Drude}}/(h/e^2)$} & \textbf{$\rho_{\text{Exact}}/(h/e^2)$}& \textbf{$\rho_{\text{Drude}}/\rho_{\text{Exact}}$} \\
\midrule
$\lambda_F$ & $2\pi$ & $1/(2\pi)$ & $0.159$ & $0.998$ \\
$a$ & $2$ & $1/2$ & $0.521$ & $0.959$\\
$1/k_F$ & $1$ & $1$ & $1.222$ & $0.818$\\
\bottomrule
\end{tabular}
\caption{Comparison of values for the Mott-Ioffe-Regel bound.}
\label{tab:Comparison-table}
\end{table}

Table~\ref{tab:Comparison-table} compares the corresponding resistivity bounds using a Drude-like relation and the exact finite-bandwidth formula~\cite{Kleger2026} (discussed further in our Supplementary Materials). The condition $\ell\sim1/k_F$ (or equivalently $\Lambda\tau\sim\hbar$) gives a resistivity closest to the quantum unit $h/e^2$. It is fascinating that none of the commonly used bounds precisely corresponds to the quantum unit, except when estimated by the Drude formula. It is also notable that the crossover between Drude-like and bad metal-like scaling in the resistivity occurs in the vicinity of this bound. In the present framework, the condition $\Lambda\tau\sim\hbar$ also marks the crossover from dispersion-dominated strange- and bad-metal behavior to a strongly incoherent regime dominated by the electronic self-energy.

For incoherent metals, the electronic bandwidth and Fermi-energy provide the relevant energy scales for determining whether the excitation lifetime is short or long. When both $\Lambda\tau\gg\hbar$ and $\varepsilon_F\tau\gg\hbar$ are satisfied, electronic transport is described by a Drude formula: $\sigma_{\text{Drude}}=e^2\varepsilon_F\tau/(\pi\hbar^2)\gg e^2/h$. In fact, for an isotropic 2d metal in the Galilean-invariant continuum limit with finite Fermi energy and infinite bandwidth (no UV cutoff), the resistivity is bounded:
\begin{equation}
\rho<\pi\times h/e^2.
\label{eq:Resistivity-bound}
\end{equation}
For an infinite electronic bandwidth, the conductivity is bounded from below. Exceeding the MIR bound without resistivity saturation therefore requires a finite bandwidth. In Ref.~\cite{Kleger2026}, we investigated electronic transport in the finite-bandwidth scenario, and found that the Drude-like result for conductivity crosses over to the form $\sigma_{\text{BM}}=2(e\Lambda\tau/\pi)^2\hbar^{-3}\ll e^2/h$ for $\Lambda\tau\ll\hbar$, describing a bad metal well beyond the MIR bound.

It is important to note that the most common definition of the MIR limit (which we refer to simply as `the MIR limit' below), cited in several notable works~\cite{Phillips2022}, is defined by the mean free path being on the order of the Fermi wavelength. This definition corresponds to the weakest bound on resistivity $\rho_{\text{MIR}}\simeq \hbar/e^2$, which is significantly below the quantum unit. Consequently, the marginal Fermi liquid solution described in the previous section is applicable both below and above the MIR bound. This provides a possible explanation of how certain critical metals maintain the $T$-linear scaling across the MIR bound, particularly without any change in slope. For systems with the separation of scales $T\ll\omega_D, \Lambda$ and $1/\tau(\omega)\ll\Lambda$, the marginal Fermi liquid solution is applicable, which allows for resistivity both below and above the MIR bound, as in Fig.~\ref{fig:Rho}. In which case, crossing the MIR bound does not coincide with a change in the physical regime, and thus a change in the resistivity slope is not concomitant with crossing the MIR bound. For certain materials, such as the cuprate superconductors LSCO and Bi-2212, we find the provided separation of scales to be applicable. 

Nevertheless, strange- and bad-metal behavior occurs in a wide variety of systems where the separation of scales required for a marginal-Fermi-liquid description may not hold. This raises the following question: how can a quantum-critical metal exhibit $T$-linear resistivity both below and well above the MIR bound? Below the MIR bound, $\hbar/\tau\ll\varepsilon_F, \Lambda$ and transport is described by a Drude formula: $\sigma_{\text{Drude}}\propto\tau$. Thus, $T$-linear resistivity requires a $T$-linear transport scattering rate, $1/\tau_{\text{tr}}\sim T$ (sometimes referred to as a Planckian scattering rate), as naturally  realized in a marginal Fermi liquid, where $\Sigma(i\omega)\sim\omega\log(1/\omega)$ so that $|\Sigma_R^{\prime\prime}(\omega)|\propto|\omega|$. However, in light of the pioneering work of Ref.~\cite{Parcollet1999} and the recent work of Ref.~\cite{Kleger2026}, it is evident that the same scaling argument does not apply for resistivity exceeding the quantum unit. Instead, $\sigma_{\text{BM}}\sim \tau^2$, and thus $\Sigma(i\omega)\propto\sqrt{\omega}$ is crucial for obtaining $T$-linear resistivity well above the MIR bound. Solutions of this kind have been discovered in the low-energy emergent conformal limit of disordered $t$-$J$~\cite{Parcollet1999} and SYK~\cite{Patel2018} models but not previously within a quantum-critical theory of interacting electrons and bosons. 

Here we find the square-root ``conformal'' solution to emerge within the critical framework of underdamped bosons. Within this framework, the conformal solution naturally arises for high temperatures or strong coupling where the fermionic self-energy dominates over the electronic dispersion, thus providing a universal description, independent of the electronic bandstructure. Indeed, we find that this solution possesses $T$-linear resistivity and emerges for $\ell\ll1/k_F$. In the following subsection, we present these results, and in Sec.~\ref{sec:Implications} we discuss the broader implications of our work for critical metals.

\begin{figure}[h]
\centering\includegraphics[width=.5\columnwidth]{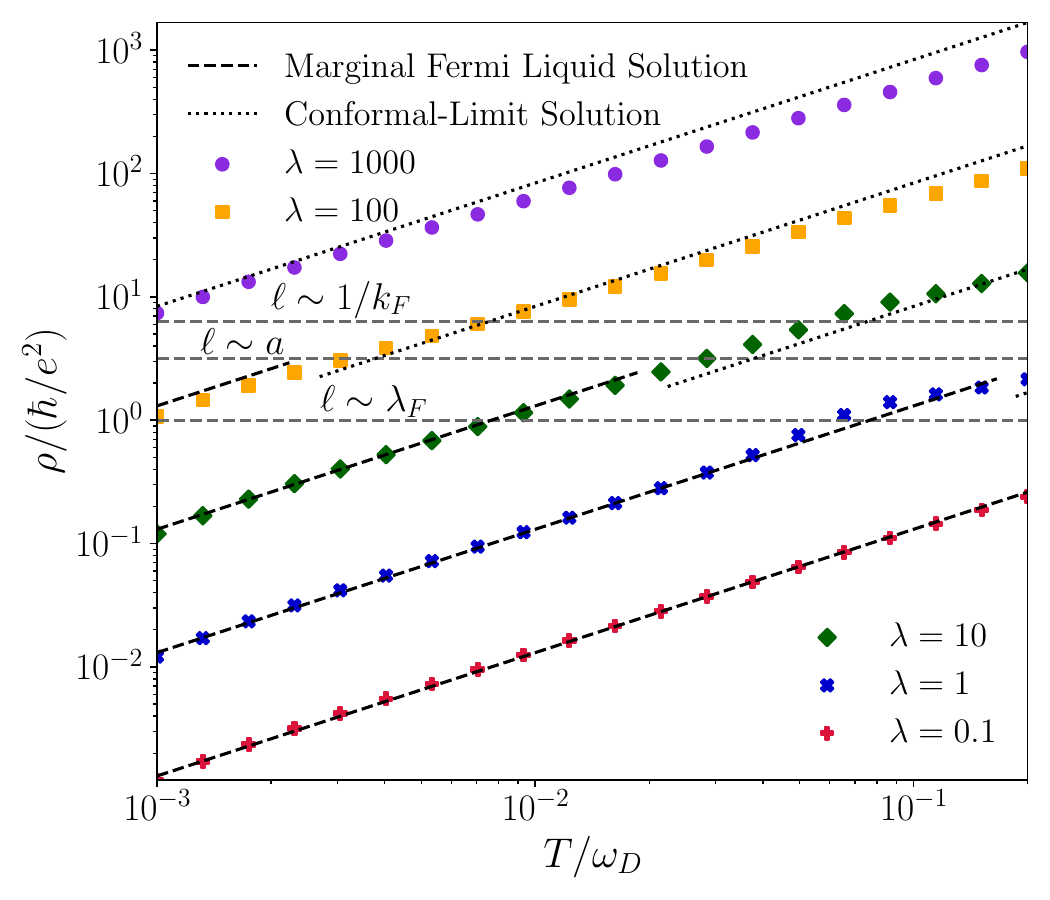}
\caption{Theoretical normal-state strange- and bad-metal resistivities as functions of temperature evaluated with a finite electronic bandwidth. We use the parameters $3\omega_D=\varepsilon_F=\Lambda/2$, $v=0$. Colored symbols denote the exact numerical results. Dashed and dotted black lines denote the infinite-bandwidth marginal Fermi liquid and finite-bandwidth conformal-limit solutions, respectively. Dashed gray lines denote various commonly used mean free paths associated with the MIR boundary.}
\label{fig:Rho2}
\end{figure}

\subsection{Strongly Incoherent Bad Metals}

At strong coupling and high temperature, where $|\Sigma(i\omega_n)|\gg|i\omega_n|,\Lambda$, the solution of the Dyson equation for the effective propagator yields $\mathcal{G}(i\omega_n)\sim -\nu\Lambda/\Sigma(i\omega_n)$ at low energies. In which case, the effective fermion scaling dimension, $\Delta_\psi$, where $\mathcal{G}(\tau)\sim \text{sgn}(\tau)/|\tau|^{2\Delta_\psi}$, directly relates to the scaling dimension of the fermionic self-energy $\Sigma(i\omega)\sim -i\text{sgn}(\omega)|\omega|^{1-2\Delta_\psi}$. Combined with the large-$N,M$ saddle-point equation for the fermionic self-energy, the fermion and boson scaling dimensions must then obey the constraint $2\Delta_{\psi}+\Delta_{\phi}=1$. This constraint was originally observed in an analogous (0+1)d fermion-boson model~\cite{Fu2017}, and indeed we find it to hold within the local critical regime of the (2+1)d Yukawa-SYK model. In the underdamped scenario, $\delta\Pi(i\Omega_m)$ may be neglected, in which case the low-energy boson effective scaling dimension is $\Delta_\phi\simeq 1/2$, and thus we obtain the desired result: $\Delta_\psi=1/4$. 

At zero-temperature, we find the low-energy result: $\Sigma(i\omega)=-i\text{sgn}(\omega)\sqrt{2\lambda\Lambda|\omega|}$ for $1/(8\lambda)\ll\omega/\Lambda\ll2\lambda$. At non-zero temperatures, this result generalizes to
\begin{equation}
\Sigma(i\omega_n)\approx -2  i \sqrt{\lambda \Lambda \pi T}\frac{\Gamma\left(\frac{3}{4}+\frac{\omega_n}{2\pi T}\right)}{\Gamma\left(\frac{1}{4}+\frac{\omega_n}{2\pi T}\right)}.  
\end{equation}
We use this approximate solution to estimate the resistivity in this strongly incoherent bad-metal regime, obtaining the $T$-linear form $\rho_{\BM} = 8\lambda (k_BT/\Lambda)(h/e^2)$. The low-temperature threshold $\Lambda/(8\pi\lambda) \ll T $, implies that $\rho_{\text{BM}}\gg (1/\pi)(h/e^2)$, confirming bad-metal behavior. Comparing with the resistivity obtained in the marginal-Fermi-liquid regime, we find $\rho_{\MFL}/\rho_{\CL}\sim\pi/2$ for $\Lambda/2\sim\varepsilon_F$. In Fig.~\ref{fig:Rho2}, we plot the exact resistivity as determined from the Kubo formula (see  Supplementary Materials). The dashed and dotted lines are the analytical expressions for the marginal Fermi liquid and conformal-limit resistivities, respectively.

\section{Crossovers and Implications of the Quantum-Critical Theory}
\label{sec:Implications}

\subsection{Comparison of the Underdamped and Overdamped-Boson Scenarios}
\label{subsec:comparison-UD-OD-maintext}

A leading theoretical model~\cite{Patel2023} of strange metals implements a disordered Yukawa interaction between quantum-critical bosons and electrons that gives rise to Landau damping of the boson. Overdamped critical bosons can mediate superconductivity and, in the case of a spatially disordered interaction, give rise to $T$-linear resistivity. In certain instances~\cite{Valentinis2026}, it has been assumed that Landau damping is necessary to achieve MFL phenomenology. As we show in Eq.~\eqref{eq:MFL-UD1-maintext}, bosonic damping is not a necessary condition to achieve MFL behavior and   $T$-linear resistivity.

\begin{figure}[t]
\centering
\begin{subfigure}{1\textwidth}
\centering
\includegraphics[width=\linewidth]{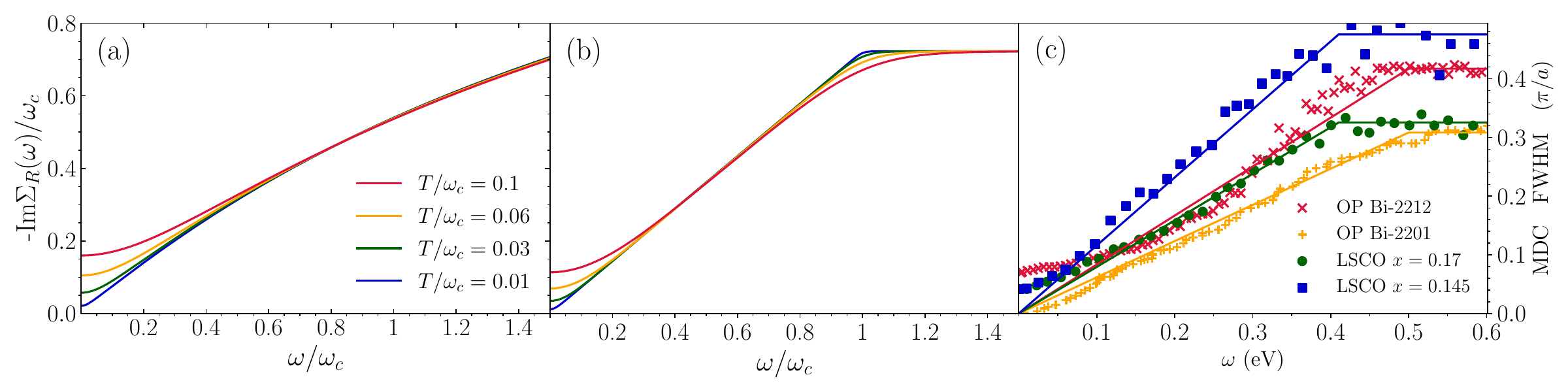}
\end{subfigure}
\caption{(a), (b) Comparison of the imaginary part of the fermionic self-energy in the scenarios with an (a) overdamped and (b) underdamped boson in the infinite-bandwidth limit. In sub-figures (a), (b), $\lambda=0.46,0.23$, and $\omega_c=\varepsilon_F/(2\pi\lambda), \omega_D$ respectively (the coupling is doubled for the overdamped case to account for the difference in slope). Sub-figure (c) is the empirical momentum distribution curve linewidth (full width at half-maximum) obtained from ARPES for various cuprate samples~\cite{Zhu2008,Meevasana2007,Chang2007,Chang2008}. The data is extracted from Fig.~(4) of Ref.~\cite{Zhu2008}.} 
\label{fig:SE-comparison-maintext}
\end{figure}

Beyond establishing that underdamped bosons produce MFL phenomenology, we demonstrate that it more accurately reflects the optical properties of strange metals than the overdamped boson scenario. The scattering rate of strange metals features $\omega/T$ scaling at intermediate frequencies $T<|\omega|<\omega_c$ up to a high-energy cutoff $\omega_c$. In particular, the scattering rate $1/(2\tau(\omega))\equiv|\Sigma_R^{\prime\prime}(\omega)|$ exhibits a relatively sharp and continuous crossover from $\sim|\omega|$ to a constant value for high frequencies $|\omega|>\omega_c$. This behavior is exhibited in the frequency-dependent optical~\cite{Marel2003} and single-particle~\cite{Kaminski2005,Zhu2008,Chang2008} scattering rates of the cuprates. In Fig.~\ref{fig:SE-comparison-maintext}(c) we provide a comparison with the empirical momentum distribution curve (MDC) linewidth from Ref~\cite{Zhu2008}, which is directly related to the single-particle scattering rate by $\Sigma_R^{\prime\prime}/v_{F,0}$, where $v_{F,0}$ is the bare Fermi-velocity~\cite{Kaminski2005,Chang2008}.

Regarding the higher-frequency regime $|\omega|\gtrsim\omega_c$, we observe significant differences between the underdamped and overdamped scenarios. For the underdamped-boson scenario (which we will assume to be completely undamped for simplicity), the frequency-dependent scattering rate exhibits the desired sharp crossover from $|\Sigma_R^{\prime\prime}(\omega)|\simeq\pi\lambda|\omega|$ to a constant value $\pi\lambda\omega_D$ for high frequencies, $|\omega|\gtrsim\omega_D$ as in Eq.~\eqref{eq:MFL-form-0T}, while the overdamped- boson scenario  crosses over to a logarithmic frequency scaling $|\Sigma_R^{\prime\prime}(\omega)|\sim\log|\omega|$ for intermediate to high frequencies. In Fig.~\ref{fig:SE-comparison-maintext} we compare the underdamped and overdamped scenarios using the exact fermionic self-energy in the infinite-bandwidth limit. For the underdamped case, we set the bosonic self-energy to zero. In the overdamped case~\cite{Kleger2026, Esterlis2021}, we include both $\Omega_m^2$ and $|\Omega_m|$ terms in the bosonic propagator and assume $k_D\sim k_F$, $N\sim M$ such that $\delta\Pi(i\Omega_m)=-2\pi\lambda(\omega_D^2/\varepsilon_F)|\Omega_m|$.

In existing theoretical treatments of quantum-critical systems~\cite{Patel2023, Patel2024, Kryhin2025, Valentinis2026}, Landau damping and overdamped bosonic modes are considered central to describing strange metals and producing MFL behavior. Numerical studies~\cite{Patel2024} suggest a low-temperature crossover to localized bosonic eigenmodes, with non self-averaging spatial disorder~\cite{Berkovits2023}. However, whether bosonic modes become strongly damped or localized within the critical regime remains an open question, particularly as the electronic mean free path approaches the lattice spacing in the MIR limit. For a weakly damped boson with a well-defined group velocity $v_g\sim c$, the Landau damping yields a bosonic inverse scattering time $\hbar/\tau_B\sim \lambda (N/M)mc^2$, corresponding to a mean free path $\ell_B \sim c\tau_{B}\sim(1/\lambda)(M/N)(\hbar/(mc))$. For $c\approx v_F$, this reduces to $\ell_B\sim (1/\lambda)(M/N)\lambda_F$. Therefore, an appropriate separation of scales, achieved by taking $M\gg N$ (near criticality), $c\ll v_F$ (for non-adiabatic energy scales $\omega_D\sim\varepsilon_F$), or $\lambda\ll1$, ensures that the boson is underdamped and has a mean free path larger than the length scales associated with the fermionic MIR bound. As a result, we obtain a framework of strongly underdamped critical bosons that do not localize.

\subsection{Strange Metal to Fermi Liquid Transition and Extended Criticality}

Assuming the ratio of the number of bosons to fermions, $M/N$, decreases away from criticality, Landau damping becomes increasingly pronounced as the system is tuned away from the critical regime (provided the variation in $M/N$ exceeds that of $\lambda$). Concomitantly, the boson mass detunes from criticality and acquires a non-critical value: $m_B\rightarrow \Delta$, which is nonzero in the zero-temperature limit~\cite{Esterlis2021}. This detuning suppresses the strange-metal phase by introducing a low-energy gap $\Delta\gg m_B$ in the boson spectrum. Landau damping, however, renormalizes the bosonic propagator resulting in a finite density of states at low energies in the bosonic modes. The combined effects of detuning and increased Landau damping results in a transition from the strange-metal phase of underdamped critical bosons with the fermion scattering rate $-\Sigma_R^{\prime\prime}(\omega)=|\omega|$ to a phase of damped non-critical bosons with the fermion scattering rate $-\Sigma_R^{\prime\prime}(\omega)\sim \omega^2$, scaling as a Fermi-liquid~\cite{Kryhin2025}. 

Anticipating enhanced bosonic damping as the system is tuned away from criticality provides a natural connection to the recent theoretical work~\cite{Kryhin2026} on the extended critical region of the cuprate phase diagram. Reference~\cite{Kryhin2026} investigates a framework of localized overdamped bosonic modes as a possible explanation of the strange metal `foot', where $T$-linear resistivity persists at low temperatures deep into the overdoped regime~\cite{Cooper2009}. Indeed, taking the parameter $N/M$ to diminish as the system is tuned away from criticality results in increasingly damped bosons in the framework we consider. In this scenario, bosons become overdamped allowing for a crossover to this regime with localized bosonic modes for finite values of $N$ and $M$. Accordingly, the quantum critical `fan' and the extended critical `foot' may be described by the respective frameworks of underdamped and overdamped bosons respectively.

\subsection{Quantum-Critical Metals in Three Spatial Dimensions}

Generalizing our analysis to three spatial dimensions reveals important implications for the theory of disordered quantum-critical metals and unconventional superconductors. In the underdamped-boson scenario, we find that the electronic scattering rate scales quadratically with frequency, consistent with a resistivity $\rho\sim T^2$, reminiscent of the Fermi liquid. Furthermore, we find that the defining signature of logarithmically enhanced low-temperature heat capacity in the strange metals is absent in three spatial dimensions. Instead, we find the leading contribution from the critical electron-boson interaction to scale linearly in temperature, renormalizing the bare Sommerfeld contribution. Finally, in the weak-coupling regime, assuming an isotropic gap, we obtain $\log(\omega_{D}/T_c) \sim 1/\lambda$, in stark contrast to the two-dimensional result $\log(\omega_{D}/T_c) \sim 1/\sqrt{\lambda}$. The former is structurally analogous to the BCS result, even though the pairing mechanism is distinct~\cite{Wang2025}, whereas the latter exhibits the anomalous $1/\sqrt{\lambda}$ exponent~\cite{Son1999, Chubukov2005, Gnezdilov2025}.

Despite tuning the boson mass to criticality, the logarithmic form $\mathcal{D}(i\Omega)\sim\log(1/\Omega)$ is absent in three spatial dimensions. Consequently, the characteristic macroscopic signatures of strange metals are absent. Instead, we find hallmark signatures of a Fermi liquid: $\sim T^2$ resistivity, $\sim T$ heat capacity, and a more conventional form of $T_c$. Nevertheless, the microscopic physics underlying superconductivity remains  unconventional: a critical boson mediating the attractive pairing interaction. In the overdamped-boson scenario, we find that the low-temperature resistivity scales as $\rho\sim T^{3/2}$, which has been observed experimentally~\cite{Pfleiderer2001}.

Together, these results may explain why strange metals are predominantly 2d materials (or in certain cases 3d with quasi-2d critical fluctuations) with some of the highest values for $T_c$. A three-dimensional disordered quantum-critical metal cannot be recognized by the defining macroscopic signatures of the strange metals. Indeed, the most notable features may be conflated with Fermi-liquid physics. Based on these results, we conjecture that there exist three-dimensional unconventional superconductors that are disordered quantum critical metals in their normal state, akin to the strange metals. However, their conventional transport and thermodynamic properties obscure their underlying quantum-critical nature, making them difficult to recognize. If this conjecture is valid, the scope of the theory of disordered quantum-critical metals is significantly broadened, introducing an avenue for discovering other unconventional superconductors in an entirely different class of systems.

\section{Conclusion}

In this paper, we demonstrate that a (2+1)-dimensional theory of fermions coupled to underdamped quantum-critical bosons through a spatially disordered interaction captures the universal features of both strange and bad metals. Within this framework, we obtain both the marginal Fermi-liquid and conformal-limit solutions, which exhibit $T$-linear resistivity both below and above the Mott-Ioffe-Regel bound, respectively. We find that underdamped critical bosons possess a constant density of states over energies $T\lesssim|\omega|\lesssim\omega_D$, suitable for obtaining the linear-in-frequency scaling and high-frequency plateau in the electronic scattering rate, which is observed in ARPES and optical transport experiments of the cuprate superconductors. Discovery of the conformal-limit solution within the critical electron-boson framework is a landmark achievement establishing a remarkable connection between quantum-critical Migdal-Eliashberg theory and the emergent conformal physics of the SYK model,  which itself has deep connections to holography. By capturing these distinct features within a single theoretical framework, our approach provides a unified description of the strange- and bad-metal regimes and offers a resolution to several long-standing questions surrounding the cuprate phase diagram.

{\it Acknowledgments.} 
 The authors thank Professor Dirk van der Marel for sharing the data from Ref.~\cite{Marel2003} contained in Fig.~(1) of our Supplementary Materials. The figures were produced with the help of ChatGPT. This research was supported by funding provided by Dartmouth College.

\bibliography{References}

\end{document}


\title{Supplementary Material: Universal Theory of Strange and Bad Metals from Spatially Random Interactions with Underdamped Quantum-Critical Bosons}

\author{Aaron Kleger}
\email{Aaron.R.Kleger.GR@dartmouth.edu}
\author{Rufus Boyack}
\email{Rufus.Boyack@dartmouth.edu}

\affiliation{Department of Physics and Astronomy, Dartmouth College, Hanover, New Hampshire 03755, USA}

\maketitle 

\tableofcontents

\setcounter{section}{0}

\newpage

\section{The Model}

\subsection{Normal-State Propagators}

We consider the Yukawa-Sachdev-Ye-Kitaev (Y-SYK) model~\cite{Esterlis2019, Wang2020} in two spatial dimensions, consisting of $N$ electrons $\psi_{i\sigma}$ coupled to $M$ bosons $\phi_i$ in the continuous limit; $\sigma=\uparrow \downarrow$ denotes spin, $i,j,(l)=1, \cdots, N (M)$ labels the fermion (boson) flavor. The Hamiltonian reads $H=H_\psi+H_\phi+V_{\psi \phi}$, where
\begin{subequations}
\begin{align}
H_\psi & =\sum_{i=1}^N \sum_{\sigma=\uparrow \downarrow} \int \psi_{i \sigma}^{\dagger}(\bm{k}) \xi_{\bm{k}}\psi_{i \sigma}(\bm{k}) \frac{d^2 k}{(2 \pi)^2}, \\
H_\phi & =\frac{1}{2} \sum_{l=1}^M \int  \phi_l(\bm{q}) \left(c^2q^2+m_{B,0}^2\right) \phi_l(-\bm{q}) \frac{d^2 q}{(2 \pi)^2}, \\
V_{\psi \phi} & =\frac{1}{\sqrt{NM}} \sum_{i, j=1}^N\sum_{l=1}^M \sum_{\sigma=\uparrow \downarrow} \int g_{i j l}(\bm{r}) \psi_{i \sigma}^{\dagger}(\bm{r}) \psi_{j \sigma}(\bm{r}) \phi_l(\bm{r}) d^2 r.
\end{align}
\end{subequations}
The electronic dispersion relation is given by $\xi_{\bm{k}}=k^2/(2m)-\mu;\;k^2/(2m)\in[0,\Lambda]$, with bandwidth $\Lambda$ and chemical potential $\mu$. The bare mass of the boson is denoted by $m_{B,0}$. Throughout, we will use Natural units: $\hbar=k_{B}=1$. The electron-boson interaction is spatially disordered~\cite{Sachdev2024}, with coupling constants $g_{i j l}$ sampled from a Gaussian orthogonal ensemble with zero mean and non-zero, local variance $\left\langle g_{i j l}(\bm{r}) g_{a b c}\left(\bm{r}^{\prime}\right)\right\rangle = 2 g^2 \delta\left(\bm{r}-\bm{r}^{\prime}\right)\delta_{ia}\delta_{jb}\delta_{lc}$. 

In the large-$N$ limit, the normal-state saddle-point equations for the fermionic and bosonic self-energies are~\cite{Gnezdilov2025}:
\begin{subequations}
\begin{align}
\Sigma\left(i \omega_n\right) & =g^2 T \sum_{\omega_{n^{\prime}}} \mathcal{D}\left(i \omega_n-i \omega_{n^{\prime}}\right) \mathcal{G}\left(i \omega_{n^{\prime}}\right), \label{eq:Sigma_saddlept} \\ 
\Pi\left(i \Omega_m\right) & =-2 g^2 \left(\frac{N}{M}\right) T \sum_{\omega_n} \mathcal{G}\left(i \omega_n\right) \mathcal{G}\left(i \omega_n+i \Omega_m\right),
\label{eq:Pi-saddlept}
\end{align}
\end{subequations}
where $\omega_n=(2 n+1) \pi T$ and $\Omega_m=2 m \pi T$ are fermionic and bosonic Matsubara frequencies, respectively, at temperature $T$. To decouple the interactions, $\Sigma$ and $\Pi$ are introduced as Lagrange multipliers that are non-local in time. Due to the local variance of the electron-boson coupling, the momentum integrals for the fermionic and bosonic self-energies factorize (rather than the usual~\cite{AGD} convolution), leading to manifestly momentum-independent self-energies.

The effective fermionic and bosonic propagators $\mathcal{G}$ and $\mathcal{D}$ are defined by
\begin{align} 
\mathcal{G}\left(i \omega_n\right) & \equiv \int G\left(i \omega_n, \bm{k}\right) \frac{d^2 k}{(2 \pi)^2} \nonumber \\  
& =\int \frac{1}{i \omega_n-\xi_{\bm{k}}-\Sigma\left(i \omega_n\right)} \frac{d^2 k}{(2 \pi)^2}. \label{eq:G-eff} \\  
\mathcal{D}\left(i \Omega_m\right) & \equiv \int D\left(i \Omega_m, \bm{q}\right) \frac{d^2 q}{(2 \pi)^2} \nonumber \\
& =\int_0^{\omega_D / c} \frac{1}{\Omega_m^2+c^2q^2+m_{B,0}^2-\Pi\left(i \Omega_m\right)} \frac{q d q}{2 \pi}.
\label{eq:Deff}
\end{align}
Here, $\omega_D$ and $c$ are analogous to the Debye frequency and the speed of sound, respectively. The ratio $\omega_D/c$ acts as a bosonic ultraviolet (UV) cutoff. Performing the momentum integration in Eq.~\eqref{eq:Deff}, the result obtained is $\mathcal{D}\left(i \Omega_m\right)=L\left(i \Omega_m\right) /\left(4 \pi c^2\right)$, where
\begin{equation}
L\left(i \Omega_m\right)=\log \left(1+\frac{\omega_D^2}{\Omega_m^2 +m_{B,0}^2-\Pi(i\Omega_m)}\right).
\label{eq:BosonL}
\end{equation}
Let $\nu=m /(2 \pi)$ be the electron density of states per spin in two-dimensions and $m$ the electron mass. The effective fermionic propagator is given by 
\begin{equation}
\mathcal{G}(i\omega_n)=\nu\int_{-\mu}^{\Lambda-\mu}\frac{1}{i\omega_n-\xi-\Sigma(i\omega_n)}d\xi.
\label{eq:G_Eff_ArbitraryBW}
\end{equation}

\subsection{Tuning to the Quantum-Critical Point} 

\label{sec:tune-QCP}

At nonzero temperatures, a nonzero boson mass is necessary to keep the thermal contribution (the resonant term with $\omega_{n^\prime}=\omega_n$ in Eq.~\eqref{eq:Sigma_saddlept}) to the fermionic self-energy finite. To ensure that the renormalized boson mass, defined by $m_B^{2}(T)\equiv m_{B,0}^2-\Pi(0)$, vanishes at $T=0$ and is nonzero for $T>0$, we introduce a constraint in the action~\cite{Esterlis2021}. The renormalized boson mass can be tuned to quantum criticality by replacing the bare bosonic mass with a Lagrange multiplier $m_{B,0}^2\rightarrow\lambda(\tau,\bm{r})$, and adding an additional term to the action~\cite{Esterlis2021}:
\begin{equation}
S_\gamma = -M\int  \frac{i\lambda(\tau,\bm{r})}{2\gamma}d\tau d^2r.
\end{equation}
The constraint yields the additional saddle-point equation
\begin{equation}
\frac{1}{\gamma} = T\sum_{\Omega_m}\mathcal{D}(i\Omega_m),
\label{eq:Constraint_saddlept}
\end{equation}
and $i\lambda(\tau,\bm{r})=m_{B,0}^2$ at the large-$N,M$ saddle-point solution. The renormalized boson mass is found by equating the right-hand side of Eq.~\eqref{eq:Constraint_saddlept} at zero and nonzero temperature (as in Eq.~\eqref{eq:Boson-constraint}), then solving for $m_{B}^2(T)$. For a specific value of $\gamma$, the renormalized boson mass vanishes at zero temperature and takes a nonzero value at nonzero temperatures. The constraint in the action gives the following additional term in the grand potential:
\begin{equation} 
\bm{\Omega}_\gamma = -M\frac{m_{B,0}^2(T)}{2\gamma}.
\label{eq:FreeEnergyConstraint}
\end{equation}

We can solve for the renormalized boson mass $m_B(T)$ at nonzero temperatures by equating the right-hand side of Eq.~\eqref{eq:Constraint_saddlept} at zero and nonzero temperatures:
\begin{equation} 
\frac{1}{2\pi}\int_{-\infty}^{\infty} \left. L(i\Omega)  d\Omega \right|_{m_B=0}= T\sum_{\Omega_m}L(i\Omega_m)|_{m_B=m_B(T)}.
\label{eq:Boson-constraint}
\end{equation}
Tuning the mass to quantum criticality, the zero-temperature renormalized bosonic mass is set to zero and we solve for $m_B(T)$. For an underdamped boson, the renormalized boson mass is given by $m_B(T)=\text{arccosh}(3/2)T$ for $T\ll\omega_D$ at critically~\cite{Kleger2026}. For an overdamped boson $m_B(T)\approx\sqrt{T\omega_D}$ up to logarithmic corrections in terms of temperature and the coupling strength~\cite{Esterlis2021}.

\section{The Strange Metal}

\subsection{Overview}

Strange metals are characterized by $T$-linear resistivity below the Mott-Ioffe-Regal limit. The strange-metal regime is therefore restricted to $\rho< h/e^2$ in two spatial dimensions~\cite{Li2024}. Consequently, the electronic self-energy of strange metals satisfies the separation of scales $|\Sigma_R^{\prime\prime}(\omega)|\ll\varepsilon_F,\Lambda$ at low temperatures. When the separation of scales $\mu, \Lambda-\mu  \gg  |i\omega_n - \Sigma(i\omega_n)|$ is satisfied, the effective electronic propagator in Eq.~\eqref{eq:G-eff} can be evaluated by extending the momentum integration limits to positive and negative infinity. In this infinite-bandwidth limit, the effective propagator depends only on frequency and is independent of the fermionic self-energy:
\begin{equation}
\mathcal{G}\left(i \omega_n\right)=\nu\int_{-\infty}^{\infty}\frac{1}{i\omega_n-\xi-\Sigma(i\omega_n)}d\xi=-i \pi \nu \operatorname{sgn}\left(\omega_n\right).
\label{eq:G-eff-IBWL}
\end{equation}

In extending the bounds of Eq.~\eqref{eq:G_Eff_ArbitraryBW} to infinity to obtain Eq.~\eqref{eq:G-eff-IBWL}, we have implicitly assumed a symmetric bandwidth, $\mu=\Lambda/2$, i.e., particle-hole symmetry. Relaxing this assumption introduces a real part to the effective fermionic propagator: $\mathcal{G}(i\omega_n) = -i\pi\nu\;\text{sgn}(\omega_n)-\nu\log(\Lambda/\mu-1), \Lambda>\mu$. This additional real contribution to the Matsubara-axis fermionic self-energy, $\Sigma(i\omega_n)$, does not contribute to the imaginary part of the real-axis fermionic self-energy, $\text{Im}\Sigma_R(\omega)$. Assuming $\text{Re}\mathcal{G}(i\omega_n)=0$ ensures that $\text{Re}\Sigma(i\omega_n)=0$. In conventional electron-phonon systems, this assumption is equivalent to the even part of the Matsubara-axis fermionic self-energy being zero~\cite{Bennemann}. 

The dynamical part of the bosonic self-energy is defined by $\delta\Pi(i\Omega_m)\equiv\Pi\left(i \Omega_m\right)-\Pi(0)$, where $\Pi(0) = m_{B,0}^2-m_B^2$. In the infinite-bandwidth limit, the dynamical part of the bosonic self-energy is determined to be $\delta\Pi(i\Omega_m)=-c_d\left|\Omega_m\right|$, where $c_d\equiv2 \pi \nu^2 g^2(N/M)$ ~\cite{Esterlis2021,Gnezdilov2025}, which provides the Landau damping of the bosonic modes. As in Ref.~\cite{Gnezdilov2025}, we define the dimensionless electron-boson coupling constant $\lambda\equiv\nu g^2 /\left(4 \pi c^2\right)$. In the infinite-bandwidth limit, the effective bosonic propagator $L(i\Omega_m)$ is given by 
\begin{equation}
L\left(i \Omega_m\right)=\log \left(1+\frac{\omega_D^2}{\Omega_m^2+m_{B}^2+c_d\left|\Omega_m\right|}\right).
\label{eq:logarithm}
\end{equation}

In the following sections, we consider a quantum-critical boson, $m_B|_{T=0}=0$, in the underdamped regime, where $|\delta\Pi(i\Omega_m)|\ll\Omega_m^2+m_{B}^2$. On the imaginary frequency axis, this condition holds for all nonzero bosonic Matsubara frequencies in the infinite-bandwidth limit when $c_d = 4\pi\lambda m c^2(N/M) \ll T$. For intermediate coupling and $\omega_D\approx\varepsilon_F$, this condition is satisfied  in the normal state for $v_F/c\gg N/M$. In Sec.~\ref{subsec:SM-0T} we compute the zero-temperature fermionic self-energy and show that underdamped bosons yield marginal-Fermi-liquid (MFL) phenomenology. In Sec.~\ref{subsec:SM-finite-T}, we show that the exact fermionic self-energy displays  $\omega/T$-scaling. Finally, in Sec.~\ref{subsec:comparison-UD-OD}, we compare the conventional approach of overdamped bosons  to our newly proposed scenario of underdamped bosons. Remarkably, unlike the overdamped scenario, the underdamped scenario does not have logarithmic corrections to linear-in-frequency scaling. Furthermore, it also displays a sharp saturation in the fermionic scattering rate in the crossover from the intermediate to high-frequency regime, in strong agreement with the scattering rates observed in the cuprates near critical doping.

\subsection{Zero-Temperature Analysis}
\label{subsec:SM-0T}

The normal-state fermionic self-energy on the imaginary axis in the infinite-bandwidth limit at zero temperature is 
\begin{align}
\Sigma(i\omega) &= g^2\int_{-\infty}^{\infty}\mathcal{G}(i\omega^{\prime})\mathcal{D}(i\omega-i\omega^{\prime})\frac{d\omega^{\prime}}{2\pi} \nonumber \\
&= -i \frac{\lambda}{2}\int_{-\infty}^{\infty}L(i\omega-i\omega^{\prime})\text{sgn}(\omega^{\prime})d\omega^{\prime} \nonumber \\
&= -i \lambda\;\text{sgn}(\omega)\int_{0}^{|\omega|}L(i\Omega)d\Omega. 
\end{align}
Here we have used the fact that the effective bosonic propagator $L(i\Omega)$ is even in $\Omega$ and tends to zero as $|\Omega|\rightarrow\infty$. Computing the fermionic self-energy for a quantum-critical underdamped boson at zero temperature, we find
\begin{align} 
\Sigma_{\text{UD}}(i\omega)&=-i \lambda\;\text{sgn}(\omega)\int_{0}^{|\omega|}\log\left(1+\left(\frac{\omega_D}{\Omega}\right)^2\right) d\Omega \nonumber\\ 
&= -i \lambda\;\text{sgn}(\omega)\left[|\omega|\log\left(1+\left(\frac{\omega_D}{\omega}\right)^2\right)+2\omega_D \arctan\left(\frac{|\omega|}{\omega_D}\right)\right] \nonumber \\ 
&\simeq -2i\lambda\omega\; \log\left(\frac{e\omega_D}{|\omega|}\right); \quad |\omega| \ll \omega_{D}.
\label{eq:MFL-UD1}
\end{align}
This form of the fermionic self-energy exhibits the desired marginal-Fermi-Liquid form $\Sigma(i\omega)\sim-i\omega\log(e\omega_c/|\omega|)$, with the MFL UV cutoff $\omega_c=\omega_D$ and the single-particle scattering rate $1/(2\tau) =\pi\lambda|\omega|$~\cite{Varma1989,Kleger2026}, for intermediate frequencies $T\ll|\omega|\ll\omega_D$.

\subsection{Nonzero-Temperature Analysis}
\label{subsec:SM-finite-T}

The exact expression for the fermionic self-energy can be obtained in the infinite-bandwidth limit by performing the Matsubara frequency summation. In terms of the Pochhammer symbol~\cite{WhittakerWatsonBook}, defined by:  $(x)_n\equiv\Gamma(x+n)/\Gamma(x)$, the exact fermionic self-energy for the underdamped boson in the infinite-bandwidth limit is given by 
\begin{equation}
\Sigma(i\omega_{n})=-i\pi\lambda T\;\text{sgn}(2n+1)\left\{ \log\left(1+\frac{\omega_D^2}{m_{B}^2}\right)+2\log\left[\frac{\left(1-i\dfrac{\sqrt{\omega_D^2+m_{B}^2}}{2\pi T}\right)_{|n|}}{\left(1-i\dfrac{m_{B}}{2\pi T}\right)_{|n|}}\frac{\left(1+i\dfrac{\sqrt{\omega_D^2+m_{B}^2}}{2\pi T}\right)_{|n|}}{\left(1+i\dfrac{m_{B}}{2\pi T}\right)_{|n|}}\right]\right\}.
\label{eq:Exact_elec_SE_IBWL}
\end{equation}
All frequency dependence in Eq.~\eqref{eq:Exact_elec_SE_IBWL} resides in the fermionic Matsubara index $n$. Performing analytical continuation $n\rightarrow-1/2-i\omega/(2\pi T)$ then yields the desired $\omega/T$-scaling of the fermionic self-energy.

For $T\ll\omega_D$, the mass of the quantum-critical boson is given by $m_B=2\pi\alpha T$, where $\alpha\equiv(\text{arcosh}(3/2))/(2\pi)$; see Sec.~\ref{sec:tune-QCP} for further details. Thus, for $T\ll\omega_D$, we find
\begin{equation}
\Sigma(i\omega_{n})\simeq-i2\pi\lambda T\;\text{sgn}(2n+1) \log\left[\frac{\omega_D}{2\pi\alpha T}\frac{\left(1-i\dfrac{\omega_D}{2\pi T}\right)_{|n|}}{\left(1-i\alpha\right)_{|n|}}\frac{\left(1+i\dfrac{\omega_D}{2\pi T}\right)_{|n|}}{\left(1+i\alpha\right)_{|n|}}\right].
\label{eq:Simplified_elec_SE_IBWL}
\end{equation}
On the real frequency axis, the imaginary part of the fermionic self-energy can be computed exactly as follows:
\begin{align}
\Sigma_R^{\prime\prime}(\omega,T) &= -\pi \lambda T \log\biggl[ \frac{\tanh^2(\omega_{D}/(2T)) }{1+\sech^2((\omega_{D}/(2T))\sinh^2(\omega/(2T))}\times \left |\alpha (1-i\alpha)_{-1/2- i \omega/(2\pi T)}(1+i\alpha)_{-1/2- i \omega/(2\pi T)}  \right|^{-2}\biggr]\nonumber \\
&= -\pi \lambda T \log\left[\frac{(3+2\cosh(\omega/T))\tanh^2(\omega_{D}/(2T)) }{1+\sech^2((\omega_{D}/(2T))\sinh^2(\omega/(2T))}\right].
\label{eq:Im-SE-SM}
\end{align}
In performing the above calculation, we have used the reflection formula $\Gamma(z)\Gamma(1-z)=\pi/\sin(\pi z)$, the duplication formula $\sqrt{\pi}\Gamma(2z)=2^{2z-1}\Gamma(z+1/2)\Gamma(z)$, the Schwarz reflection principle, the identity $|\Gamma(i \zeta)|^2=\pi/[\zeta \sinh(\pi \zeta)]$, for $\zeta\in\mathbb{R}$, and the identity $|\Gamma(1/2+iz)|^2=\pi \sech(\pi z)$. Let us now consider the two cases where $|\omega|<\omega_{D}$ and $|\omega|\geq\omega_{D}$. Taking the limit $T/\omega_{D}\ll1$ in the above expression, we obtain 
\begin{equation}
\Sigma_R^{\prime\prime}(\omega) = -\pi \lambda
\begin{cases}
T \log\left(3+2\cosh(\omega/T)\right), \quad |\omega|<\omega_{D}, \label{eq:Im-SE-SM-simplified} \\
\omega_{D},  \quad |\omega|\geq\omega_{D}.
\end{cases}
\end{equation}
Thus, $\Sigma_R^{\prime\prime}(\omega,T)$, the electronic scattering rate, exhibits exact Planckian characteristics: $T$-linear and $\omega/T$-scaling within the intermediate scaling regime. For $|\omega|\ll T\ll\omega_D$, Eq.~\eqref{eq:Im-SE-SM-simplified} reduces  to $\Sigma_R(\omega,T)=  -i\pi\log(5) \lambda T$. In the zero-temperature limit, $T\ll|\omega|,\omega_{D}$, Eq.~\eqref{eq:Im-SE-SM-simplified} gives the fermionic self-energy in Eq.~(9) of the main text.

\subsection{Comparison of Underdamped- and Overdamped-Boson Scenarios}
\label{subsec:comparison-UD-OD}

One of the foremost theoretical models for the strange-metal phase is based on the premise of a disordered Yukawa interaction between quantum-critical bosons and electrons~\cite{Patel2023, Li2024, Valentinis2026, Cichutek2024, Kryhin2025}. In this framework, it is assumed that the boson is renormalized by the electrons, giving rise to Landau damping. Within this framework, it is conventionally understood that overdamped critical bosons intermediate superconductivity, and in the case of a spatially disordered interaction, give rise to $T$-linear resistivity. In certain instances~\cite{Valentinis2026}, it has been explicitly assumed that Landau damping is necessary to achieve MFL phenomenology. As we show in Eq.~\eqref{eq:MFL-UD1}, bosonic damping is not a necessary condition to achieve MFL phenomeonology (and thus $T$-linear resistivity) within this framework.

Beyond establishing that the underdamped-boson scenario can reproduce MFL phenomenology, we demonstrate that it provides a better description of the observed optical properties of various strange metals than the overdamped-boson scenario. Characteristic of the scale invariance of a state with local criticality, the scattering rate of strange metals features $\omega/T$ scaling at intermediate frequencies $T<|\omega|<\omega_c$ up to a high-energy cutoff $\omega_c$. In particular, the scattering rate $1/(2\tau(\omega))\equiv|\Sigma_R^{\prime\prime}(\omega)|$ exhibits a relatively sharp and continuous crossover from $\sim|\omega|$ to a constant value for high frequencies $|\omega|>\omega_c$. This behavior is exhibited in the frequency-dependent optical~\cite{Marel2003} and single-particle~\cite{Kaminski2005} scattering rates of the optimally doped cuprate Bi-2212.

At higher frequencies, $\omega\gtrsim\omega_c$, the underdamped and overdamped scenarios differ significantly. For an underdamped boson (which we will assume to be entirely undamped for simplicity), the scattering rate exhibits the desired sharp crossover from $\sim|\omega|$ to a constant at high frequencies, while the overdamped-boson scenario exhibits a crossover from $\sim|\omega|$ to $\sim\log|\omega|$. The former can be observed in the high-frequency regime of Eq.~\eqref{eq:MFL-UD1}.

\begin{figure}[t]
\centering\includegraphics[width=1\columnwidth]{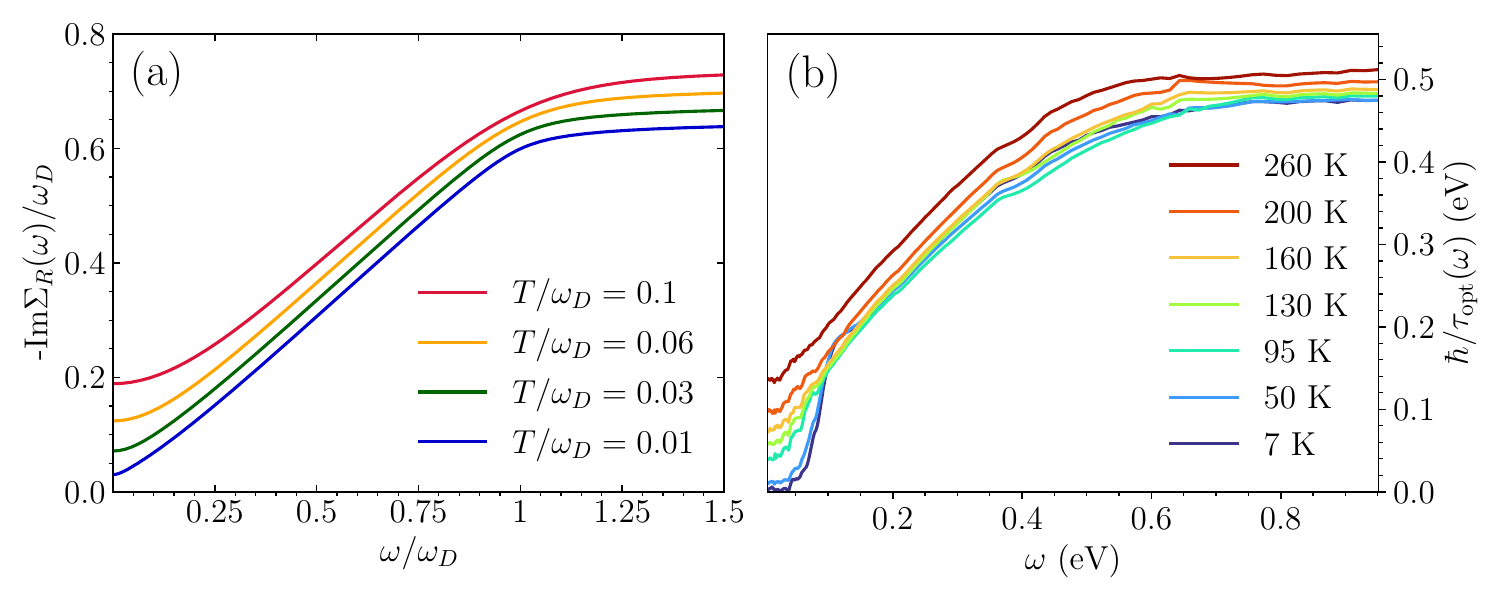}
\caption{(a) Imaginary part of the normal-state fermionic self-energy for a slightly damped boson in the infinite-bandwidth limit. Here, we set the boson mass to the value $m_B^2=0.5T^2$, yielding a scattering rate that shifts upwards with temperature. The relative strength of the boson damping is controlled by $N\omega_D/(M\varepsilon_F)=0.1$, and we set $\lambda=0.23$. (b) Empirical optical scattering rate for the optimally doped cuprate Bi-2212 from Ref.~\cite{Marel2003}.}
\label{fig:ad-hoc}
\end{figure}

Upon analytical continuation of Eq.~\eqref{eq:MFL-UD1}, one finds that the frequency-dependent scattering rate exhibits a sharp crossover from $1/(2\tau(\omega))\simeq\pi\lambda|\omega|$ to $1/(2\tau(\omega))\simeq\pi\lambda\omega_D$ for $\omega\gtrsim\omega_D$. Neglecting the $\Omega^2$ term in the effective bosonic propagator for the overdamped case, we find
\begin{align}
\Sigma_{\text{OD}}(i\omega)&=-i \lambda\;\text{sgn}(\omega)\int_{0}^{|\omega|}\log\left(1+\frac{\omega_D^2}{c_d \Omega }\right)d\Omega \nonumber \\
&=-i\lambda\;\text{sgn}(\omega)\left\{\left(|\omega|+\frac{\omega_D^2}{c_d}\right)\log\left(1+\frac{c_d|\omega|}{\omega_D^2}\right)-|\omega|\log\left(\frac{c_d|\omega|}{\omega_D^2}\right)\right\}.
\label{eq:MFL-SE1}
\end{align}
For $|\omega|\ll \omega_D^2/c_d$ we obtain MFL phenomenology with cutoff $\omega_c=\omega_D^2/c_d$ and scattering rate $1/(2\tau(\omega))=\pi\lambda|\omega|/2$, half that of the underdamped case. At higher frequencies $|\omega|\gtrsim\omega_D^2/c_d$, the scattering rate crosses over to $1/(2\tau(\omega))\sim \lambda (\omega_D^2/c_d)\log(ec_d|\omega|/\omega_D^2)$. In Fig.~(4) in the main text, we compare the underdamped and overdamped scenarios using the exact fermionic self-energy in the infinite-bandwidth limit. In Figs.~\ref{fig:ad-hoc} and~(4) of the main text, we assume $k_D\approx k_F$ such that $c_d=2\pi\lambda (N/M)\omega_D^2/\varepsilon_F$. The underdamped case neglects the bosonic self-energy, while the overdamped case retains both the $\Omega_m^2$ and $|\Omega_m|$ terms in the bosonic propagator. An exact evaluation of the latter can be found in Ref.~\cite{Kleger2026}. In Fig.~(4) of the main text, the boson mass is evaluated using the constraint method of Sec.~\ref{sec:tune-QCP}, yielding $m_B^2\sim\text{arccosh}^2(3/2)T^2$ and $m_B^2\simeq T\omega_D$ for the underdamped and overdamped cases, respectively. The relative damping strength is controlled by $N\omega_D/(M\varepsilon_F)=0$ or 1, respectively.

We also consider a third option, displayed in Fig.~\ref{fig:ad-hoc}, where the boson is ``slightly damped'' with an alternative parametrization of the quantum-critical boson mass. Allowing for some boson damping smooths out the transition from the high to low-frequency regimes without introducing discernible logarithmic corrections at high frequencies. Here we used an ad hoc scaling for the boson mass: $m_B^2=0.5 T^2$, with a smaller coefficient than determined by the constraint method. We find that this ad hoc value for the boson mass (with a smaller coefficient) increases the scaling with temperature of the low-frequency electronic scattering rate. Note that the empirical scattering rate of Ref.~\cite{Marel2003}, presented in Fig.~\ref{fig:ad-hoc}, is defined by the optical conductivity, rather than a direct measure of the single-particle scattering rate defined by the fermionic self-energy, $1/(2\tau)\equiv|\Sigma_R^{\prime\prime}|$ (as can be accessed with ARPES).

For the scenario with an underdamped boson, $\omega_c=\omega_D$. The energy scale $\omega_D$ and dimensionless coupling $\lambda$ can be extracted from ARPES or optical scattering rates of a strange metal. Together with the Fermi-energy, $\varepsilon_F$, and electronic bandwidth, $\Lambda$, these define all energy scales in the underdamped-boson framework.

\subsection{Crossover Physics: Strange Metal to Fermi Liquid and Extended Criticality}

In the Fermi-surface regime, the boson damping rate is given by $c_d=4\pi \lambda m c^2 (N/M)$. Assuming that the relative number of bosons to fermions diminishes away from criticality, then the Landau damping parameter's magnitude will increase as the system is tuned away from the critical regime. Concomitantly, the boson mass is detuned from criticality taking on a larger non-critical value $\Delta$, which does not vanish in the zero-temperature limit~\cite{Esterlis2021}. The zero-temperature fermionic self-energy in the damped and detuned regime is given by
\begin{align} 
\Sigma_{\text{Detuned}}(i\omega)&=-i \lambda\;\text{sgn}(\omega)\int_{0}^{|\omega|}\log\left(1+\frac{\omega_{D}^2}{\Omega^2+\Delta^2+c_d\Omega}\right)d\Omega \nonumber \\ 
&= -i \lambda\;\text{sgn}(\omega)\left(|\omega|\log\left(1+\omega_{D}^2/\Delta^2\right) - \frac{c_d \omega_{D}^2}{\Delta^2(\Delta^2+\omega_{D}^2)}\omega^2 + O((\omega/\omega_D)^3)\right).
\end{align}

Detuning the boson mass from its critical value, $m_B\sim T$, suppresses the strange-metal phase by opening a low-energy gap $\Delta\gg m_B$ in the bosonic spectrum. However, Landau damping renormalizes the bosonic propagator, generating finite spectral weight at low energies. The combined effects of mass detuning and enhanced Landau damping drive a crossover from a strange-metal phase with underdamped bosons and a linear fermion scattering rate, $|\Sigma_R^{\prime\prime}(\omega)|=\pi\lambda|\omega|$, to a phase of damped bosons with the fermion scattering rate $|\Sigma_R^{\prime\prime}(\omega)|\sim\lambda(c_d\omega_D^2/(2\Delta^4))\omega^2$, which has the leading-order $\sim\omega^2$ scaling of the Fermi-liquid~\cite{Kryhin2025}.

\section{The Strongly Incoherent Bad Metal}

\subsection{Overview}

The strongly incoherent bad-metal regime is characterized by $T$-linear resistivity above the Mott-Ioffe-Regel limit. As noted in Ref.~\cite{Parcollet1999}, realizing this behavior requires the fermionic self-energy to be of the form $\Sigma(i\omega_{n})\sim\sqrt{\omega_{n}}$. This solution has fermionic scaling dimension $\Delta_{\psi}=1/4$, as found in the SYK model~\cite{Maldacena2016}. In Ref.~\cite{Kleger2026}, we focused on the impurity regime of the Y-SYK model, which corresponds to $\Delta_{\psi}=1/2$. The pertinent question is how to obtain the solution $\Delta_{\psi}=1/4$ in the (2+1)-dimensional Y-SYK model. Indeed, this result was realized~\cite{Marcus2019} in an analogous (0+1)-dimensional fermion-boson model, where it was shown that in the limit $M/N\rightarrow\infty$, where $M, N$ are the respective numbers of bosons and fermions, the fermion scaling dimension coincides with that of the SYK model. 

Here we establish that in (2+1)d and in the limit $M/N\rightarrow\infty$, the Y-SYK model admits an approximate IR solution with $\Delta_{\psi}=1/4$, realizing a bad metal. This solution resembles the emergent conformal solutions found in the SYK model. Here, the limit $N/M\ll1$ is used to suppress $\delta\Pi(i\Omega_m)$, leaving the effective boson scaling dimension determined by the undamped critical boson propagator. This scenario differs from the solutions discovered in Refs.~\cite{Marcus2019, ClassenChubukov2021}, where $\delta\Pi(i\Omega_m)$ is crucial, and the fermion and boson scaling dimensions are tuned by the ratio $N/M$. In the strongly incoherent bad-metal regime, the fermionic self-energy  determines the scaling dimension of the effective fermionic propagator and therefore the fermionic saddle-point equation enforces $2\Delta_{\psi}+\Delta_{\phi}=1$~\cite{Marcus2019}. This implies that the strongly incoherent bad-metal regime is characterized by $\Delta_{\psi}=1/4, \Delta_{\phi}=1/2$, which we will confirm. 

Thus, here we consider the (2+1)d Y-SYK model in the limit $N\Lambda/(M\varepsilon_F)\ll1$, corresponding to a scenario in which the bosonic self-energy is subdominant: $|\delta\Pi(i\Omega_m)| \ll \Omega_m^2,\omega_D^2$. The effective bosonic propagator thus takes on its bare form and, evaluated at criticality, becomes $\mathcal{D}(i\Omega_m)\simeq\log(1+\omega_D^2/\Omega_m^2)/(4\pi c^2)$, as in the case of the strange-metal regime. Focusing on the intermediate-temperature regime above the quantum critical point, we seek self-consistent solutions to the normal-state saddle-point Eqs.~\eqref{eq:Sigma_saddlept} and~\eqref{eq:Pi-saddlept} where the fermionic self-energy is strongly renormalized, satisfying $|\Sigma(i\omega_n)|\gg|\omega_n|,\Lambda$ at low frequencies. In this regime, the effective fermionic propagator is given by $\mathcal{G}(i\omega_n)\approx-\nu\Lambda/\Sigma(i\omega_n)$.

\subsection{Zero-Temperature Analysis}
\label{subsec:BM-0T}

\subsubsection*{Power-law solutions in (2+1)-dimensions}

At zero temperature, we solve the saddle-point equations self consistently at low energies by assuming local criticality, that is, the effective (momentum-integrated) propagators have the power-law form:
\begin{equation}
\mathcal{G}_{\psi}(\tau)\sim \text{sgn}(\tau)\frac{1}{|\tau|^{2\Delta_{\psi}}}, \quad \mathcal{D}_{\phi}(\tau)\sim \frac{1}{|\tau|^{2\Delta_{\phi}}}. 
\end{equation}
Here, $\Delta_\psi$ and $\Delta_\phi$ are the scaling dimensions of the effective fermionic and bosonic propagators, respectively. For $\Delta_{\psi}>0$, $\mathcal{G}$ is dominated by $\Sigma$ at low energies such that $\mathcal{G}(i\omega)=-\nu\Lambda/\Sigma(i\omega)$. In this regime, we define
\begin{equation}
\Sigma(i\omega) = -ia_{\Delta}\text{sgn}(\omega)|\omega|^{1-2\Delta_{\psi}}.
\label{eq:power-law}
\end{equation}
In the imaginary time domain, the effective fermionic propagator is~\cite{Esterlis2019}:
\begin{equation}
\mathcal{G}(\tau) =- \frac{\nu\Lambda}{\pi a_{\Delta}}\Gamma(2\Delta_{\psi})\text{sin}(\pi\Delta_{\psi})\text{sgn}(\tau)|\tau|^{-2\Delta_{\psi}}.
\label{eq:Im-time-Geff}
\end{equation}
The Fourier transform of Eq.~\eqref{eq:power-law} is given by
\begin{equation}
\Sigma(\tau) = -\frac{a_{\Delta}}{\pi}\Gamma(2-2\Delta_{\psi})\sin(\pi\Delta_{\psi})\text{sgn} (\tau)|\tau|^{2\Delta_{\psi}-2}.
\label{eq:Im-time-sigma}
\end{equation}

\subsubsection*{Undamped boson tuned to criticality}

In terms of the dynamical part of the bosonic self-energy, $\delta\Pi(i\Omega_m)\equiv\Pi(i\Omega_m)-\Pi(0)$, and the renormalized boson mass $m_{B}$, defined via $m_{B}^2\equiv m_{B,0}^2-\Pi(0)$, the exact bosonic propagator is given by
\begin{equation}
\mathcal{D}(i\Omega_m)=\frac{1}{4\pi c^2}\log\left(1+\frac{\omega_D^2}{\Omega_m^2+m_{B}^2-\delta\Pi(i\Omega_m)}\right). 
\end{equation}
The exact closed-form expression for the imaginary-time propagator $\mathcal{D}(\tau)$ is challenging to obtain. Neglecting the dynamical contribution to the bosonic self-energy allows $\mathcal{D}(\tau)$ to be evaluated analytically. In general, $\delta \Pi$ can be neglected when it is subdominant to the free part of the bosonic propagator. This frequently occurs within electron-phonon systems when there is a small adiabatic ratio between the bosonic and electronic energy scales~\cite{Gnezdilov2025}: $\omega_D/\varepsilon_F\ll1$. Alternatively, $\delta \Pi$ can be neglected in a controlled manner, without assuming a particular separation of fermionic and bosonic energy scales, by letting the number of flavors of bosons greatly exceed that of the fermions, i.e., $M\gg N$. In the analysis that follows, we assume $\delta \Pi\approx0$ with this justification in mind.

At zero temperature, the boson mass is tuned to criticality: $m_{B}(T)|_{T=0}=0$. The effective bosonic propagator is then given by $\mathcal{D}(i\Omega)=\log\left(1+\omega_D^2/\Omega^2\right)/(4\pi c^2)$, which in the imaginary time domain becomes (see Sec.~\ref{App:Boson_FT}):
\begin{equation}
\mathcal{D}(\tau)=\frac{1}{4\pi c^2}\frac{1-e^{-|\tau|\omega_D}}{|\tau|}\sim \frac{1}{4\pi c^2}\frac{1}{|\tau|}, \quad |\tau|\omega_{D}\gg1. 
\label{eq:Im-time-D-no-Pi}
\end{equation}
At nonzero temperatures, $\mathcal{D}(\tau)$ is defined over the imaginary time interval $\tau\in(0,1/T)$. Therefore, we require $T\ll\omega_D$ to ensure the long-time asymptotic behavior can be reached. This condition is trivially satisfied at zero temperature.

We thus observe that the effective propagator for the underdamped quantum-critical boson has the low-energy scaling dimension $\Delta_\phi=1/2$. By equating the product of the imaginary-time effective propagators in the long-time limit for the fermions and bosons given by Eqs.~\eqref{eq:Im-time-Geff} and~\eqref{eq:Im-time-D-no-Pi} respectively with the imaginary-time expression for the fermionic self-energy, Eq.~\eqref{eq:Im-time-sigma}, we obtain the constraints on the IR fermion scaling dimension and coefficient
\begin{equation}
2\Delta_{\psi}-2 = -2\Delta_{\psi}-1, \quad a_{\Delta}=\sqrt{\lambda\Lambda \Gamma(2\Delta_{\psi})/\Gamma(2-2\Delta_{\psi})},
\end{equation}
Solving for the fermion scaling dimension, we find
\begin{equation}
\Delta_{\psi}=1/4;\quad a_{1/4}=\sqrt{2\lambda\Lambda};\quad\text{thus} \quad\Sigma(i\omega)\simeq-i\text{sgn}(\omega)\sqrt{2\lambda\Lambda|\omega|}.
\label{eq:0T-BM-soln}
\end{equation}
Note that, here we find that in (2+1)d the fermion and boson scaling dimensions obey the constraint $2\Delta_{\psi}+\Delta_{\phi}=1$, which was originally noted in an analogous (0+1)d fermion-boson model~\cite{Fu2017}. In order for this solution to be valid, we require $|\Sigma(i\omega)|\gg|\omega|,\Lambda/2$. This restricts the frequency interval for the solution to $1/(8\lambda)\ll|\omega|/\Lambda\ll2\lambda$, which is well-defined for sufficiently strong coupling. The IR constraint is required by $\Sigma(i\omega)\gg\Lambda/2$, which may be restated as a sufficient condition (for the leading-order expansion of the effective fermionic propagator) in terms of the lowest Matsubara frequency $\omega_0=\pi T$, that $8\pi \lambda T/\Lambda\gg1$, alternatively in terms of a threshold temperature: $T_{\CL}\gg \Lambda/(8\pi\lambda)$. Combining this with the condition $T\ll\omega_D$, we obtain $\Lambda/(8\pi\lambda)\ll T_{\CL}\ll \omega_D$.

\subsection{Nonzero Temperatures} 
\label{subsec:BM-finite-T}

Let $\Lambda/2,|i\omega_n|\ll|\Sigma(i\omega_n)|$ such that $\mathcal{G}(i\omega_n)\sim \nu\Lambda G(i\omega_n)|_{\xi=0}\sim -\nu\Lambda/\Sigma(i\omega_n)$. The $\xi_{\bm{k}}$-dependence is not pertinent in this limit. Furthermore, let us continue to assume that $\delta\Pi(i\Omega_m)$ may be neglected and that $T\ll\omega_D$, so that for $\tau\gg1/\omega_D$ the bad-metal solution with the fermion scaling dimension $\Delta_{\psi}=1/4$ is valid and $L(\tau)\equiv4\pi c^2\mathcal{D}(\tau)\sim1/|\tau|$ at zero temperature. In the imaginary time domain, the fermionic Dyson and large-$N$ saddle-point equations are respectively given by 
\begin{equation}
\int_0^\beta  \Sigma(\tau_{1},\tau_{2})G(\tau_{2},\tau_{3})d\tau_{2} =-\delta(\tau_{1}-\tau_{3});\quad \quad \Sigma(\tau_{1},\tau_{2})= \lambda\Lambda L(\tau_{1},\tau_{2})G(\tau_{2},\tau_{1}).
\label{eq:Dsyon-and-saddle-pt}
\end{equation}
Here we have written $\Sigma,G$, and $L$ as bilocal fields. In the undamped scenario, the effective bosonic propagator $L(\tau)\sim1/|\tau|$ has the low-energy scaling dimension $\Delta_\phi=1/2$ at zero temperature. In the limit $M\gg N$, we may neglect the dynamical part of the bosonic self-energy, and thus the bosonic Dyson and large-$N,M$ saddle-point equations are trivial. At nonzero temperatures, the bosonic propagator is thus given by
\begin{equation}
L(\tau) = T\sum_{\Omega_m} e^{-i\Omega_m\tau} \log\left(1+\omega_D^2/(m_B^2+\Omega_m^2)\right).
\label{eq:BosonL-tau-finite-T-exact}
\end{equation}

\begin{figure}[t]
\centering\includegraphics[width=0.6\columnwidth]{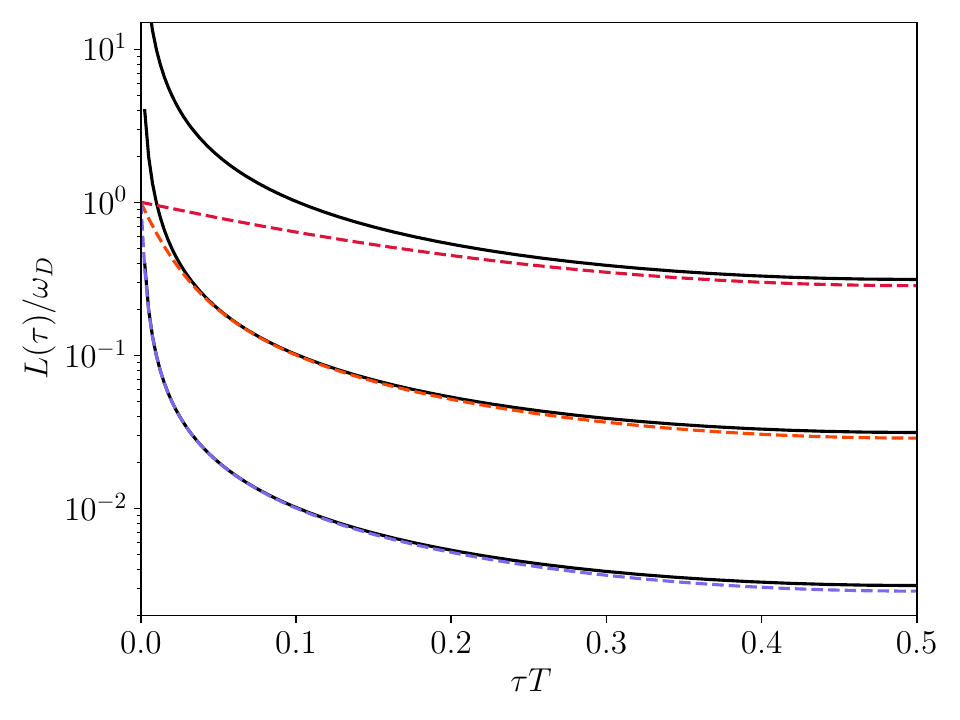}
\caption{Comparison of the exact (dimensionless) bosonic propagator $L(\tau)/\omega_D$ with the conformal solution Eq.~\eqref{eq:conformal-boson-prop}, in the imaginary time domain. The exact propagator for temperatures $T/\omega_D = 1/1000, 1/100, 1/10$ is given by the dashed blue, orange, and red lines respectively, which are plotted along side the conformal solution, given by the solid black lines.}
\label{fig:FT-comparison}
\end{figure}

In analogy to the SYK model, we search for symmetries within Eqs.~\eqref{eq:Dsyon-and-saddle-pt} and~\eqref{eq:BosonL-tau-finite-T-exact} to facilitate the generalization of the previous solution to nonzero temperature. Indeed, dispersive effects present in the bosonic propagator inhibit the recovery of the full emergent reparameterization and U(1) gauge invariance found in the SYK model. Nevertheless, we observe an emergent conformal covariance~\cite{Parcollet1998, Parcollet1999, Sachdev2015, Maldacena2016, Sarosi2017, Fu2017, Marcus2019, Chowdhury2022} in the effective propagators and fermionic self-energy at zero temperature as well as an approximate reparameterization covariance under a particular imaginary-time reparameterization. The effective propagators and fermionic self-energy of the (2+1)d Y-SYK model with local disorder are approximately covariant under the following transformation:
\begin{align}
\tau &= f(\sigma), \nonumber\\  
G(\tau_{1},\tau_{2}) &= [f^\prime(\sigma_1) f^\prime(\sigma_2)]^{-1/4} \widetilde{G}(\sigma_1,\sigma_2), \nonumber\\ 
\Sigma(\tau_{1},\tau_{2}) &= [f^\prime(\sigma_1)f^\prime(\sigma_2)]^{-3/4}\widetilde{\Sigma}(\sigma_1,\sigma_2), \nonumber\\
L(\tau_{1},\tau_{2}) &= [f^\prime(\sigma_1)f^\prime(\sigma_2)]^{-1/2}\widetilde{L}(\sigma_1,\sigma_2),
\label{eq:transformation}
\end{align}
for $f(\sigma)=(a\sigma+b)/(c\sigma+d)$, $ad-bc=1$ at zero temperature and $f(\sigma)=\tan(\pi T\sigma)/(\pi T)$ at finite temperatures, within the $\Delta_\psi=1/4,\; \Delta_\phi=1/2$ `conformal' regime. The former transformation corresponds to the SL$(2,\mathbb{R})$ (conformal) transformation. Under the latter transformation, the zero-temperature solution $L(\tau)=1/|\tau|$ is mapped to the finite-temperature `conformal' solution 
\begin{equation}
L(\tau)=\left|\frac{\pi T}{\sin(\pi T \tau)}\right|, |\tau|<1/T.
\label{eq:conformal-boson-prop}
\end{equation}
Quite fortuitously, we find that, when tuned to criticality, the exact imaginary time bosonic propagator and the conformal solution have significant agreement in the IR for $T\ll\omega_D$. In Fig.~\ref{fig:FT-comparison} we compare the exact imaginary time bosonic propagator and the conformal solution with $\Delta_{\phi}=1/2$ for various temperatures. 

From this comparison, it is apparent that the strongest agreement between the conformal solution and the exact boson propagator occurs at intermediate values of $\tau$ for $T\ll\omega_D$. At zero temperature, the exact bosonic propagator is $\sim1/|\tau|$ as $|\tau|\omega_{D}\rightarrow\infty$ and thus asymptotically approaches the conformal solution in the IR. However, at finite temperature, from the comparison provided in Fig.~\ref{fig:FT-comparison}, it is clear that there is a small persistent discrepancy between the conformal solution and the exact propagator in the deep IR ($\tau T\rightarrow1/2$). The precision of the approximate conformal solution can be assessed directly from the numerical analysis provided in Sec.~\ref{sec:NumericalAnalysis}. In Appendix~\ref{App:Boson_FT}, we compare the exact and conformal propagators in the limit $T/\omega_{D}\ll1$. 

In order to simplify our nonzero-temperature analysis, we leverage the observed approximate emergent ``thermal'' reparameterization symmetry given by Eq.~\eqref{eq:transformation}. Applying the thermal reparametrization transformation to the zero-temperature fermionic self-energy, $\Sigma(\tau) = -C\text{sgn}(\tau)|\tau|^{-3/2}$, where $C=\sqrt{\lambda\Lambda/\pi}/2$, we find
\begin{align}\nonumber
\widetilde{\Sigma}(\sigma_1,\sigma_2) &= [f^\prime(\sigma_1)f^\prime(\sigma_2)]^{3/4}\Sigma(f(\sigma_1),f(\sigma_2))\\ \nonumber
&= -C\frac{[\sec^2(\pi T \sigma_1)\sec^2(\pi T \sigma_2)]^{3/4}}{|\tan(\pi T \sigma_1)-\tan(\pi T \sigma_2))|^{3/2}} (\pi T)^{3/2}\\ \nonumber
&= -C [\sin(\pi T \sigma_1)\cos(\pi T \sigma_2) - \sin(\pi T \sigma_2) \cos(\pi T \sigma_1)]^{-3/2} (\pi T)^{3/2}\\
& = -C \left[\frac{\pi T}{\sin(\pi T(\sigma_1-\sigma_2))}\right]^{3/2}, \quad 0<\sigma<1/T.
\end{align}
Now we drop the tilde notation and use $\tau$ for the imaginary-time variable. Applying the Kubo-Martin-Schwinger (KMS) condition $\Sigma(\tau+1/T)=-\Sigma(\tau)$ (and utilizing a single imaginary-time argument), we obtain
\begin{equation}
\Sigma(\tau) = -\frac{1}{2}\sqrt{\frac{\lambda \Lambda}{\pi}}\text{sgn}(\tau)\left|\frac{\pi T}{\sin(\pi T \tau)}\right|^{3/2}; \quad T,|\tau|^{-1}\ll\omega_D.
\end{equation}
On the Matsubara frequency axis, the fermionic self-energy becomes
\begin{align}
\Sigma(i\omega_n) &= \int_{\tau_D}^{1/T-\tau_D}e^{i\omega_n\tau}\Sigma(\tau)d\tau \nonumber\\  
&= -\frac{1}{2}\sqrt{\frac{\lambda\Lambda}{\pi}}\int_{\tau_D}^{1/T-\tau_D}e^{i\omega_n\tau}\left|\frac{\pi T}{\sin(\pi T \tau)}\right|^{3/2}d\tau \nonumber \\ 
&= -2  i \sqrt{\lambda \Lambda \pi T}\left[\frac{\Gamma\left(\frac{3}{4}+\frac{\omega_n}{2\pi T}\right)}{\Gamma\left(\frac{1}{4}+\frac{\omega_n}{2\pi T}\right)} + O(\sqrt{T\tau_D})\right].
\label{eq:approx-conformal-soln-BM}
\end{align}
Here $\tau_{D}$ is a cut off in the time domain that arises from the UV cutoff for frequencies $\omega_{D}$ where $\tau_{D}\sim1/\omega_{D}$. The fermionic self-energy is independent of the cutoff at leading order in $T\tau_D$. Upon analytical continuation, we find
\begin{equation}
\Sigma_R(\omega,T)\simeq-2  i \sqrt{\lambda \Lambda \pi T}\frac{\Gamma\left(\frac{3}{4}-i\frac{\omega}{2\pi T}\right)}{\Gamma\left(\frac{1}{4}-i\frac{\omega}{2\pi T}\right)}.
\label{eq:Sigma_R-conformalSoln}
\end{equation}
The zero-temperature limit of this expression can be deduced using the asymptotic expansion of the Gamma function~\cite{WhittakerWatsonBook}: $\log\Gamma(z+\alpha)=(z+\alpha-1/2)\log(z)-z+(1/2)\log(2\pi)+o(1)$. Thus, $\Gamma(z+\alpha)/\Gamma(z+\beta)=z^{\alpha-\beta}(1+O(1/z))$. Hence, the zero-temperature limit (for fixed, non-zero $\omega$) is 
\begin{equation}
\lim_{T\rightarrow0}\Sigma_R(\omega,T) = -2  i \sqrt{\lambda \Lambda \pi T}\sqrt{-i\frac{\omega}{2\pi T}} = -i\sqrt{\lambda\Lambda|\omega|}(1-i\text{sgn}(\omega)).
\end{equation}
This result agrees with the analytical continuation of the zero-temperature result obtained in Eq.~\eqref{eq:power-law}.

\subsubsection*{Fermionic propagator at nonzero temperatures}

The effective fermionic propagator at zero temperature is obtained by setting $\Delta_{\psi}=1/4$ in Eq.~\eqref{eq:Im-time-Geff}. Applying the thermal reparameterization transformation for arbitrary temperatures, we find 
\begin{equation}
\mathcal{G}(\tau)=-\frac{\nu}{2}\sqrt{\frac{\Lambda}{\pi \lambda}}\text{sgn}(\tau)\left|\frac{\pi T}{\sin(\pi T\tau)}\right|^{1/2}.
\label{eq:finite-T-conformal-bm}
\end{equation}

\subsubsection*{Bosonic self-energy at nonzero temperatures}

Employing the imaginary-time saddle-point equation $\Pi(\tau)=-2g^2(N/M)\mathcal{G}(\tau)\mathcal{G}(-\tau)$ and the approximate finite-temperature conformal solution, Eq.~\eqref{eq:finite-T-conformal-bm}, the bosonic self-energy is given by 
\begin{equation}
\Pi(\tau) =-\left(\frac{N}{M}\frac{\Lambda}{\varepsilon_F}\right)\frac{\omega_D^2}{2\pi}\left|\frac{\pi T}{\sin(\pi T \tau)}\right|, \quad \tau_D<|\tau|<1/T-\tau_D.
\end{equation}
For self-consistency, we assume that $\delta\Pi$ may be neglected in comparison to the (inverse) bare propagator. The solution presented in this section is therefore contingent on the separation of scales $(N/M)(\Lambda/\varepsilon_F)\ll1$.

\subsection{The Fully Conformal Scenario and Comparison with Previous Work} 
\label{subsec:Fully-conformal-case}

In the case of a large bosonic self-energy $|\delta\Pi(i\Omega_m)|\gg\omega_D^2,\Omega_m^2$, the logarithm in $\mathcal{D}(i\Omega_m)$ can be expanded to leading order: $\mathcal{D}(i\Omega_m)\sim (\omega_D^2/(4\pi c^2))/(m_{B}^2-\delta\Pi(i\Omega_m))$. Assuming a finite-bandwidth and a large fermionic self-energy too, the electron and boson propagators then have analogous forms to the (0+1)-dimensional theory~\cite{Esterlis2019}. The normal-state saddle-point equations then simplify to
\begin{align}
\label{eq:Sigma-low-E}
\Sigma(i\omega_n) &= -\lambda \omega_D^2\Lambda T\sum_{\omega_{m}}\frac{1}{m_{B,0}^2-\Pi(i\omega_{n}-i\omega_{m})}\frac{1} {\Sigma(i\omega_{m})}, \\
\Pi(i\Omega_m)&=-2(g\nu\Lambda)^2 \left(\frac{N}{M}\right)T\sum_{\omega_{n}}\frac{1} {\Sigma(i\omega_{n})\Sigma(i\omega_{n}+i\Omega_m)}.
\label{eq:Pi-low-E}
\end{align}

\subsubsection*{Zero-temperature analysis}

Earlier works~\cite{Esterlis2019, ClassenChubukov2021} have evaluated Eqs.~\eqref{eq:Sigma-low-E},~\eqref{eq:Pi-low-E} in the vicinity of the QCP, where the renormalized boson mass is zero at zero temperature. The saddle-point equations are solved at low energies by assuming the solutions to have the power-law form as in Sec.~\ref{subsec:BM-0T}. In the absence of a boson mass, the saddle-point Eq.~\eqref{eq:Sigma-low-E} for the electron self-energy is homogeneous, leaving the coefficient $a_{\Delta}$ of the electron self-energy unconstrained. References~\cite{Esterlis2019, ClassenChubukov2021} fix the value of $a_{\Delta}$ by imposing the zero-temperature constraint $m_{B,0}^2=\Pi(0)$, ensuring that the renormalized boson mass is zero. The values of $a_{\Delta}$ and $\Delta_\psi$ are then readily obtained at zero temperature. The self-consistent equation that determines the scaling dimension of the fermionic self-energy is determined in Refs.~\cite{Esterlis2019, ClassenChubukov2021}; when modified to account for finite electronic bandwidth, we obtain: 
\begin{equation}
\frac{N}{M}\frac{\Lambda}{\varepsilon_F} = \frac{1}{2}\left(\frac{4\Delta_\psi-1}{4\Delta_\psi-2}\right)\left(\cot^2(\pi\Delta_{\psi})-1\right).
\end{equation}
Here we have simplified the coefficient to $N/M$ by assuming $k_D\approx k_F$ such that $\omega_D^2/(2mc^2)=\varepsilon_F$. Due to the presence of the prefactor $\Lambda/\varepsilon_{F}$, the scaling of the fermionic self-energy is tuned by non-universal parameters in the theory. In the limit $(N/M)(\Lambda/\varepsilon_F)\rightarrow0$, $\Delta_\psi\rightarrow1/4$ as found in Sec.~\ref{subsec:BM-0T}. However, as $\Delta_\psi\rightarrow1/4$, $a_\Delta\propto\sqrt{\Delta_\psi-1/4}$, so that the coefficient of the fermionic self-energy tends to zero when tuned towards $\Delta_\psi=1/4$ at $T=0$.

\subsubsection*{Nonzero-temperature analysis}

For $T>0$, the renormalized boson mass is nonzero, thereby breaking the homogeneity of the saddle-point equation of the fermionic self-energy and fixing the value of $a_{\Delta}$. In this regime, the renormalized boson mass has been found to self-tune to criticality~\cite{Esterlis2019, ClassenChubukov2021, Kleger2026}, limiting to zero as $T\rightarrow0$ for arbitrary bare bosonic mass. This approach stands in contrast to the constraint method employed in Refs.~\cite{Li2024,Kleger2026,Esterlis2021} and discussed in Sec.~\ref{sec:tune-QCP}, which relies on fixing the bosonic density to its zero-temperature value at criticality. This approach is used in the present work.

In Ref.~\cite{Kleger2026}, it was found that when the (2+1)-d saddle-point equations are solved without a constraint, i.e., by relying on the self-tuning feature, the strong-coupling regime admits a regime with strong impurity physics, i.e. $\Delta_\psi=1/2$, analogous to the finite-temperature strong-coupling regime found in Ref.~\cite{Esterlis2019}. However, the bosonic self-energy in this regime is not large enough to justify the expansion of the effective bosonic propagator as in Eqs.~\eqref{eq:Sigma-low-E}.

\section{Entropy}

Logarithmically enhanced low-temperature heat capacity is a hallmark of strange-metal phenomenology. Here, we demonstrate its presence in the underdamped scenario by computing the low-temperature entropy density. In Sec.~\ref{subsec:Action} we derive the large-$N,M$ action and saddle-point equations (including both normal and anomalous components). In Sec.~\ref{subsec:normal-state-grand-potential} we derive the low-temperature entropy-density in the normal state for the underdamped scenario.

\subsection{Large-\texorpdfstring{$N, M\ $}\  Action}
\label{subsec:Action}

The action describing $N$ electrons interacting with $M$ bosons in $(2+1)$d through a spatially disordered Y-SYK interaction is given by $S = S_{\psi} + S_{\phi} + S_{\psi\phi}$, where the individual contributions are 
\begin{align}
S_{\psi} & = \int_x \, \sum_{i=1}^N \sum_{\sigma=\uparrow\downarrow} \psi_{i\sigma}^\dag(x) \left(\partial_\tau - \tfrac{\nabla_{\bm r}^2}{2m}-\mu\right) \psi_{i\sigma}(x); \quad 
S_{\phi}  =  \int_x \, \sum_{l=1}^M \frac{1}{2} \phi_l(x) (-\partial_\tau^2 -c^2 \nabla_{\bm r}^2 + m_{B,0}^2) \phi_l(x), \label{S_0} \\
S_{\psi\phi} & =  \frac{1}{\sqrt{NM}} \sum_{i,j=1}^N  \sum_{l=1}^M \int_x  g_{ijl}({\bm r}) \sum_{\sigma=\uparrow\downarrow}\psi_{i\sigma}^\dag(x) \psi_{j\sigma}(x) \phi_l(x) = \frac{1}{\sqrt{NM}} \sum_{i,j=1}^N  \sum_{l=1}^M   \int_x g_{ijl}({\bm r}) {\cal O}_{ij}(x)\phi_l(x).
\label{S_YSYK} 
\end{align}
Here, $\psi_{i\sigma}$, $\psi_{i\sigma}^\dag$ are fermions with spin $\sigma=\uparrow\downarrow$, mass $m$, and a chemical potential $\mu$; $\phi_l$ are bosons with bare mass $m_{B,0}$, and $c$ is the speed of sound. The variable $x$ denotes $x = (\tau,{\bm r})$ and the integration notation is $\int_x\equiv\int_0^{\beta}d\tau\int d^2r$, where $\beta = 1/T$ is the inverse temperature. We define ${\cal O}_{ij}(x) = \sum_{\sigma} \psi_{i\sigma}^\dag(x) \psi_{j\sigma}(x)$. To allow for anomalous propagators in the limit $N,M\rightarrow\infty$~\cite{Esterlis2019, ClassenChubukov2021}, we draw the coupling constants $g_{ijl}$ in the Y-SYK term \eqref{S_YSYK} independently from a Gaussian orthogonal ensemble with zero mean and non-zero variance $\langle g_{ijl}({\bm r}) g_{ijl} ({\bm r}')\rangle = 2 g^2 \delta({\bm r}-{\bm r}')$.

Performing the disorder average, we obtain 
\begin{equation}
\langle e^{-S_{\psi\phi}}\rangle = \exp\!\bigg[\! \frac{g^2}{4NM} \!\int_{x,x'}\!\! \delta({\bm r}-{\bm r}') \!\sum_{l}^M\!\phi_l(x) \phi_l(x')   \!\sum_{i,j}^N({\cal O}^\dag_{ij}(x) {\cal O}^\dag_{ij}(x') \!+\! {\cal O}^\dag_{ij}(x) {\cal O}_{ij}(x') \!+\! {\cal O}_{ij}(x) {\cal O}^\dag_{ij}(x') \!+\! {\cal O}_{ij}(x) {\cal O}_{ij}(x') )\!\bigg]. 
\label{dav1}
\end{equation}
To decouple the interactions in Eq.~\eqref{dav1}, we use the following resolutions of unity in terms of auxiliary bilocal fields:
\begin{align} 
    {\cal I} &= \int \! \mathcal{D}\Sigma\mathcal{D}G \exp\bigg[  \int_{x,x'}  \sum_{\sigma=\uparrow\downarrow} \Sigma_{\sigma}(x-x') (N G_{\sigma}(x'\!-x) - \sum_{i=1}^N \psi^\dag_{i\sigma}(x)\psi_{i\sigma}(x') ) \bigg], \label{G_def}\\ 
    {\cal I} &=  \int \! \mathcal{D}\Phi\mathcal{D}F^+ \exp\bigg[  \int_{x,x'}\Phi_{\uparrow\downarrow}(x-x') (N F^+_{\downarrow\uparrow}(x'\!-x) - \sum_{i=1}^N \psi^\dag_{i\uparrow}(x)\psi^\dag_{i\downarrow}(x') ) \bigg], \label{F+_def}\\ 
    {\cal I}  &= \int \! \mathcal{D}\Phi^+ \mathcal{D}F \exp\bigg[  \int_{x,x'} \Phi^+_{\downarrow\uparrow}(x-x') (N F_{\uparrow\downarrow}(x'\!-x) - \sum_{i=1}^N \psi_{i\downarrow}(x)\psi_{i\uparrow}(x') ) \bigg],  \label{F_def} \\
    {\cal I}  &= \int \! \mathcal{D}\Pi\mathcal{D}D \exp\bigg[ - \frac{1}{2}\int_{x,x'}  \Pi(x-x') (M D(x'\!-x) - \sum_{l=1}^M \phi_{l}(x)\phi_{l}(x') ) \bigg]. \label{D_def}
\end{align}  
We assume that the Cooper pairing occurs in the spin-singlet channel and the normal components of the fermionic propagator are spin diagonal. Using Eqs.~(\ref{G_def}-\ref{F_def}), we have 
\begin{align} 
   \frac{1}{N} \sum_{i,j}^N{\cal O}_{ij}(x) {\cal O}^\dag_{ij}(x') &= \frac{1}{N} \sum_{i,j}^N \sum_{\sigma\sigma'}\psi_{i\sigma}^\dag(x) \psi_{j\sigma}(x) \psi_{j\sigma'}^\dag(x') \psi_{i\sigma'}(x')  
   = - N \sum_{\sigma=\uparrow\downarrow}  G_{\sigma}(x-x') G_{\sigma}(x'\!-x), \label{OOdag}\\ 
   \frac{1}{N} \sum_{i,j}^N{\cal O}_{ij}^\dag(x) {\cal O}_{ij}(x') &= \frac{1}{N} \sum_{i,j}^N \sum_{\sigma\sigma'}\psi_{j\sigma}^\dag(x) \psi_{i\sigma}(x) \psi_{i\sigma'}^\dag(x') \psi_{j\sigma'}(x')  
   = - N \sum_{\sigma=\uparrow\downarrow}  G_{\sigma}(x-x') G_{\sigma}(x'\!-x), \label{OdagO}\\ \nonumber
   \frac{1}{N} \sum_{i,j}^N{\cal O}_{ij}(x) {\cal O}_{ij}(x') &= \frac{1}{N} \sum_{i,j}^N (\psi_{i\uparrow}^\dag(x) \psi_{j\uparrow}(x) \psi_{i\downarrow}^\dag(x') \psi_{j\downarrow}(x') + \psi_{i\downarrow}^\dag(x) \psi_{j\downarrow}(x) \psi_{i\uparrow}^\dag(x') \psi_{j\uparrow}(x'))\\ 
   &= N(F_{\uparrow\downarrow}(x-x')F^+_{\downarrow\uparrow}(x'\!-x)+ F^+_{\downarrow\uparrow} (x-x')F_{\uparrow\downarrow}(x'\!-x)), \label{OO}\\ \nonumber
   \frac{1}{N} \sum_{i,j}^N{\cal O}_{ij}^\dag(x) {\cal O}_{ij}^\dag(x') &= \frac{1}{N} \sum_{i,j}^N ( \psi_{j\uparrow}^\dag(x) \psi_{i\uparrow}(x) \psi_{j\downarrow}^\dag(x') \psi_{i\downarrow}(x') + \psi_{j\downarrow}^\dag(x) \psi_{i\downarrow}(x) \psi_{j\uparrow}^\dag(x') \psi_{i\uparrow}(x') ) \\ 
   &= N(F_{\uparrow\downarrow}(x-x')F^+_{\downarrow\uparrow}(x'\!-x)+ F^+_{\downarrow\uparrow} (x-x')F_{\uparrow\downarrow}(x'\!-x)). \label{OdagOdag}
\end{align}
For a spin-diagonal normal channel and spin-singlet pairing, we set $\Sigma_{\uparrow\uparrow} = \Sigma_{\downarrow\downarrow}=\Sigma$, $G_{\uparrow\uparrow} = G_{\downarrow\downarrow}=G$, $\Phi_{\uparrow\downarrow}=-\Phi_{\downarrow\uparrow}=\Phi$, $\Phi^+_{\downarrow\uparrow}=-\Phi^+_{\uparrow\downarrow}=\Phi^*$, $F_{\uparrow\downarrow}=-F_{\downarrow\uparrow}=F$, $F^+_{\downarrow\uparrow}=-F^+_{\uparrow\downarrow}=F^*$.
From Eqs.~(\ref{S_0}-\ref{dav1},\ref{OOdag}-\ref{OdagOdag}), we note that the action $S$ is quadratic in the fermionic and bosonic fields, and thus they can be integrated out exactly, leading to 
\begin{align}
S &= S_{\psi\Sigma} + S_{\phi\Pi} + S_{\Sigma G} + S_{\Pi D} + S_{DG}, \label{S_eff}\\ 
S_{\psi\Sigma} &= -N \, {\rm Tr}\log  
  \begin{pmatrix}
  (\partial_\tau - \frac{\nabla_{\bm r}^2}{2m} - \mu)\delta(x-x')  + \Sigma(x-x') & \Phi(x-x') \\ \Phi^*(x-x') &  (\partial_\tau + \frac{\nabla_{\bm r}^2}{2m} + \mu)\delta(x-x')  - \Sigma(x'\!-x) 
  \end{pmatrix},\\ 
   S_{\phi\Pi} &= \frac{M}{2} \, {\rm Tr}\log\big((-\partial_\tau^2 - c^2 \nabla_{\bm r}^2 +m_{B,0}^2)\delta(x-x') - \Pi(x-x')\big) ,\\ 
  S_{\Sigma G} &= -N \int_{x,x'} ( 2  \Sigma(x-x') G(x'\!-x) + \Phi(x-x') F^*(x'\!-x) + \Phi^*(x-x') F(x'\!-x) ),\\
  S_{\Pi D} &= \frac{M}{2} \int_{x,x'} \Pi(x-x')  D(x'\!-x), \\ 
  S_{DG} &= \frac{N g^2}{2} \int_{x,x'}  \delta(\bm{r}-\bm{r}^\prime) ( 2 G(x-x') G(x'\!-x)  - F(x-x')F^*(x'\!-x) - F^*(x-x')F(x'\!-x) ) D(x'\!-x). \label{S_DG}
\end{align}

In the large-$N,M$ limit, we obtain the saddle-point equations for $G, \Sigma, F, \Phi, \Pi, D$. To do so, we set the variation of the effective action \eqref{S_eff} with respect to the bilocal fields equal to zero. Performing these variations of $S$ gives
\begin{align}
G(i\omega_n, {\bm k}) &= \frac{i\omega_n + \xi_{-\bm k}+ \Sigma(-i\omega_n)}{(i\omega_n - \xi_{\bm k} - \Sigma(i\omega_n))(i\omega_n + \xi_{-\bm k}+ \Sigma(-i\omega_n))-|\Phi(i\omega_n)|^2}, \\
F(i\omega_n, {\bm k}) &= \frac{\Phi(i\omega_n)}{(i\omega_n - \xi_{\bm k} - \Sigma(i\omega_n))(i\omega_n + \xi_{-\bm k}+ \Sigma(-i\omega_n))-|\Phi(i\omega_n)|^2}, \\
D(i\Omega_m, {\bm q}) &= \frac{1}{\Omega_m^2 + c^2\bm{q}^2 + m_{B,0}^2 - \Pi(i \Omega_m)}, \\
    \Sigma(i\omega_n) &= g^2 T \sum_{\omega_{n'}} \int \! D(i\omega_n-i\omega_{n'},{\bm q}) G(i\omega_{n'},{\bm k}') \frac{d^2q}{(2\pi)^2}\frac{d^2k'}{(2\pi)^2}, \label{Sigma_eq}  \\
\Phi(i\omega_n) &= -g^2 T \sum_{\omega_{n'}} \int \! D(i\omega_n-i\omega_{n'},{\bm q}) F(i\omega_{n'},{\bm k}') \frac{d^2q}{(2\pi)^2}\frac{d^2k'}{(2\pi)^2},\label{Phi_eq} \\ \nonumber
\Pi(i\Omega_m) &= -g^2 \left(\frac{N}{M}\right)T \sum_{\omega_n} \int \!  \bigg( 2 G(i\omega_n, {\bm k}) G(i\omega_n+i\Omega_m, {\bm k}') \\ &\quad - F(i\omega_n, {\bm k})F^*(i\omega_n+i\Omega_m, {\bm k}') - F^* (i\omega_n, {\bm k})F(i\omega_n+i\Omega_m, {\bm k}') \bigg) \frac{d^2k}{(2\pi)^2}\frac{d^2k'}{(2\pi)^2}, \label{Pi_eq}
\end{align}
where $\omega_n = (2n+1)\pi T$ and $\Omega_m = 2m\pi T$ are fermionic and bosonic Matsubara frequencies, respectively. The momentum integrals in Eqs.~(\ref{Sigma_eq}-\ref{Pi_eq}) factorize because of the spatial Dirac delta function in Eq.~\eqref{S_DG}.

\subsection{Normal-State Grand Potential}
\label{subsec:normal-state-grand-potential}

In the normal state, the Y-SYK action (including the constraint) is
\begin{align} 
S &=-2 N \int_k \log\left(-i\omega_n + \xi_{\bm k} + \Sigma(k)\right) + \frac{M}{2} \int_q \log\left(\Omega_m^2 +c^2\bm{q}^2 +m_{B,0}^2- \Pi(q)\right) - M \int \frac{i\lambda(\tau,\bm{r})}{2\gamma}d\tau d^2r \nonumber \\
&\quad -2 N \int_k  \Sigma(k) G(k) +\frac{M}{2}\int_q  \Pi(q) D(q) +N g^2 \int_{k,k',\bm{q}} G(k) D(i\omega_n-i\omega_{n^\prime},\bm{q}) G(k') .
\label{eq:S}
\end{align}
This expression has the same general form as in Luttinger and Ward's~\cite{Luttinger1960} and Eliashberg's papers~\cite{Eliashberg1963}. At the saddle-point level, the second and third terms in the second line cancel each other due to the variational condition $\delta S/\delta D=0$. In particular, both terms describe the exact same skeleton diagram in the grand potential~\cite{Zhang2023}. Thus, only the first term in the second line contributes to the saddle-point action. 

Using the definitions $\mathcal{G}(i\omega_{n})\equiv \int_{\bm{k}} G(k)$, $\mathcal{D}(i\Omega_{m})\equiv \int_{\bm{q}} D(q)$, and $\lambda=\nu g^2 /\left(4 \pi c^2\right)$ and substituting the saddle-point self-energies from Eqs.~\eqref{eq:Sigma_saddlept} and~\eqref{eq:Pi-saddlept} into the action, along with the mass constraint $i\lambda(\tau,\bm{r})=m_{B,0}^2$, we obtain the normal-state grand canonical potential:
\begin{equation}
\bm{\Omega} = N\left[-2T\int_{\bm{k}}\sum_{\omega_n} \log (-G^{-1}(k)) -2 T \sum_{\omega_n}\mathcal{G}(i\omega_n)\Sigma(i\omega_n)\right]+ M\left[\frac{1}{2} T \int_{\bm{q}}\sum_{\Omega_m} \log (D^{-1}(q)) -\frac{m_{B,0}^2(T)}{2\gamma}\right]. \label{eq:FreeEnergy}
\end{equation}

\subsection{Fermionic Contribution}

We will use Eq.~\eqref{eq:FreeEnergy} to evaluate the entropy in the infinite-bandwidth limit, where $|i\omega_n-\Sigma(i\omega_n)|\ll \Lambda/2$.  In the infinite-bandwidth limit, the interacting contribution to $\log(-G^{-1})$ cancels  the $\mathcal{G}\Sigma$ term in Eq.~\eqref{eq:FreeEnergy}~\cite{Esterlis2021, Zhang2023}. The details of this result can be found in Refs.~\cite{Esterlis2021, Kleger2026}.  Thus, we are left with only the free-fermion grand potential 
\begin{equation}
\bm{\Omega}_{F}=-2NT\int_{\bm{k}}\log\left(1+e^{- \xi_{\bm{k}}/T}\right).
\end{equation}
The fermionic entropy-density is given by $s_{F} = -\frac{1}{N}(\partial_T \bm{\Omega}_F)_{V,\mu}$, which yields the Sommerfeld result $s_F=(2\pi^2/3)\nu T$. For consistency, we normalize all contributions to the entropy-density by $N$.

\subsection{Bosonic Contribution}

The bosonic contribution to the grand potential is
\begin{equation}
\bm{\Omega}_B = \frac{M}{2}\left[T\int_{\bm{q}}\sum_{\Omega_m} \log (D^{-1}(q)) - \frac{1}{\gamma}m_{B}^2(T)\right], 
\label{eq:F_boson}
\end{equation}
where the constraint term has been re-expressed in terms of the renormalized boson mass using the temperature independence of $\Pi(0)$ in the infinite-bandwidth limit. Entropy is determined from $s_{B} = -\frac{1}{N}(\partial_T \bm{\Omega}_B)_{V,\mu}$. Performing the Matsubara frequency summation using a contour integral, and using the constraint equation in Eq.~\eqref{eq:Constraint_saddlept}, the bosonic contribution to the entropy density is
\begin{align}
s_{B} &= -\frac{1}{2}\frac{M}{N}\left[\partial_T \int_{\bm{q}}\int_{-\infty}^{\infty}n_B(\Omega)\text{Im}\log\left(D_{R}^{-1}(\Omega,\bm{q})\right)\frac{d\Omega}{\pi}-\frac{1}{\gamma}\partial_T m_{B}^2(T)\right] \nonumber \\ \nonumber
 &= -\frac{1}{2}\frac{M}{N}\Bigg\{ \int_{\bm{q}}\int_{-\infty}^{\infty}\Bigg[\left(\partial_T n_B(\Omega)\right)\text{Im}\log\left(D_{R}^{-1}(\Omega,\bm{q})\right)+(\partial_T m_B^2(T))n_B(\Omega)\text{Im}D_R(\Omega,\bm{q})\Bigg]\frac{d\Omega}{\pi}- \frac{1}{\gamma}\partial_T m_B^2\Bigg\} \nonumber \\
  &= -\frac{1}{2\pi}\frac{M}{N}\int_{\bm{q}}\int_{-\infty}^{\infty}\left(\partial_T n_B(\Omega)\right)\text{Im}\log\left(D_{R}^{-1}(\Omega,\bm{q})\right) d\Omega. 
 \label{eq:BosonEntropy1}
\end{align}

In the infinite-bandwidth limit, $\partial_T \Pi_R(\Omega)=0$, so the temperature derivative acts only on the distribution function; an analogous result is found in Sec.~19.5 of Ref.~\cite{AGD}. Substituting $c^2q^2=x\omega_{D}^2$, Eq.~\eqref{eq:BosonEntropy1} can then be expressed as 
\begin{equation}
s_{B} = -\frac{\omega_D^2}{8\pi^2 c^2}\frac{M}{N}\int_0^{1}\left[\int_{-\infty}^{\infty}  \left(\partial_T n_B(\Omega)\right) \text{Im}\log\left(-\Omega^2+x\omega_{D}^2+m_B^2-i c_d\Omega\right)d\Omega\right]dx.
\end{equation}
Next, we carefully evaluate the imaginary part of the logarithm, accounting for which quadrant of the complex plane the logarithm lies in, and find that the entropy becomes
\begin{align}
s_{B} &= \frac{\omega_D^2}{8\pi^2 c^2}\frac{M}{N}\int_0^{1}\left[\int_{-\infty}^{\infty} \left(\partial_T n_B(\Omega)\right)\arctan\left(\frac{c_d\Omega}{-\Omega^2+x\omega_{D}^2+m_B^{2}}\right) 
d\Omega \right]dx \nonumber \\ 
 &\quad + \frac{\omega_D^2}{4\pi c^2}\frac{M}{N}\int_0^{1}\left[\int_{\sqrt{x\omega_{D}^2+m_B^2}}^{\infty} \left(\partial_T n_B(\Omega)\right)
d\Omega \right]dx.
\label{eq:BosonEntropy2}
\end{align}
Since $c_d\propto \lambda$, if we set $\lambda=0$, then the first term vanishes. Thus, the second term is the entropy density for a free boson with quantum-critical mass, $m_B$, in (2+1) dimensions. Note that the mass is held fixed, since the temperature dependence of the mass was shown above to cancel with the constraint term. Let $\omega_{B}=\sqrt{\omega_D^{2}+m_B^2}$. The free contribution is evaluated in Ref.~\cite{Kleger2026}:
\begin{align}
s_{B,0}	&= \frac{\omega_D^2}{2\pi c^2}\frac{M}{N}\biggl\{\left(\frac{\omega_{B}}{\omega_{D}}\right)^{2}\log\left(1-e^{-\omega_{B}/T}\right)-\left(\frac{m_B}{\omega_{D}}\right)^{2}\log\left(1-e^{-m_B/T}\right) \nonumber\\
&\quad +3\left(\frac{T}{\omega_{D}}\right)^{2}\left[\frac{m_B}{T}\textrm{Li}_{2}(e^{-m_B/T})-\frac{\omega_{B}}{T}\textrm{Li}_{2}(e^{-\omega_{B}/T})+\textrm{Li}_{3}(e^{-m_B/T})-\textrm{Li}_{3}(e^{-\omega_{B}/T})\right]\biggr\},
\label{eq:boson-free-entropy}
\end{align}
Here, $\text{Li}_{s}(z)\equiv\sum_{k=1}^{\infty}z^{k}/k^{s}$ is the polylogarithm. Let $m_B=\text{arccosh}(3/2)T$. The free-boson contribution to the entropy density has the following low-temperature limit 
\begin{equation} 
s_{B,0} = \frac{6\zeta(3)}{5\pi}\frac{M}{N}\left(\frac{\omega_D}{c}\right)^2\left(\frac{T}{\omega_D}\right)^2;\quad T\ll \omega_D.
\label{eq:impurity-entropy}
\end{equation}

The contribution to the entropy arising from the electron-boson interaction is given by the first term in Eq.~\eqref{eq:BosonEntropy2}. For low temperatures and weak boson damping, we can approximate this contribution to the entropy as follows:
\begin{align}
s_{\text{Int}}&=
\frac{\omega_D^2}{8\pi^2c^2}\frac{M}{N}\int_{0}^{1}\left[\int_{-\infty}^{\infty} y\ \text{cosech}^{2}(y) \arctan\left(\frac{c_d(2Ty)}{-(2Ty)^2+x\omega_{D}^2+m_B^{2}}\right) 
dy \right]dx \nonumber \\
&\approx \frac{\omega_D^2}{4\pi^2c^2}\frac{M}{N}c_d T \int_{0}^{1}\frac{1}{x\omega_{D}^2 + m_B^2}dx \int_{-\infty}^{\infty}y^2 \text{cosech}^2(y) dy \nonumber\\ 
&=\frac{1}{12}\frac{M}{N}\frac{c_d}{c^2} T \log\left(1+\frac{\omega_D^2}{m_B^2}\right)\nonumber\\
&=\frac{2\pi^2}{3}\lambda \nu T \log\left(1+\frac{\omega_D^2}{m_B^2}\right).
\label{eq:s_int}
\end{align}
The integral over $y$ can be evaluated using the digamma function and gives $\pi^2/3$, and the integral over $x$ is elementary. In the final line of Eq.~\eqref{eq:s_int} we have used the definitions of $c_d$ and $\nu$. The interacting bosonic contribution is the dominant contribution to the entropy density at low temperature for finite values of $N/M$. For a boson mass $m_B\simeq T$, the low-temperature interacting bosonic contribution to the entropy density has the expected form $\sim T\log T$ as in the marginal Fermi liquid~\cite{Crisan1996}:
\begin{equation}
s_{\text{Int}}\approx \frac{4\pi^2}{3} \lambda \nu T\log\left(\frac{\omega_D}{T}\right).
\end{equation}
In the infinite-bandwidth limit, all contributions to the entropy scale with $\sim N$, except for the free boson contribution, which scales as $\propto M$. Although the free-boson contribution is dominant in the limit $N\ll M$, it is subdominant for finite values of $N/M$ at low temperatures, since $s_{B,0}\sim T^2$ and $s_{\text{Int}}\sim T\log T$.

The heat capacity can be evaluated using $C=T(\partial s/\partial T)_V$. The leading contribution at low temperatures is 
\begin{equation}
C\sim \frac{4\pi^2}{3}\lambda\nu T \log\left(\frac{\omega_D}{eT}\right).
\end{equation}

\section{Electrical Conductivity}
\subsection{Formula and Overview}

\subsubsection*{Kubo Formula and Spectral Representation}

For a system with nonzero dissipation, the longitudinal optical conductivity is defined by~\cite{Li2024}: 
\begin{align} 
\sigma(\Omega) &= \frac{i}{\Omega}\left[K^{xx}_R(\Omega,0)-\lim_{\Omega\rightarrow 0} \text{Re}K^{xx}_R(\Omega,0)\right], \nonumber\\
K^{xx}_R(\Omega,0) &= i \int_{0}^{+\infty} d t \int e^{i \Omega t} \left\langle\left[\hat{J}_{x}(\bm{r}, t), \hat{J}_{x}(\bm{0}, 0)\right]\right\rangle d^2r.
\end{align}
The components of the current operator are given by $\hat{J}_x$. Using the Matsubara formalism, we express the Fourier transform of the time-ordered correlation function as
\begin{equation}
K^{xx}(i\Omega_m,0) = 2e^2T\sum_{\omega_n}\int \left(\frac{k_x}{m}\right)^2 G(i\omega_n, \bm{k}) G(i\omega_n+i\Omega_m, \bm{k})\frac{d^2k}{(2\pi)^2}.
\end{equation}
Using the spectral representation $G(z) = \int_{\mathbb{R}} \frac{A(\omega)}{\omega-z}\frac{d\omega}{2\pi}$, where $\text{Im}z\neq0$ and $A(\omega) = -2 \text{Im}G_R(\omega)$, we obtain
\begin{equation}
K^{xx}(i\Omega_m,0) = 2e^2T\sum_{\omega_n}\int \biggl[\int \left(\frac{k_x}{m}\right)^2 \frac{A(\omega, \bm{k})}{i \omega_n -\omega}\frac{A(\omega^\prime , \bm{k})}{i \omega_n + i \Omega_m - \omega^\prime}\frac{d^2 k}{(2\pi)^2}\biggr] \frac{d \omega d \omega^\prime}{(2\pi)^2}.
\end{equation}
Evaluating the Matsubara frequency summation then performing the analytical continuation of the external frequency $i\Omega_m \rightarrow \Omega+i0^+$ leads to
\begin{equation}
K^{xx}_R(\Omega,0) = 2e^2\int \left[\int \left(\frac{k_x}{m}\right)^2 A(\omega, \bm{k})A(\omega^\prime, \bm{k})\frac{n_F(\omega)-n_F(\omega^\prime )}{\omega-\omega^\prime+\Omega+i0^+}\frac{d^2 k}{(2\pi)^2}\right]\frac{d \omega d \omega^\prime}{(2\pi)^2}.
\label{eq:full-AA-K-expression}
\end{equation}
For the dispersion relation $\xi_{\boldsymbol{k}}=k^2 /(2 m)-\varepsilon_F$, where $\xi_{\boldsymbol{k}} \in[-\varepsilon_F, \Lambda -\varepsilon_F]$, and for an arbitrary function $F(k)$, we have
\begin{equation}
    \int\left(\frac{k_x}{m}\right)^2 F(k) \frac{d^2 k}{(2 \pi)^2}=\frac{1}{(2 \pi m)^2} \int_{k_{\min }}^{k_{\max }} k^3 F(k) d k \int_0^{2 \pi} \cos ^2 \theta d \theta=\int_{-\varepsilon_F}^{\Lambda -\varepsilon_F}(\xi+\varepsilon_F) F(\xi) \frac{d \xi}{2 \pi} .
\end{equation}

\subsubsection*{Optical conductivity}

The $\xi$-integration in Eq.~\eqref{eq:full-AA-K-expression} can be evaluated exactly. In the limit of infinite bandwidth, $\xi\in[-\infty,\infty]$,
\begin{align}
K^{xx}_R(\Omega,0) &= 2e^2\int \left[\int_{-\infty}^{\infty} \left(\xi+\varepsilon_F\right) A(\omega, \xi)A(\omega^\prime, \xi)\frac{n_F(\omega)-n_F(\omega^\prime )}{\omega-\omega^\prime+\Omega+i0^+}\frac{d\xi}{2\pi}\right]\frac{d \omega d \omega^\prime}{(2\pi)^2} \nonumber \\
&= 4e^2\int \frac{(\varepsilon_{F}-\Sigma^{\prime}_{R}(\omega^{\prime})+\omega^{\prime})\Sigma^{\prime\prime}_{R}(\omega)+(\varepsilon_{F}-\Sigma^{\prime}_{R}(\omega)+\omega)\Sigma^{\prime\prime}_{R}(\omega)}{(\omega-\omega^{\prime}-\Sigma_R^{\prime}(\omega)+\Sigma_R^{\prime}(\omega^{\prime}))^2 + (\Sigma_R^{\prime\prime}(\omega)+\Sigma_R^{\prime\prime}(\omega^\prime))^2}
\frac{n_F(\omega)-n_F(\omega^\prime )}{\omega-\omega^\prime+\Omega+i0^+}\frac{d \omega d \omega^\prime}{(2\pi)^2}.
\end{align}
To leading order in the Fermi energy, we obtain 
\begin{align}
K^{xx}_R(\Omega,0) 
&= -4e^2\varepsilon_F\int \text{Im}\left[\frac{1}{\omega-\omega^{\prime}+\Sigma_R(\omega^{\prime})-\Sigma_A(\omega)}\right]\frac{n_F(\omega)-n_F(\omega^\prime )}{\omega-\omega^\prime+\Omega+i0^+}\frac{d \omega d \omega^\prime}{(2\pi)^2}.
\label{eq:full-AA-K-expression3}
\end{align}
Thus, the imaginary part of the response kernel becomes
\begin{equation}
\text{Im}K^{xx}_R(\Omega,0) = e^2\varepsilon_F\int_{-\infty}^{\infty} \left(n_F(\omega)-n_F(\omega+\Omega )\right)\text{Im}\left[\frac{1}{-\Omega+\Sigma_R(\omega+\Omega)-\Sigma_A(\omega)}\right]\frac{d \omega }{\pi}.
\label{eq:full-AA-K-expression4}
\end{equation}
Using $\text{Re}\sigma(\Omega)=-\text{Im}K_R^{xx}(\Omega,0)/\Omega$, we obtain the real part of the optical conductivity:
\begin{equation}
\text{Re}\sigma(\Omega) 
= e^2\varepsilon_F\text{Re}\left[\int_{-\infty}^{\infty} \frac{n_F(\omega)-n_F(\omega+\Omega)}{i\Omega}\frac{1}{-\Omega+\Sigma_R(\omega+\Omega)-\Sigma_A(\omega)}\frac{d \omega }{\pi}\right].
\label{eq:full-AA-K-expression5}
\end{equation}
The optical conductivity $\sigma(\Omega)$ is a holomorphic function of $\Omega$ in the upper half of the complex plane and obeys $\lim_{R\rightarrow\infty}\sup_{|\Omega|=R,\text{Im}(\Omega)\geq0}|\sigma(\Omega)|=0$. Therefore, provided the real-axis Cauchy principal-value integrals converge, $\sigma(\Omega)$ satisfies the Kramers-Kronig relations. Complex differentiability of $\sigma(\Omega)$ is demonstrated in Appendix A of Ref.~\cite{Berthod2013}. The unique function which satisfies both Eq.~\eqref{eq:full-AA-K-expression5} and the Kramers-Kronig relations is thus~\cite{Allen2015, Berthod2013, NormanChubukov2006} 
\begin{equation}
\sigma(\Omega) = e^2\varepsilon_F\int_{-\infty}^{\infty} \frac{n_F(\omega)-n_F(\omega+\Omega )}{i\Omega}\frac{1}{-\Omega+\Sigma_R(\omega+\Omega)-\Sigma_A(\omega)}\frac{d\omega}{\pi}.
\label{eq:full-AA-K-expression6}
\end{equation}

\subsubsection*{DC conductivity} 
\label{subsec:DC-conductivity}

The DC conductivity for the isotropic dispersion in two spatial dimensions with symmetric bandwidth is
\begin{equation}
\sigma =e^2\bigintssss_{-\infty}^{\infty} \left[\bigintssss_{-\varepsilon_F}^{\Lambda-\varepsilon_F}  \left(\xi+\varepsilon_F\right)A^2(\omega,\xi)\ \frac{d\xi}{2\pi}\right] \left(-\frac{\partial n_F}{\partial\omega}\right) \frac{d\omega}{2\pi}.
\label{eq:conductivity-arb-bw}
\end{equation}

\subsection{Infinite-Bandwidth Limit}

\subsubsection*{Exact result}

In the infinite-bandwidth limit, the $\xi$-integration evaluates to 
\begin{equation}
J(\omega)\equiv\int_{-\varepsilon_F}^{\infty}(\xi
+\varepsilon_F)A^2(\omega,\xi)\frac{d\xi}{2\pi}=\frac{1}{\pi} + \frac{\varepsilon_F+\omega-\Sigma_R^{\prime}(\omega)}{|\Sigma_R^{\prime\prime}(\omega)|}\left(\frac{1}{2}+ \frac{1}{\pi}\arctan\left[\frac{\varepsilon_F+\omega-\Sigma_R^{\prime}(\omega)}{|\Sigma_R^{\prime\prime}(\omega)|}\right]\right).
\end{equation} 
Expanding in $1/\varepsilon_F$, $J(\omega)=(\varepsilon_F+\omega-\Sigma_R^{\prime}(\omega))/|\Sigma_R^{\prime\prime}(\omega)|+O(1/\varepsilon_F^2)$. In the particle-hole symmetric case $\Lambda/2=\varepsilon_F$ and the contribution from $(\omega-\Sigma_R^{\prime}(\omega))/|\Sigma_R^{\prime\prime}(\omega)|$ vanishes because it is an odd function integrated over symmetric bounds. To leading order in the Fermi energy, Eq.~\eqref{eq:conductivity-arb-bw} simplifies to the well-known result
\begin{equation} 
\sigma =\frac{e^2\varepsilon_F}{4\pi}\int_{-\infty}^{\infty} \frac{\sech^2(x)}{|\Sigma_R^{\prime\prime}(2Tx)|} dx.
\label{eq:cond-inf-bw}
\end{equation}
In the infinite-bandwidth limit, the imaginary part of the fermionic self-energy is given by Eq.~\eqref{eq:Im-SE-SM-simplified}. For $T\ll \omega_{D}$, only the region $|\omega|<\omega_D$ contributes to the integral. The conductivity is then $\sigma_{\MFL}= e^2\varepsilon_F I_{\MFL}/(4\pi^2\lambda T)$, where $I_{\MFL}$ is given by
\begin{equation}
I_{\MFL} =\int_{-\infty}^{\infty}\frac{\sech^2(x)}{\log(3+2\cosh(2x))}dx=2\int_{0}^{1}\frac{1}{\log(5-x^2)-\log(1-x^2)}dx\approx1.0072.
\end{equation}
Approximating $I_{\MFL}\approx1$, we obtain $\rho_{\MFL}\simeq4\pi^2\lambda T/(e^2\varepsilon_F)$. Restoring $\hbar, k_{B}$, we obtain the result given in the main text: $\rho_{\MFL}\simeq 2\pi\lambda (k_{B}T/\varepsilon_F)(h/e^2)$.

\subsubsection*{Conformal-limit approach}

The fermionic self-energy is evaluated exactly in the infinite-bandwidth limit, which enables the electrical resistivity to be determined exactly. In the finite-bandwidth limit, the large-$N,M$ saddle-point equation for the fermionic self-energy is a (discretized) integral equation that must be solved self-consistently. In Sec.~\ref{sec:Aux_Res_BM} we apply approximate conformal methods to determine the fermionic self-energy in this regime. Here, we apply the same approach for the infinite-bandwidth limit and compare with the exact result to facilitate a qualitative measure of the accuracy of the approximate conformal approach. 

In the infinite-bandwidth limit, the effective fermionic and (undamped quantum-critical) bosonic propagators are dominated by their dispersion. In $(2+1)$-d they share the effective scaling dimension $\Delta_{\psi}=\Delta_\phi=1/2$, such that
\begin{equation}
4\pi c^2\mathcal{D}(\tau)=-\mathcal{G}(\tau)/\nu =  \pi T/\sin(\pi T \tau),\quad \tau_D<\tau<1/T-\tau_D.
\end{equation}
Thus, the fermionic self-energy is $\Sigma(\tau)=g^2\mathcal{D}(\tau)\mathcal{G}(\tau)\simeq - \lambda (\pi T /\sin(\pi T \tau))^2$. In terms of Matsubara frequencies,
\begin{equation}
\Sigma(i\omega_n) = 2i\lambda \omega_{n}\left[\psi\left(\frac{\omega_n}{2\pi T}\right)+\frac{\pi T}{\omega_{n}}+\log(2\pi e^{\gamma-1} T \tau_D) + O((\tau_D T)^2)\right].
\end{equation}
Here, $\psi$ is the digamma function and $\gamma=-\psi(1)$ is the Euler-Mascheroni constant. This result for $\Sigma(i\omega_{n})$ is analogous to the MFL solution found in Ref.~\cite{Patel2018}. Thus, $\Sigma_R^{\prime\prime}(\omega) \simeq 2 \lambda \left[\omega\text{Im}\psi(-i \omega/(2\pi T))+ \pi T\right]$ is a function of $x$ alone and is independent of $\tau_D$. Let $x\equiv\omega/(2T)$. The conductivity is given by
\begin{equation}
\sigma =\frac{e^2\varepsilon_F}{4\pi}\int_{-\infty}^{\infty} \frac{\sech^2(x)}{|\Sigma_R^{\prime\prime}(2Tx)|} dx = \frac{e^2\varepsilon_F}{8\pi \lambda T}\int_{-\infty}^{\infty} \frac{\sech^2(x)}{|2x\text{Im}\psi(-i x/\pi)+ \pi|}  dx 
=\frac{e^2\varepsilon_F}{8\pi \lambda T}\frac{14\zeta(3)}{\pi^3}.
\label{eq:conductivity-sm}
\end{equation}
A similar integral was approximately evaluated in Ref.~\cite{Patel2018}. Here we have obtained the exact result; see Sec.~\ref{sec:Aux_Res_SM} for the derivation. The approximate conformal solution slightly overestimates the exact resistivity by a factor of $\pi^2/(7\zeta(3))\approx1.17$.

\subsubsection*{Infinite-bandwidth Resistivity bound}

For a strongly incoherent metal, we consider the limit $|\Sigma_{R}^{\prime\prime}|\gg\varepsilon_{F},|\Sigma_{R}^\prime|, T$. To leading order, $J(\omega)=1/\pi + (\varepsilon_F+\omega-\Sigma^{\prime}_R(\omega))/(2|\Sigma_R^{\prime\prime}(\omega)|)+O(1/|\Sigma_R^{\prime\prime}(\omega)|^2)$. This places an upper bound on electronic transport for a system with finite Fermi energy and infinite bandwidth (i.e., no UV cutoff). In two spatial dimensions, we find
\begin{equation}
\rho<\pi\times h/e^2.
\end{equation}
In this scenario, the absence of a UV cutoff in the electronic dispersion imposes a lower bound on the conductivity. To exceed the MIR bound without resistivity saturation, the electronic dispersion must therefore have a UV cutoff, corresponding to a finite bandwidth. In Ref.~\cite{Kleger2026} we investigate electronic transport in this finite-bandwidth scenario. We find that the Drude-like conductivity crosses over to $\sigma_{\BM}=2(e\Lambda\tau/\pi)^2\hbar^{-3}\ll e^2/h$ for $\Lambda\tau\ll\hbar$, necessarily describing a bad metal with a resistivity well above the MIR bound. In the following section we compute conductivity using the conformal-limit solution in this regime.

\subsection{Finite-Bandwidth Limit}

\label{subsec:rho-FBWL}

In the strongly incoherent bad-metal regime, we find low-energy solutions with local criticality, with the electronic spectral function given by $A(\omega) = -2\text{Im}G_R(\omega) = 2\text{Im}(1/\Sigma_R(\omega))$. The $\xi$-integral contributes the factor 
\begin{equation}
\int_{-\varepsilon_F}^{\Lambda-\varepsilon_F}(\xi+\varepsilon_F)d\xi = \Lambda^2/2.
\end{equation}
Let $x=\omega/(2T)$, DC conductivity is given by 
\begin{align}
\sigma_{\CL} &=\frac{e^2\Lambda^2}{16\pi^2}\int_{-\infty}^{\infty} \sech^2(x) A^2(x) dx = \frac{e^2\Lambda}{16\pi^3 \lambda T}\int_{-\infty}^{\infty} \sech^2(x) \left(\text{Re}\left[\frac{\Gamma(\frac{1}{4}-i\frac{x}{\pi})}{\Gamma(\frac{3}{4}-i\frac{x}{\pi})}\right]\right)^2 dx
= \frac{e^2\Lambda}{16\pi \lambda T}.
\label{eq:conductivity-bm}
\end{align}
The integral above is evaluated in Sec.~\ref{sec:Aux_Res_BM}. We thus obtain the resistivity $\rho_{\CL} = 16\pi \lambda T/(e^2\Lambda)$ for $T\gg\Lambda/(8\pi\lambda)$. From this lower-temperature threshold, we obtain $\rho_{\CL}\gg (1/\pi)(h/e^2)$, confirming the bad-metal nature of this solution. Compared with the resistivity obtained for the strange metal, we find $\rho_{\MFL}/\rho_{\CL}\sim\pi/2$ for $\Lambda/2\sim\varepsilon_F$.

Many strange metals have a large electronic bandwidth such that the separation of scales required to access the ``conformal'' bad-metal solutions occurs only at exceptionally high temperatures. This allows $T$-linear resistivity to persist through the MIR bound without a discernible change in slope. However, other strange metals have significantly smaller electronic bandwidth, such as $\text{Sr}_3\text{Ru}_2\text{O}_7$. This material exhibits~\cite{Bruin2013} $T$-linear resistivity at low temperatures, when tuned to the critical magnetic field $\mu_0H_c= 7.9 T$, as well as at high temperatures for arbitrary field strength. At temperatures above 100K, the resistivity is again linear but the slope decreases by approximately 30$\%$ throughout the crossover between these two regimes. Although a direct comparison with our analysis is not straightforward, this reduction is qualitatively consistent with the $\sim2/\pi$ crossover predicted by our theoretical analysis. We leave a detailed investigation of this feature, incorporating the band structure of $\text{Sr}_3\text{Ru}_2\text{O}_7$, to future work.

\subsection{Mott-Ioffe-Regel Limit and Associated Bounds on Resistivity}

Several common definitions of the Mott-Ioffe-Regel limit can be specified in terms of the mean free path $\ell$ (defined as the average distance traversed between collisions) of the electronic excitation. Delineated, they are as follows~\cite{Hussey2004}: (i) $\ell\sim\lambda_F$, (ii) $\ell\sim a$, and (iii) $\ell\sim1/k_F$, in terms of the Fermi wavelength, lattice spacing, and inverse Fermi wavevector respectively. The scattering time is defined via the electronic self-energy $\tau\equiv1/(2|\Sigma_R^{\prime\prime}|)$. In certain instances, the transport and scattering times are equivalent, $\tau_{\text{tr}}=\tau$. It is more convenient to re-express these bounds in terms of a scattering time $\tau$, using the relation $\ell=v_F\tau$. Thus, we obtain (i) $\Lambda\tau\sim 2\pi\hbar$, (ii) $\Lambda\tau\sim 2\hbar$, and (iii) $\Lambda\tau\sim \hbar$ respectively. To derive (ii), we assume $\hbar v_F/a=\Lambda/2$, for an electronic bandwidth $\Lambda$ and hopping amplitude $t$, related to the Fermi energy by $\Lambda/2=\varepsilon_F=4t$. Conditions (i) and (iii) are related by $k_F\equiv2\pi/\lambda_F$. As an aside, we note that there can be variation in the value of $v_F$ along the Fermi surface~\cite{Kaminski2005}, as well as a separation of scales between the electronic bandwidth and Fermi energy (i.e., $\Lambda\neq2\varepsilon_F$), particularly for strong particle-hole asymmetry. 

Finally, each criterion can be used to determine the corresponding resistivity bound. When the electronic bandwidth is the largest energy scale, $\Lambda\gg T,1/\tau$, the resistivity is given by a Drude-like formula, which in two spatial dimensions we write as $\rho_{\text{Drude}}=h/(e^2\Lambda\tau)$~\cite{Kleger2026}. For simplicity, we approximate $1/(2\tau)\approx|\Sigma_R^{\prime\prime}(\omega=0)|$, although the exact transport time involves an average over frequencies $\omega\sim T$. When $|\Sigma_R|\gtrsim\Lambda$, both the real and imaginary parts can contribute to the resistivity, leading to non-Boltzmann transport. Neglecting $\Sigma_R^{\prime}$ for simplicity, we obtain a formula that interpolates between the Drude $\rho_{\text{Drude}}\sim1/\tau$ and bad-metal $\rho_{\BM}\sim1/\tau^2$ regimes:
\begin{equation}
\rho_{\text{Exact}} = \frac{h}{e^2s}\frac{\pi}{2}\left[\arctan(s)+\frac{s}{s^2+1}\right]^{-1},
\label{eq:resistivity-interpolation}
\end{equation}
where $s\equiv\Lambda\tau/\hbar$. This equation is derived from Eq.~\eqref{eq:conductivity-arb-bw} under the assumption of low temperatures $-n_F^{\prime}(\omega)\approx\delta(\omega)$, such that $A^2(\omega,k)\approx A^2(0,k)$, and $|\Sigma_R^{\prime}|\ll |\Sigma_R^{\prime\prime}|,\Lambda$ such that $\Sigma_R^{\prime}$ may be neglected. The previously stated criterion and their equivalent forms are summarized in Table~(1) of the main text. We also include an estimate of the respective resistivity bounds using the Drude formula and the interpolation formula Eq.~\eqref{eq:resistivity-interpolation}, denoted ``Exact'', as well as their respective ratios.

We find that the criterion $\ell\sim\lambda_F/(2\pi)$ (equivalently $\Lambda\tau\sim\hbar$) yields a resistivity closest to the quantum unit, $h/e^2$. Interestingly, none of the commonly used bounds precisely correspond to this quantum unit, except when estimated using the Drude formula. It is also notable that the crossover between Drude-like and bad-metal-like scaling of the resistivity occurs in the vicinity of this bound. This observation is particularly relevant in the context of the present work, where the crossover between the strange-metal regime (where dispersive effects are dominant) and the bad-metal regime (where the electronic self-energy dominates over dispersive effects), also occurs roughly at $\Lambda\tau\sim\hbar$.

\begin{figure}[t]
\centering\includegraphics[width=.45\columnwidth]{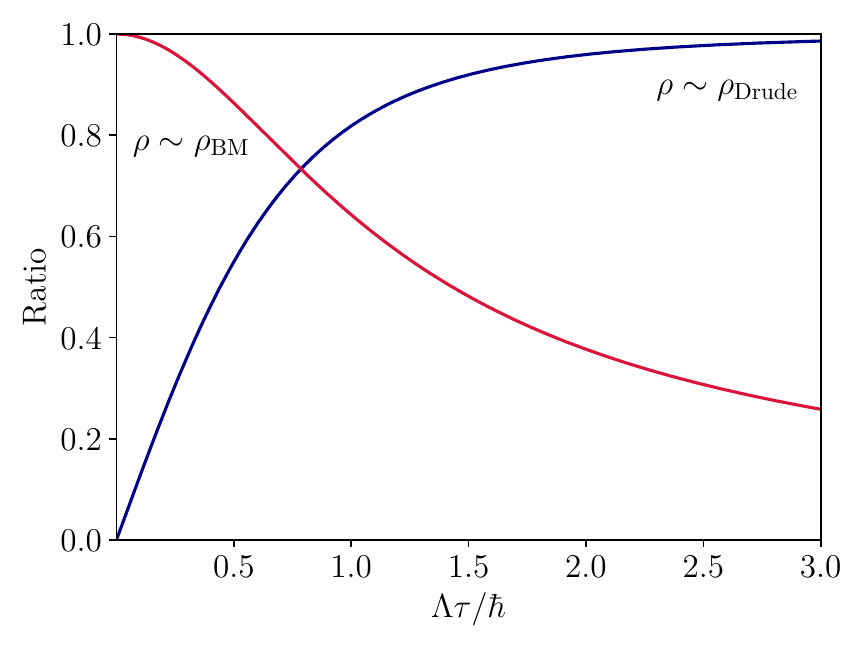}
\caption{$\rho_{\text{Drude}}/\rho_{\text{Exact}}$ and $\rho_{\BM}/\rho_{\text{Exact}}$ in blue and red respectively, versus $\Lambda\tau/\hbar$.}
\label{fig:rho-comparison}
\end{figure}

Finally, Fig~\ref{fig:rho-comparison} compares the Drude and bad-metal resistivities, normalized by the exact formula given in Eq.~\eqref{eq:resistivity-interpolation}, as a function of $\Lambda\tau$. We find that the crossover point, at which the two approximation have equal errors, occurs for $\Lambda\tau\approx0.785\hbar$, where both formulae underestimate the resistivity by approximately $0.73\%$.

\section{Numerical Analysis}
\label{sec:NumericalAnalysis}

\begin{figure}[h]
\centering\includegraphics[width=.45\columnwidth]{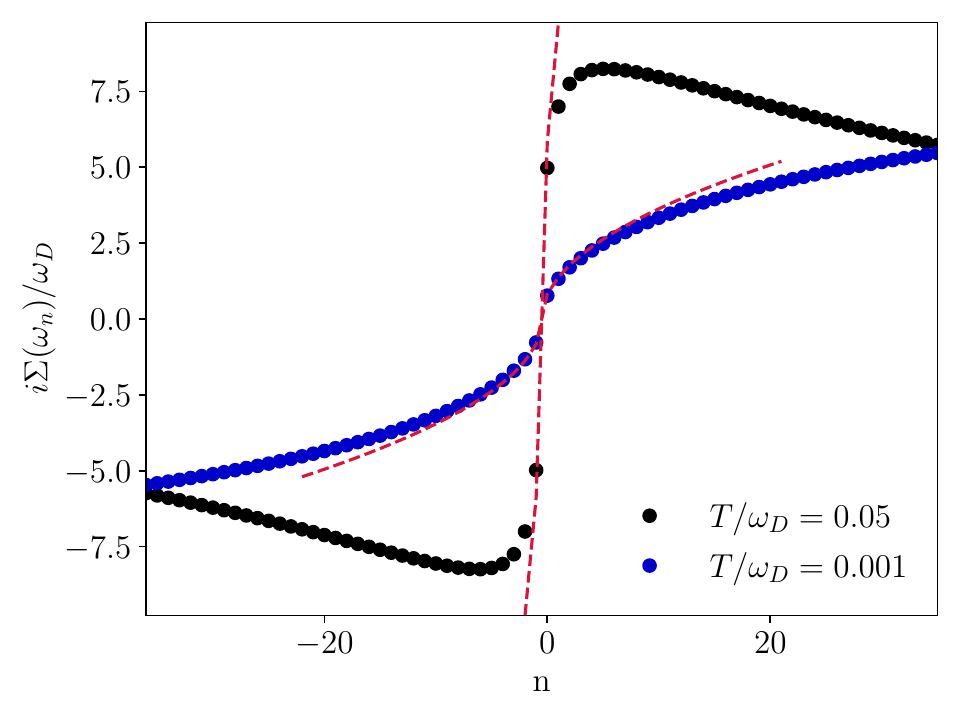}
\caption{Comparison of the numerically exact (black and blue dotted lines) and approximate ``conformal'' solution (dashed red) for the fermionic self-energy. Parameters $\Lambda=\omega_D$ and $\lambda=100$ are used.}
\label{fig:Im-axis-numerical-SE}
\end{figure}

The large-$N,M$ saddle-point equation for the fermionic self-energy in \eqref{eq:Sigma_saddlept} is solved self-consistently with a finite electronic bandwidth using the quantum-critical boson mass described in Sec.~\ref{sec:tune-QCP}. We use a fixed-point iterative method, summarized as follows. First, we truncate the fermionic Matsubara frequencies $\omega_n=(2n+1)\pi T;\; n\in [-N,N-1]$, where $N\in\mathbb{Z}$ is chosen to be sufficiently large to ensure that the solution converges within the frequency regime of interest. The algorithm begins with an initial guess for $\Sigma(i\omega_n)$, which we take to be zero, i.e., $\forall n\in[-N,N-1]:\Sigma_0(i\omega_n)=0$, where the subscript ``$0$'' denotes the first iteration of the algorithm. The effective fermionic propagator is a functional of the fermionic self-energy, $\Sigma\mapsto\mathcal{G}[\Sigma]$, and the effective bosonic propagator is known exactly in the undamped scenario. The fermionic self-energy is solved iteratively as indicated below
\begin{align}
\label{eq:alg1}
\widetilde{\Sigma}_{i+1}(i\omega_n)&= g^2 T\sum_{{n^\prime}\in[-N,N-1]}\mathcal{D}\left(i \omega_n-i \omega_{n^{\prime}}\right) \mathcal{G}\left[\Sigma_i\right]\left(i \omega_{n^{\prime}}\right),\\ 
\label{eq:alg2}
\Sigma_{i}(i\omega_n)&=\alpha\widetilde{\Sigma}_{i}(i\omega_n) + (1-\alpha)\Sigma_{i-1}(i\omega_n).
\end{align}
Upon evaluating Eq.~\eqref{eq:alg1}, the resulting solution of the current iteration is then substituted into Eq.~\eqref{eq:alg2}, where it is combined with the solution of the previous iteration through a weighted average controlled by a mixing parameter $\alpha\in(0,1)$. The resulting weighted combination is then substituted in Eq.~\eqref{eq:alg1} in the next iteration of the algorithm. This process is repeated until the error between successive iterations, quantified by $\epsilon=\sum_{\omega_n}|\Sigma_i(\omega_n)-\Sigma_{i-1}(\omega_n)|$, is below a prescribed tolerance $\epsilon<\epsilon_{*}$. Figure~\ref{fig:Im-axis-numerical-SE} is obtained by implementing the iterative algorithm described above.

In Sec.~\ref{subsec:DC-conductivity}, real-axis expressions for the fermionic self-energy are used to compute transport properties in the strange and bad-metal regimes. The real-axis expressions in Eqs.~\eqref{eq:Simplified_elec_SE_IBWL} and~\eqref{eq:Sigma_R-conformalSoln} are obtained by analytical continuation of the imaginary-axis fermionic self-energy in the infinite- and finite-bandwidth cases, respectively. The former is exact in the infinite-bandwidth limit for $T\ll\omega_D$, whereas the latter is an approximation, as it is derived from the emergent ``conformal'' solution. In the latter case, the approximation is made on the imaginary axis prior to analytical continuation, making a comparison with numerically exact real-axis solutions essential.

The real-axis equations are obtained by using the spectral representations of the effective propagators:
\begin{equation}
\mathcal{D}(i\omega_n-i\omega_{n^\prime})=\int \frac{\mathcal{D}_R^{\prime\prime}(\Omega)}{\Omega-(i\omega_n-i\omega_{n^\prime})}\frac{d\Omega}{\pi};\quad 
\mathcal{G}(i\omega_{n^\prime})=\int \frac{\mathcal{G}_R^{\prime\prime}(\omega^{\prime})}{\omega^\prime-i\omega_{n^\prime}}\frac{d\omega^\prime}{\pi}.
\end{equation}
Performing the Matsubara frequency summation using the Eliashberg contour method~\cite{AGD}, we obtain the following
\begin{align}
\Sigma_R(\omega) &= g^2\int \left[\mathcal{G}_R(\omega-\Omega)\mathcal{D}_R^{\prime\prime}(\Omega)\coth\left(\frac{\Omega}{2T}\right) + \mathcal{G}_R^{\prime\prime}(\Omega)\mathcal{D}_R(\omega-\Omega)\tanh\left(\frac{\Omega}{2T}\right)\right]\frac{d\Omega}{2\pi} \nonumber \\ 
&= g^2\int \left[\mathcal{G}_R(-\Omega)\mathcal{D}_R^{\prime\prime}(\Omega+\omega)\coth\left(\frac{\Omega+\omega}{2T}\right) + \mathcal{G}_R^{\prime\prime}(\Omega)\mathcal{D}_R(\omega-\Omega)\tanh\left(\frac{\Omega}{2T}\right)\right]\frac{d\Omega}{2\pi} \nonumber\\ 
&= \lambda\int \Bigg[\text{Im}\left(\text{arctanh}\left(\frac{\Lambda/2}{\Omega-\Sigma_R(\Omega)}\right)\right)L_R(\omega-\Omega)\tanh\left(\frac{\Omega}{2T}\right) \nonumber\\ 
&\quad\quad\quad\quad -\text{arctanh}\left(\frac{\Lambda/2}{\Omega+\Sigma_R(-\Omega)}\right)L_R^{\prime\prime}(\Omega+\omega)\coth\left(\frac{\Omega+\omega}{2T}\right) \Bigg]\frac{d\Omega}{\pi}.
\label{eq:partial-spectral-rep}
\end{align}
In the undamped scenario $\delta\Pi_R(\Omega)=0$, and assuming particle-hole symmetric symmetry, $\Lambda/2=\varepsilon_{F}$, the real-axis effective fermionic and bosonic propagators are respectively given by
\begin{equation}
L_R(\Omega) = \log \left(1+\frac{\omega_D^2}{m_B^2-(\Omega+i0^+)^2}\right);\quad \mathcal{G}_R(\omega) = 2\nu\ \text{arctanh}\left(\frac{\Lambda/2}{\omega-\Sigma_R(\omega)}\right).
\end{equation}

\begin{figure}[h]
\centering\includegraphics[width=1\columnwidth]{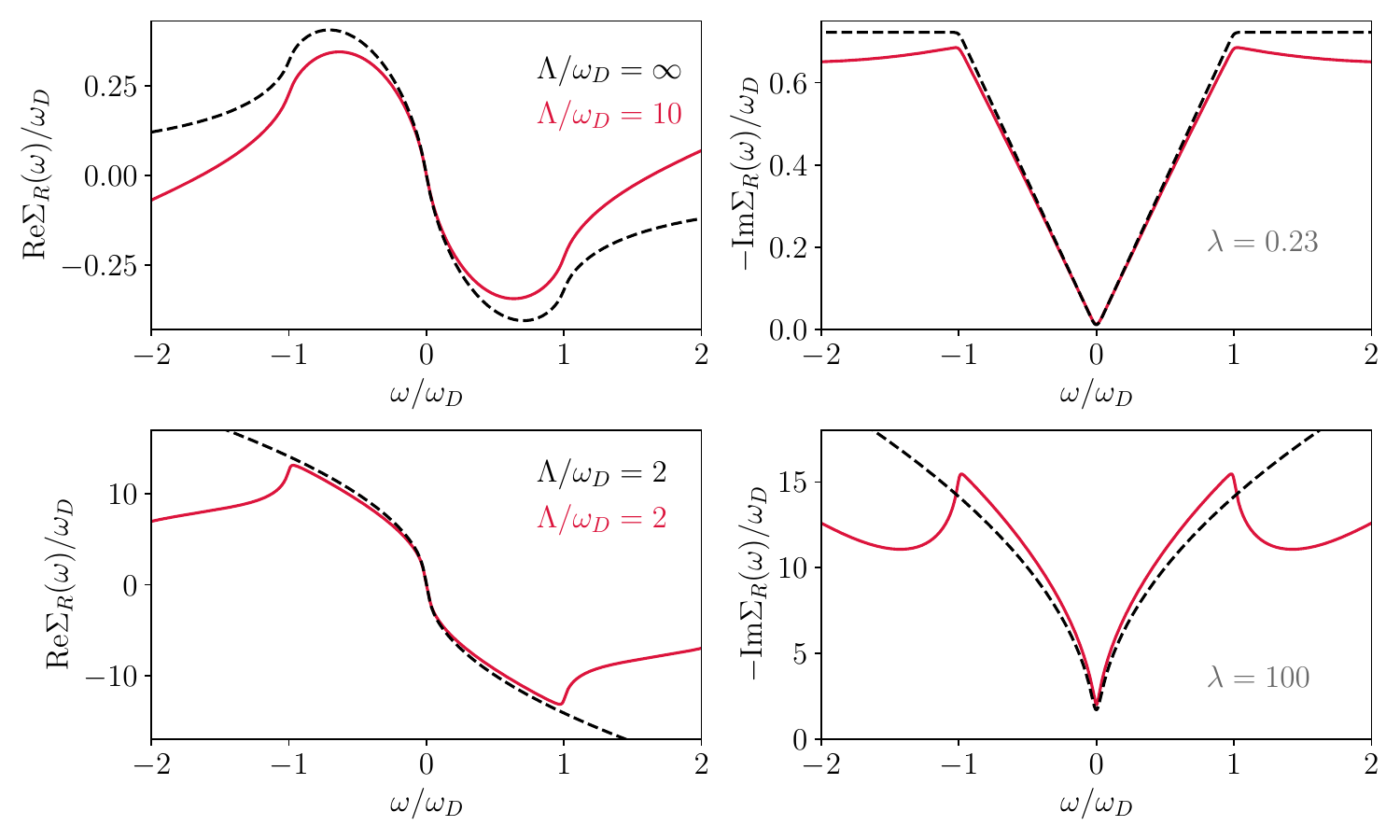}
\caption{(a, b) Comparison of the numerically exact (red) fermionic self-energy with finite electronic bandwidth and the exact infinite-bandwidth solution. (c, d) Comparison of the numerically exact (red) fermionic self-energy and the approximate ``conformal'' solution for the fermionic self-energy (both are evaluated for $\Lambda=2\omega_{D}$). Temperature $T=0.01\omega_D$.}
\label{fig:Re-axis-numerical-SE}
\end{figure}

The numerical analysis on the real-axis is carried out using an analogous algorithm to that on the imaginary-axis described above. We begin by discretizing the real-axis into $N$ points over an interval in frequency $\omega\in[-xT,xT]$, where $x\in\mathbb{R}_{>0}$ is chosen to be sufficiently large to include the UV structure of the solution. We include an arbitrarily small imaginary part to ensure that the analytic structure of the propagators is preserved. The algorithm begins with an initial guess for $\Sigma_{R}(\omega)$, which we take to be a negative imaginary constant $\Sigma_{R,0}(\omega)=-i\Gamma$; $\Gamma\ll\Lambda$, after which Eq.~\eqref{eq:partial-spectral-rep} is evaluated iteratively using the same weighting procedure described for the imaginary-axis protocol. We use this  algorithm to obtain the numerical results shown in Fig.~\ref{fig:Re-axis-numerical-SE}.

\section{Superconducting Critical Temperature}

\label{sec:Tc}

The linearized gap equation for the auxiliary function $\Phi\left(i \omega_n\right)$ in the Y-SYK model in (2+1)d reads~\cite{Gnezdilov2025}
\begin{align}\label{eq:gapeq1}
\Phi(i\omega_n)=g^2T_c\sum_{\omega_{n^\prime}}\mathcal{D}({i\omega_{n}-i\omega_{n^\prime}})\mathcal{F}(i\omega_{n^{\prime}}).
\end{align}
For a quadratic dispersion with $\xi\in[-\Lambda/2,\Lambda/2]$ and finite bandwidth in two spatial dimensions, the linearized component of the effective anomalous Gor’kov propagator is
\begin{equation} 
\mathcal{F}(i\omega_{n}) \equiv\int G\left(i \omega_n, \bm{k}\right) G\left(-i \omega_n,-\bm{k}\right) \Phi\left(i \omega_n\right) \frac{d^2 k}{(2 \pi)^2}=\frac{2\nu}{\omega_n Z(i\omega_{n})}\arctan\left(\frac{\Lambda/2}{\omega_n Z(i\omega_n)}\right)\Phi(i\omega_n).
\label{eq:F-gapeq}
\end{equation}
The even-frequency renormalization function is defined as $Z(i\omega_n)\equiv1+i\Sigma(i\omega_n)/\omega_n$~\cite{Gnezdilov2025}. Inserting this result into the linearized gap equation and defining $\Delta(i\omega_n)\equiv\Phi(i\omega_n)/Z(i\omega_n)$, we obtain
\begin{equation}
\Delta(i\omega_n)Z(i\omega_n)=2\lambda T_c \sum_{\omega_{n^\prime}} \frac{L(i\omega_n-i\omega_{n^\prime})}{\omega_{n^\prime}}\arctan\left(\frac{\Lambda/2}{\omega_{n^\prime}Z(i\omega_{n^\prime})}\right)\Delta(i\omega_{n^{\prime}}).
\label{eq:Linearized-gap-eq-1}
\end{equation}
The dimensionless bosonic propagator (in the underdamped scenario) is given by $L(i\Omega_m)\equiv\log(1+\omega_D^2/(m_B^2+\Omega_m^2))$. The renormalization function is given by
\begin{equation}
Z(i\omega_n) = 1 + \frac{2\lambda T_c}{\omega_n}\sum_{\omega_{n^{\prime}}}L(i\omega_n-i\omega_{n^{\prime}})\arctan\left(\frac{\Lambda/2}{\omega_{n^{\prime}}Z(i\omega_{n^{\prime}})}\right),
\end{equation}
Substituting this expression into Eq.~\eqref{eq:Linearized-gap-eq-1}, we then obtain
\begin{equation}
\Delta(i\omega_n)=2\lambda T_c \sum_{\omega_{n^\prime}} \left[\Delta(i\omega_{n^{\prime}})-\frac{\omega_{n^{\prime}}}{\omega_n}\Delta(i\omega_{n})\right]\frac{L(i\omega_n-i\omega_{n^\prime})}{\omega_{n^\prime}}\arctan\left(\frac{\Lambda/2}{\omega_{n^\prime}Z(i\omega_{n^\prime})}\right).
\label{eq:Linearized-gap-eq-2}
\end{equation}

As in conventional Migdal-Eliashberg theory~\cite{Chubukov2020}, the term in square brackets in Eq.~\eqref{eq:Linearized-gap-eq-1} vanishes when $n^{\prime}=n$, allowing the summation to be restricted to terms where $n^{\prime}\neq n$. However, for a finite electronic bandwidth $\Lambda$, the presence of the renormalization function $Z(i\omega_{n})$ prevents the complete cancellation of thermal contributions. Indeed, we can express the fermionic self energy as $\Sigma_{n}=g^2 T\mathcal{D}_0\mathcal{G}_{n}+g^2 T\sum_{n^{\prime}\neq n}\mathcal{D}_{n-n^{\prime}}\mathcal{G}_{n^\prime}$, where the first and second terms represent the thermal and quantum contributions, respectively. Through the inclusion of $Z(i\omega_{n})$, thermal fluctuations affect pairing at finite bandwidth. In contrast, in the infinite-bandwidth limit, $\arctan((\Lambda/2)/(|\omega_n|Z(i\omega_n)))\rightarrow\pi/2$, and the thermal contribution to the linearized gap equation drops out completely~\cite{Gnezdilov2025}. This result is analogous to Anderson's theorem, which states that time-reversal-symmetric impurities do not affect the superconducting critical temperature. At finite electronic bandwidth, the thermal contribution is responsible for the saturation of $T_c(\lambda)$ in the strong-coupling impurity regimes of both the $(0+1)$-dimensional~\cite{Esterlis2019} and $(2+1)$-dimensional~\cite{Kleger2026} Y-SYK models. This mechanism is distinct from the saturation of $T_{c}(\lambda)$ found in the infinite-bandwidth (2+1)-dimensional Y-SYK model, which instead originates from the bosonic self-energy~\cite{Gnezdilov2025}. 

For $|\Sigma|\ll\Lambda$, the pairing of incoherent fermions is protected by a mechanism analogous to Anderson's theorem for disordered superconductors~\cite{Esterlis2019}, as the thermal contribution of the fermionic self-energy does not affect $T_c$. Conventional superconductors, particularly in the weak-coupling regime, satisfy this separation of scales and therefore obey Anderson's theorem. Indeed, non-magnetic impurities enter the linearized gap equation through the resonant contribution to $Z(i\omega_n)$, which cancels in this regime. In strongly incoherent metals, however, the separation of scales $|\Sigma|\ll\Lambda$ breaks down, allowing the thermal contribution to affect pairing. This contrasts with the conventional expectation that only the `quantum' part of $\Sigma$ contributes to $T_c$. Notably, in certain quantum-critical pairing scenarios, the thermal contribution is responsible for the strong-coupling saturation of $T_c$~\cite{Esterlis2019, Kleger2026}. The energy scales $\Lambda$ and $\Sigma$ thus provide a framework for understanding the applicability of Anderson's theorem and when it can be violated.

\begin{figure}[t]
\centering
\begin{subfigure}{0.58\textwidth}
\centering
        \includegraphics[width=\linewidth]{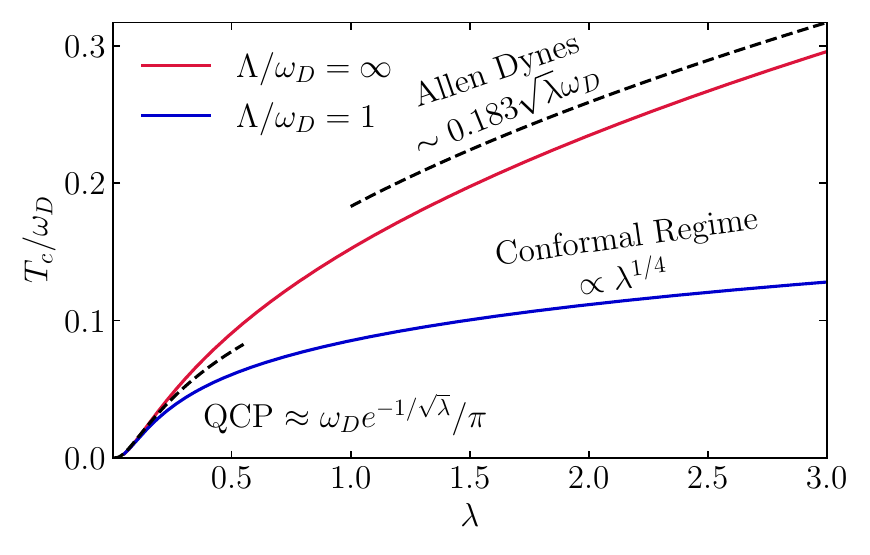}
        \caption{Superconducting critical temperature in units of $\omega_D$. Red and blue curves denote the cases of infinite and finite electronic bandwidth respectively. Dashed black lines denote the weak coupling (finite and infinite bandwidth) and strong coupling (infinite bandwidth) limits. }
\label{fig:Tc}
    \end{subfigure}
\begin{subfigure}{0.33\textwidth}
\centering
        \includegraphics[width=\linewidth]{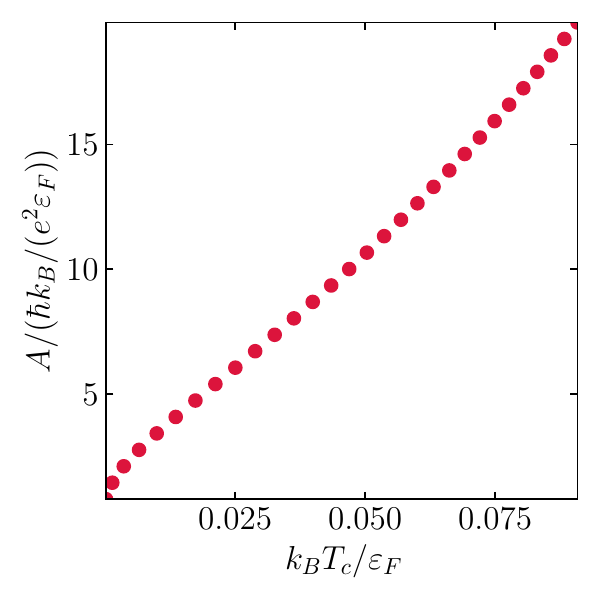}
        \caption{Coefficient of the $T$-linear resistivity $\rho_{\MFL}=AT$ (in $d=2$ spatial dimensions) versus $T_c$. Evaluated for $\lambda\in(0,0.5)$ in the infinite-bandwidth limit with $\omega_D=\varepsilon_F$ and for the quantum-critical boson mass.}
        \label{fig:Acoeff}
    \end{subfigure}
    \caption{Compilation of the superconducting critical temperature.}
\end{figure}

For numerical convenience, we re-express the linearized gap equation in terms of positive frequencies as
\begin{equation}
Z(i\omega_n)\Delta(i\omega_n)=\frac{\lambda}{\pi}\sum_{n^\prime=0}^{\infty}\left[L(i\omega_{n}-i\omega_{n^{\prime}})+L(i\omega_{n}+i\omega_{n^{\prime}})\right]\arctan\left(\frac{\Lambda/2}{\omega_{n^\prime}Z(i\omega_{n^\prime})}\right)\frac{\Delta(i\omega_{n^{\prime}})}{n^{\prime}+1/2}.
\label{eq:full-pairing-problem}
\end{equation}
Following Ref.~\cite{Gnezdilov2025}, $T_c$ can be determined by evaluating the zero of the determinant $\operatorname{det}[\mathcal{M}-\mathcal{I}]=0$, where $\mathcal{I}$ is the identity matrix in the space of positive frequencies and 
\begin{equation}
\mathcal{M}_{n,n^{\prime}}\equiv\frac{\lambda}{\pi}\frac{L(i\omega_{n}-i\omega_{n^{\prime}})+L(i\omega_{n}+i\omega_{n^{\prime}})}{Z(i\omega_n)(n^{\prime}+1/2)}\arctan\left(\frac{\Lambda/2}{\omega_{n^\prime}Z(i\omega_{n^\prime})}\right).
\label{eq:full-pairing-problem-2}
\end{equation}
Equation~\eqref{eq:full-pairing-problem-2} is solved numerically using the numerically evaluated imaginary-axis solutions described in Sec.~\ref{sec:NumericalAnalysis} and displayed in Fig.~\ref{fig:Tc}. In Fig.~\ref{fig:Acoeff}, we plot the coefficient of the $T$-linear resistivity versus $T_{c}$, showing that, as expected~\cite{Hussey2011, Taillefer2010}, the two quantities scale with each other. 

We can obtain an analytical expression for the superconducting critical temperature in the weak-coupling limit by taking the infinite-bandwidth limit and approximating $Z(i\omega_n)\simeq1$ and $L(i\Omega_m)\simeq\log(1+\omega_D^2/\Omega_m^2)$ for $|m|>0$. The superconducting critical temperature is then determined by the weak-coupling analysis of Ref.~\cite{Gnezdilov2025}. Approximating $L\left(i \omega_n \pm i \omega_{n^{\prime}}\right)$ by $L\left(i \omega_n \pm i \omega_0\right)$ and taking the continuous limit, the low-temperature equation for $T_c$ is given by~\cite{Gnezdilov2025}
\begin{align}
1 &= \lambda\left[L(2i\omega_0) + \int_{\omega_1}^{\infty}\frac{L(i\omega)}{\omega}d\omega \right] \nonumber\\
  &= \lambda\left[L(2i\omega_0) + \sum_{k=1}^{\infty} \frac{(-1)^{k+1}}{k}\int_{\omega_1/\omega_{D}}^{\infty}\frac{1}{x^{2k+1}}dx \right] \nonumber\\ 
 &= \lambda\left[L(2i\omega_0) -\frac{1}{2}\text{Li}_2\left(-\left(\frac{\omega_D}{\omega_1}\right)^2\right) \right].
\end{align}
Thus, at the weakest coupling (and temperature),  where the electronic self-energy is an insignificant energy scale compared to the electronic bandwidth, $|\Sigma(i\omega_n)|\ll\Lambda$, we obtain $1/\sqrt{\lambda}\sim\log(\omega_D/(\pi T_c))$. This result is parametrically larger than the BCS result~\cite{Chubukov2005,Gnezdilov2025,Son1999,Kryhin2026}. For $\lambda=0.23$, this estimate yields $T_c\approx\omega_De^{-1/\sqrt{0.23}}/\pi\simeq 0.040\omega_D$, close to the exact numerical values $T_c\simeq 0.042\omega_D$ and $T_c\simeq 0.035\omega_D$ in the infinite-bandwidth $\Lambda/\omega_D = \infty$ and finite-bandwidth $\Lambda/\omega_D=1$ cases, respectively.

In the strong-coupling limit, we find two distinct scaling laws for $T_c$. For an infinite electronic bandwidth, the strong-coupling result is given by the well-known Allen-Dynes expression, $T_c\sim0.183\sqrt{\lambda}\omega_D$. Consequently, the weak- and strong-coupling regimes correspond to the $\gamma=0$ and $\gamma=2$ limits, respectively, of the $\gamma$-model of quantum-critical pairing. In contrast, for a finite electronic bandwidth $\Lambda\sim\omega_D$, the strong-coupling growth of the superconducting critical temperature is suppressed to $T_c\propto\lambda^{1/4}$. This result arise from a crossover to the strongly incoherent metal solution characterized by $Z(i\omega_n)\propto\sqrt{\lambda}T^0$, which substantially  reduces the strong-coupling growth of $T_c$.

\section{Critical Solutions in Three Spatial Dimensions} 
\label{sec:ArbDim}

\subsection{Effective Bosonic and Fermionic Propagators}

We now extend the analysis of the previous two sections to $d=3$ spatial dimensions. Analogously, there are two regimes for the effective (momentum-integrated) fermionic propagator:
\begin{align}
\mathcal{G}(i\omega_n)&\sim -\nu(0)\begin{cases}
i \pi  \text{sgn}(\omega_n); \quad \Lambda\gg|i\omega_n -\Sigma(i\omega_n)|\quad \text{(FS)}\\ 
 \Lambda /\Sigma(i\omega_n); \quad |\Sigma(i\omega_n)|\gg \Lambda, |\omega_n|\quad \; \text{(CL)}.
\end{cases}
\label{eq:G-arb}
\end{align}
The Fermi-surface (FS) solution is obtained by performing a Fermi-surface approximation~\cite{Patel2018}: $(2\pi)^{-3}\int F(k)d^{3}k\approx\nu(0)\int F(\xi) d\xi$ 
and taking the limits of integration in the $\xi$ integral to be infinite. This approximation is exact for $d=2$ in the limit $\Lambda\gg|i\omega_n -\Sigma(i\omega_n)|$ in the Galilean continuum for the particle-hole symmetric case. For $d=3$, the density of states is approximated by its value at the Fermi-surface $\nu(0)$. The conformal limit (CL) solution results from the factorization $\int (i\omega_n-\xi-\Sigma(i\omega_n))^{-1}d^{3}k\sim (-\Sigma(i\omega_n))^{-1}\int d^{3}k$ for $|\Sigma(i\omega_n)|\gg \Lambda, |\omega_n|$. Here, we will consider the regime described by the FS form of the effective propagator, as given in Eq.~\eqref{eq:G-arb}. This is a generalization of the marginal Fermi liquid solution for the strange- and bad-metal regimes.

In three spatial dimensions at zero temperature the effective bosonic propagator for an undamped critical boson is
\begin{equation}
\mathcal{D}(i\Omega)=\frac{1}{2\pi^{2}}\int_{0}^{\omega_{D}/c}\frac{q^{2}}{\Omega^{2}+c^{2}q^{2}}dq = \frac{1}{2\pi^{2}}\frac{\omega_{D}}{c^{3}}\left[1-\frac{\Omega}{\omega_{D}}\arctan\left(\frac{\omega_{D}}{\Omega}\right)\right].
\end{equation}
In three spatial dimensions, we define the dimensionless electron-boson coupling as $\lambda_{\ThreeD} \equiv \nu(0)g^2\omega_D/(2\pi^2 c^3)$, and the dimensionless effective bosonic propagator as $L_{\ThreeD}(i\Omega)\equiv 1-(\Omega/\omega_D)\arctan(\omega_D/\Omega)$. Despite being at criticality ($m_B=0$), the  propagator remains finite in the low-frequency limit: $\lim_{\Omega\rightarrow0}L_{\ThreeD}(i\Omega)=1$, in direct contrast with the logarithmically divergent propagator in two spatial dimensions:  $\Omega\rightarrow0: L_{\text{2d}}(i\Omega)\sim2\log(\omega_D/|\Omega|)$. Consequently, the anomalous properties characterizing the strange metals -- $T$-linear resistivity, $T\log(1/T)$ heat capacity, and superconducting critical temperature scaling as $\sim e^{-1/\sqrt{\lambda}}$ -- are absent in three spatial dimensions at criticality. Instead, we obtain conventional Fermi-liquid behavior $\rho_{\ThreeD}\sim T^2$, $C_{\ThreeD}\sim T$, and $T_{c}^{\ThreeD}\propto e^{-1/\lambda_{\ThreeD}}$.

In the FS regime the effective fermionic propagator is $\mathcal{G}(i\omega_n)=-i\pi\nu(0)\text{sgn}(\omega_n)$. The zero-temperature fermionic self-energy is then given by 
\begin{align}
\Sigma(i\omega) & = -ig^{2}\nu(0)\text{sgn}(\omega)\int_{0}^{|\omega|}\mathcal{D}(i\Omega)d\Omega \nonumber\\ &= -i\lambda_{\ThreeD}\text{sgn}(\omega)\int_{0}^{|\omega|}\left[1-\frac{\Omega}{\omega_{D}}\arctan\left(\frac{\omega_{D}}{\Omega}\right)\right]d\Omega\nonumber \\
 &= -i\lambda_{\ThreeD}\text{sgn}(\omega)\omega_{D}\frac{1}{2}\left[|\omega|/\omega_{D}-(\omega/\omega_{D})^{2}\arctan(\omega_{D}/|\omega|)+\arctan(|\omega|/\omega_{D})\right].
\end{align}
The retarded part of the fermionic self-energy is given by 
\begin{equation}
\Sigma_{R}(\omega)=-\lambda_{\ThreeD}\omega_{D}\frac{1}{2}\left[\omega/\omega_{D}-(\omega/\omega_{D})^{2}\text{arctanh}(\omega_{D}/\omega)+\text{arctanh}(\omega/\omega_{D})\right]
\end{equation}
Thus, the low-frequency imaginary part of the fermionic self-energy becomes $\Sigma_{R}^{\prime\prime}(\omega)=-\lambda_{\ThreeD}\omega_{D}(\pi/4)(\omega/\omega_{D})^{2}$. This result is analogous to the Fermi-liquid behavior. In two-spatial dimensions, the spatially disordered quantum-critical metal has $T$-linear resistivity, and in three spatial dimensions, resistivity scales quadratically with temperature, reminiscent of the Fermi-liquid. This result may have profound practical consequences, as it seems to provide a reason for why the strange metals are primarily (quasi) 2d or layered materials. 

Our analysis shows that a three-dimensional spatially disordered quantum-critical metal with an underdamped critical boson exhibits Fermi-liquid-like resistivity, $\rho\sim T^2$, obscuring its non-Fermi-liquid nature. As shown below, its heat capacity and superconducting critical temperature also display conventional scaling. We therefore conjecture that certain unconventional superconductors may, in fact, be quantum-critical metals in three spatial dimensions. Their conventional transport and thermodynamic properties obscure their underlying non-Fermi-liquid nature, making them difficult to distinguish from conventional Fermi-liquids. If valid, this conjecture broadens the scope of disordered quantum-critical metals and suggests new avenues for discovering unconventional superconductors.

Within the Fermi-surface approximation and the infinite-bandwidth limit, the fermionic  heat capacity follows the Sommerfeld result: $C_{F}=2\pi^2\nu(0)T/3$, while the free-boson contribution is higher-order in temperature. The interacting entropy contribution, responsible for the $\sim T\log(1/T)$ scaling in the 2d quantum critical metal, scales as
\begin{equation}
s_{\text{int}}\sim T\int_0^1 \frac{\sqrt{x}}{x+m_B^2/\omega_D^2}dx\sim T\left(1+O(m_B/\omega_D)\right).
\end{equation}
At low temperatures, near criticality, $m_B\ll\omega_D$ and we observe that the leading contribution to the heat capacity from interactions is $\sim T$ which renormalizes the Sommerfeld result. For weak coupling, we can apply the same analysis as in Sec.~\ref{sec:Tc} to obtain 
\begin{equation}
1 = \lambda_{\ThreeD}\left[L_{\ThreeD}(2i\omega_0) + \int_{\omega_1}^{\infty}\frac{L_{\ThreeD}(i\omega)}{\omega}d\omega \right] \sim \lambda_{\ThreeD}\log\left(\frac{\omega_D}{3\pi T}\right).
\end{equation}
Therefore, $T_{c}^{\ThreeD}\approx \omega_D e^{-1/\lambda_{\ThreeD}}/(3\pi)$.

\subsection{Bosonic Mass}

Applying the constraint approach, the self-consistent equation for the boson mass is:
\begin{equation}
\int_{-\infty}^{\infty}\left.\mathcal{D}(i\Omega)\frac{d\Omega}{2\pi}\right|_{m_{B}=0}=T\sum_{m\in\mathbb{Z}}\left.\mathcal{D}(i\Omega_{m})\right|_{m_{B}=m_{B}(T)}.
\end{equation}
The effective bosonic propagator for underdamped bosons in three spatial dimensions is given by 
\begin{equation}
\mathcal{D}(i\Omega_{m}) = \frac{1}{2\pi^{2}}\int_{0}^{\omega_{D}/c}\frac{q^2}{\Omega_{m}^{2}+c^{2}q^{2}+m_{B}^{2}}dq
 = \frac{1}{2\pi^{2}}\frac{\omega_{D}}{c^{3}}\left[1-\sqrt{\frac{\Omega_{m}^{2}+m_{B}^{2}}{\omega_{D}^{2}}}\arctan\left(\sqrt{\frac{\omega_{D}^2}{\Omega_{m}^{2}+m_{B}^{2}}}\right)\right].
\end{equation}
The mass constraint, for underdamped bosons, becomes
\begin{align}
\frac{1}{2\pi}\int_{-\infty}^{\infty}\left.\mathcal{D}(i\Omega)d\Omega\right|_{m_{B}=0} &= \frac{1}{2\pi^{2}}\frac{\omega_{D}}{c^{3}}\omega_{D}\int_{0}^{\infty}\left(1-x\arctan(1/x)\right)\frac{1}{\pi}dx\nonumber \\
 &= \frac{1}{8\pi^{2}}\frac{\omega_{D}^{2}}{c^{3}}\nonumber \\
 &= T\sum_{m\in\mathbb{Z}}\left.\mathcal{D}(i\Omega_{m})\right|_{m_{B}=m_{B}(T)}.
\end{align}
For the right-hand side, we interchange the order of momentum integration and frequency summation (which is valid in this instance) to obtain 
\begin{align}
\frac{1}{8\pi^{2}}\frac{\omega_{D}^{2}}{c^{3}} &= T\sum_{m\in\mathbb{Z}}\frac{1}{2\pi^{2}}\int_{0}^{\omega_{D}/c}\frac{1}{\Omega_{m}^{2}+c^{2}q^{2}+m_{B}^{2}}q^{2}dq \nonumber \\
 &= \frac{1}{2\pi^{2}}\int_{0}^{\omega_{D}/c}\left[T\sum_{m\in\mathbb{Z}}\frac{1}{\Omega_{m}^{2}+c^{2}q^{2}+m_{B}^{2}}\right]q^{2}dq\nonumber \\
 &= \frac{1}{4\pi^{2}}\int_{0}^{\omega_{D}/c}\frac{1}{\sqrt{c^{2}q^{2}+m_{B}^{2}}}{\rm coth}\left(\frac{\sqrt{c^{2}q^{2}+m_{B}^{2}}}{2T}\right)q^{2}dq.
\end{align}
Let $q=(\omega_{D}/c)x, \overline{m}\equiv m_{B}/\omega_{D}$, and $\overline{t}=T/\omega_{D}$, then we obtain the mass constraint as the following integral equation
\begin{equation}
\int_{0}^{1}\frac{2x^{2}}{\sqrt{x^{2}+\overline{m}^{2}}}{\rm coth}\left(\frac{\sqrt{x^{2}+\overline{m}^{2}}}{2t}\right)dx=1.
\end{equation}

\section{Auxiliary Derivations}

\subsection{Fourier Transforms}
\label{App:Boson_FT}

At zero temperature, the frequency- and imaginary-time-domain propagators are related as follows:
\begin{align}
\mathcal{H}(i\omega) &= \int_{-\infty}^{\infty}e^{i\omega\tau}\mathcal{H}(\tau)d\tau, \\
\mathcal{H}(\tau) &= \frac{1}{2\pi}\int_{-\infty}^{\infty}e^{-i\omega\tau}\mathcal{H}(i\omega) d\omega.
\end{align} 
The following identities are useful for performing Fourier-transform calculations~\cite{Marcus2019}:
\begin{align}
\int_{-\infty}^{\infty}e^{i\omega\tau}\frac{\text{sgn}(\tau)}{|\tau|^{2\Delta}}d\tau &= 2i\cos(\pi\Delta)\Gamma(1-2\Delta)\text{sgn}(\omega)|\omega|^{2\Delta-1}, \\
\int_{-\infty}^{\infty}e^{i\omega\tau}\frac{1}{|\tau|^{2\Delta}}d\tau &= 2\sin(\pi\Delta)\Gamma(1-2\Delta)|\omega|^{2\Delta-1}.
\end{align}
These results can be deduced from the following identities. For $b>0$~\cite{WhittakerWatsonBook}:
\begin{align}
\int_{0}^{\infty}\frac{\sin(bx)}{x^s}dx &= \frac{\pi}{2}\frac{b^{s-1}}{\Gamma(s)}\text{cosec}\left(\frac{\pi s}{2}\right), \quad 0<s<2, \\
\int_{0}^{\infty}\frac{\cos(bx)}{x^s}dx &= \frac{\pi}{2}\frac{b^{s-1}}{\Gamma(s)}\sec\left(\frac{\pi s}{2}\right), \quad 0<s<1. 
\end{align}

Let us evaluate the Fourier transform of the effective bosonic propagator at zero temperature. This result can be derived straightforwardly using integration by parts. On the frequency axis, the dimensionless boson propagator is $L(i\Omega)=\log(1+\omega_{D}^2/\Omega^2)$, and therefore in the imaginary-time domain
\begin{align}
L(\tau) &= \frac{1}{2\pi}\int_{-\infty}^{\infty}e^{-i\Omega\tau}L(i\Omega) d\Omega \nonumber\\
&= \frac{\omega_{D}}{2\pi}\int_{-\infty}^{\infty}\cos(x\omega_{D}\tau)\log\left(1+\frac{1}{x^2}\right)dx \nonumber\\
&= \frac{1}{2\pi\tau}\int_{-\infty}^{\infty}\left(\frac{d}{dx}\sin(x\omega_{D}\tau)\right)\log\left(1+\frac{1}{x^{2}}\right)dx \nonumber\\ 
&= \frac{1}{\pi\tau}\mathcal{P}\int_{-\infty}^{\infty}\frac{\sin(x\omega_{D}\tau)}{x(1+x^{2})}dx \nonumber\\ 
&= \frac{1}{|\tau|}(1-e^{-|\omega_{D}\tau|}).
\end{align}
Here, $\mathcal{P}$ denotes the Cauchy Principal value integral. 

At non-zero temperatures, the Fourier transform of the effective bosonic propagator is computed by performing the Matsubara frequency summation. We express $L(\tau)$ as given below:
\begin{equation}
L(\tau)=T\sum_{m\in\mathbb{Z}}e^{-i\Omega_{m}\tau}\log\left(1+\frac{\omega_{D}^{2}}{\Omega_{m}^{2}+m_{B}^{2}}\right).
\end{equation}
In the limit $T/\omega_{D}\ll1$, $m_{B}=2\pi\alpha T$ where $\alpha=(\text{arccosh}(3/2))/(2\pi)$.
Thus, if we define $t=T/\omega_{D}$, then 
\begin{equation}
L(\tau)/\omega_{D}=t\sum_{m\in\mathbb{Z}}e^{2\pi imt\tau\omega_{D}}\log\left(1+\frac{1/t^{2}}{(2\pi m)^{2}+(2\pi\alpha)^{2}}\right).
\end{equation}
Let $b=(2\pi m)^{2}+(2\pi\alpha)^{2}+1/t^{2}$ and $a=(2\pi m)^{2}+(2\pi\alpha)^{2}$ and apply Frullani's theorem~\cite{WhittakerWatsonBook}. The logarithm can then be written as
\begin{equation}
\log\left(1+\frac{1/t^{2}}{(2\pi m)^{2}+(2\pi\alpha)^{2}}\right)=\int_{0}^{\infty}(e^{-ax}-e^{-bx})\frac{1}{x}dx.
\end{equation}
Therefore, the dimensionless propagator becomes
\begin{equation}
L(\tau)/\omega_{D} =  t\sum_{m\in\mathbb{Z}}\left[\int_{0}^{\infty}e^{2\pi imt\tau\omega_{D}}e^{-(2\pi m)^{2}x}\left(e^{-(2\pi\alpha)^{2}x}-e^{-((2\pi\alpha)^{2}+1/t^{2})x}\right)\frac{1}{x}\right]dx.
\end{equation}
Next, we interchange the integration and frequency summation, and then apply the Poisson summation formula~\cite{WhittakerWatsonBook}:
\begin{equation}
\sum_{m\in\mathbb{Z}}e^{-m^{2}\pi/\alpha+2m\pi i\beta}=\sqrt{\alpha}e^{-\pi\alpha\beta^{2}}\sum_{m\in\mathbb{Z}}e^{-m^{2}\pi\alpha-2m\pi\alpha\beta}.
\end{equation}
Thus, after again interchanging integration and summation, we now have 
\begin{equation}
L(\tau)/\omega_{D}  =  t\frac{1}{2\sqrt{\pi}}\sum_{m\in\mathbb{Z}}\int_{0}^{\infty}e^{-(m+t\tau\omega_{D})^{2}/(4x)}\left(e^{-(2\pi\alpha)^{2}x}-e^{-((2\pi\alpha)^{2}+1/t^{2})x}\right)\frac{1}{x^{3/2}}dx.
\end{equation}
The $x$ integration can be performed using the identity 
\begin{equation}
\int_{0}^{\infty}\frac{1}{x^{3/2}}e^{-ax-b/x}dx=\sqrt{\frac{\pi}{b}}e^{-2\sqrt{ab}}.
\end{equation}
Simplifying this result, we find that the exact bosonic propagator is given by
\begin{equation}
L(\tau)/\omega_{D}=t\sum_{m\in\mathbb{Z}}\frac{1}{|m+t\tau\omega_{D}|}\left(e^{-2\pi\alpha|m+t\tau\omega_{D}|}-e^{-|m+t\tau\omega_{D}|\sqrt{(2\pi\alpha)^{2}+1/t^{2}}}\right).
\end{equation}
Now suppose $T/\omega_{D}=t\ll1$ and $|\tau|\omega_{D}$ is fixed. Then, $t\sqrt{(2\pi\alpha)^{2}+1/t^{2})}\approx1+(2\pi\alpha)^{2}t^{2}/2$.
Hence, for $t=T/\omega_{D}\ll1$,
\begin{align}
L(\tau)/\omega_{D} &= \frac{1}{|\tau\omega_{D}|}\left(e^{-2\pi\alpha t|\tau\omega_{D}|}-e^{-|\tau\omega_{D}|(1+(2\pi\alpha)^{2}t^{2}/2)}\right)\nonumber \\
 &\quad +t\sum_{m\in\mathbb{Z}/\{0\}}\frac{1}{|m+t\tau\omega_{D}|}\left(e^{-2\pi\alpha|m+t\tau\omega_{D}|}-e^{-|m+t\tau\omega_{D}|\sqrt{(2\pi\alpha)^{2}+1/t^{2}}}\right)\nonumber \\
 &= \frac{1}{|\tau\omega_{D}|}\left(1-2\pi\alpha t|\tau\omega_{D}|+\frac{1}{2}(2\pi\alpha t\tau\omega_{D})^{2}-e^{-|\tau\omega_{D}|}\left(1-\frac{1}{2}|\tau\omega_{D}|(2\pi\alpha)^{2}t^{2}\right)\right)\nonumber \\
 &\quad +t\sum_{m\in\mathbb{Z}/\{0\}}\frac{1}{|m+t\tau\omega_{D}|}\left(e^{-2\pi\alpha|m+t\tau\omega_{D}|}-e^{-|m+t\tau\omega_{D}|\sqrt{(2\pi\alpha)^{2}+1/t^{2}}}\right).
\end{align}
For the last term, we find the following:
\begin{equation}
 t\sum_{m\in\mathbb{Z}/\{0\}}\frac{1}{|m+t\tau\omega_{D}|}\left(e^{-2\pi\alpha|m+t\tau\omega_{D}|}-e^{-|m+t\tau\omega_{D}|\sqrt{(2\pi\alpha)^{2}+1/t^{2}}}\right)
 \approx  2t\sum_{m=1}^{\infty}\frac{1}{m}e^{-2\pi\alpha m} = 2\pi\alpha t.
\end{equation}
In the last step, we used $2\pi\alpha=\text{arccosh}(3/2)$. In summary, the expansions of the conformal-limit and exact imaginary-time bosonic propagators (for $T/\omega_{D}\ll1$) are given by
\begin{align}
L_{\CL}(\tau)/\omega_{D} &= \left|\frac{1}{\tau\omega_{D}}+\frac{1}{6}(\pi t)^{2}\tau\omega_{D}+\dots\right|.\\
L(\tau)/\omega_{D} &= \frac{1}{|\tau\omega_{D}|}\left(1-e^{-|\tau\omega_{D}|}\right)+\frac{1}{2}(2\pi\alpha t)^{2}\left(|\tau\omega_{D}|+e^{-|\tau\omega_{D}|}\right)+\dots.
\end{align}

\subsection{Resistivity}

\subsubsection*{Strange Metal}
\label{sec:Aux_Res_SM}

Here we present a derivation of Eq.~\eqref{eq:conductivity-sm}. We need to evaluate the following integral:
\begin{equation}
I=\int_{-\infty}^{\infty}\dfrac{\sech^{2}(x)}{\left|2x\text{Im}\psi(-ix/\pi)+\pi\right|}dx.
\end{equation}
The digamma function is defined by~\cite{WhittakerWatsonBook}:
\begin{equation}
\psi(z)=-\gamma+\sum_{k=0}^{\infty}\left(\frac{1}{k+1}-\frac{1}{k+z}\right).
\end{equation}
Let $z=-ix/\pi$. Then, 
\begin{equation}
\psi(-ix/\pi)=-\gamma+\sum_{k=0}^{\infty}\left(\frac{1}{k+1}-\frac{k}{k^{2}+(x/\pi)^{2}}\right)-i\frac{x}{\pi}\sum_{k=0}^{\infty}\frac{1}{k^{2}+(x/\pi)^{2}}.
\end{equation}
Taking the imaginary part, we find 
\begin{equation}
\text{Im}\psi(-ix/\pi)=-\frac{x}{\pi}\sum_{k=0}^{\infty}\frac{1}{k^{2}+(x/\pi)^{2}}=-\frac{\pi}{2x}\left[1+x\coth(x)\right].
\end{equation}
Thus, $\left|2x\text{Im}\psi(-ix/\pi)+\pi\right|=\pi x\coth(x)$ and
so the integral reduces to 
\begin{equation}
I=\frac{1}{\pi}\int_{-\infty}^{\infty}\dfrac{\sech^{2}(x)}{x\coth(x)}dx=\frac{1}{\pi}\int_{-\infty}^{\infty}\dfrac{\sech^{2}(x)\tanh(x)}{x}dx.
\end{equation}
Using integration by parts, we then obtain 
\begin{equation}
I=\frac{1}{2\pi}\int_{-\infty}^{\infty}\left(\dfrac{\tanh(x)}{x}\right)^{2}dx.
\end{equation}
The remaining integral can be evaluated in numerous ways, using either real-variable or  complex-variable methods. We will use the latter for simplicity. Consider 
\begin{equation}
\widetilde{I}=\oint_{C}\left(\dfrac{\tanh(z)}{z}\right)^{2}dz,
\end{equation}
where $C$ is the closed contour consisting of a semi circle in
the upper half plane (of radius $R$), concatenated with the contour along the real axis (indented about the origin by a semi circle of radius $r$). Thus, 
\begin{equation}
\widetilde{I}=\lim_{R\rightarrow\infty}\mathcal{P}\int_{-R}^{R}\left(\dfrac{\tanh(x)}{x}\right)^{2}dx+I_{\Gamma_{R}}+I_{\Gamma_{r}}.
\end{equation}
Here, $\mathcal{P}$ denotes the Cauchy principal value integral. In the limits of $R\rightarrow\infty$ and $r\rightarrow0$, the latter two integrals tend to zero. The integrand has poles within $C$ when $z=z_{0}$, where $z_{0}=(2n+1)i\pi/2$, $n\in\mathbb{Z}_{\geq0}$. The residue is $-2/z_{0}^{3}$ and so from the residue theorem, we find
\begin{equation}
\widetilde{I}=\frac{32}{\pi^{2}}\sum_{n=0}^{\infty}\frac{1}{(2n+1)^{3}}=\frac{28}{\pi^{2}}\zeta(3).
\end{equation}
The sum of the series is easily obtained using the definition of the Riemann-zeta function~\cite{WhittakerWatsonBook}. Here, $\zeta(3)$ is known as Ap\'ery's constant. Thus, we have the result given in Eq.~\eqref{eq:conductivity-sm}:
\begin{equation}
I=\frac{14}{\pi^{3}}\zeta(3).
\end{equation}

\subsubsection*{Bad Metal}
\label{sec:Aux_Res_BM}

Here we present a derivation of Eq.~\eqref{eq:conductivity-bm}. We need to evaluate the following integral:
\begin{equation}
I=\int_{-\infty}^{\infty}\sech^{2}(x)\left[\text{Re}\left(\dfrac{\Gamma(1/4-ix/\pi)}{\Gamma(3/4-ix/\pi)}\right)\right]^{2}dx.
\end{equation}
Let $z=3/4-ix/\pi$. Using the reflection formula, $\Gamma(z)\Gamma(1-z)=\pi/\sin(\pi z)$
along with the Schwarz reflection principle $\Gamma(z^{*})=[\Gamma(z)]^{*}$,
we find 
\begin{equation}
\Gamma(z)\Gamma(1-z) = \frac{\pi}{\sin\left(3\pi/4-ix\right)}= \frac{\sqrt{2}\pi}{\cosh(x)+i\sinh(x)}.
\end{equation}
Hence, 
\begin{equation}
\dfrac{\Gamma(1/4-ix/\pi)}{\Gamma(3/4-ix/\pi)}=\frac{\cosh(x)+i\sinh(x)}{\sqrt{2}\pi}\left|\Gamma(1/4+ix/\pi)\right|^{2}.
\end{equation}
It follows that the integrand is given by 
\begin{equation}
\sech^{2}(x)\left[\text{Re}\left(\dfrac{\Gamma(1/4-ix/\pi)}{\Gamma(3/4-ix/\pi)}\right)\right]^{2}=\frac{1}{2\pi^{2}}\left|\Gamma(1/4+ix/\pi)\right|^{4}. \end{equation}
Thus, the integral now becomes 
\begin{equation}
I=\frac{1}{2\pi}\int_{-\infty}^{\infty}\left|\Gamma(1/4+ix)\right|^{4}dx.
\end{equation}
Next, we apply Barnes' lemma~\cite{Barnes1908, WhittakerWatsonBook}:
\begin{equation}
\frac{1}{2\pi i}\int_{-i\infty}^{i\infty}\Gamma(\alpha+s)\Gamma(\beta+s)\Gamma(\gamma-s)\Gamma(\delta-s)ds=\frac{\Gamma(\alpha+\gamma)\Gamma(\beta+\gamma)\Gamma(\alpha+\delta)\Gamma(\beta+\delta)}{\Gamma(\alpha+\beta+\gamma+\delta)}.
\end{equation}
Thus, we have the result given in Eq.~\eqref{eq:conductivity-bm}:
\begin{equation}
I=[\Gamma(1/2)]^{4}/\Gamma(1)=\pi^{2}.
\end{equation}

\bibliography{References}